\documentclass[manuscript,screen,nonacm=true]{acmart}
\usepackage{lscape}

\AtBeginDocument{%
  \providecommand\BibTeX{{%
    \normalfont B\kern-0.5em{\scshape i\kern-0.25em b}\kern-0.8em\TeX}}}

\setcopyright{none}
\begin{document}

\title{A Bayesian Agent-Based Framework for Argument Exchange Across Networks}


\author{Leon Assaad}
\authornote{Order of authors bar last author is alphabetical}
\email{leon.assaad@googlemail.com}
\affiliation{%
  \institution{Munich Center for Mathematical Philosophy, LMU}
  \streetaddress{Geschwister Scholl Platz 1}
  \city{Munich}
  \state{Bavaria}
  \country{Germany}
  \postcode{80539}
}

\author{Rafael Fuchs}
\authornotemark[1]
\affiliation{%
  \institution{Graduate School of Systemic Neuroscience, LMU}
  \streetaddress{Großhadernerstraße 2 }
  \city{Planegg-Martinsried}
  \state{Bavaria}
  \country{Germany}
  \postcode{82152}
}

\author{Ammar Jalalimanesh}
\authornotemark[1]
\author{Kirsty Phillips}
\authornotemark[1]
\affiliation{%
  \institution{Birkbeck College, University of London}
  \streetaddress{Malet Street}
  \city{London}
  \country{United Kingdom}
  \postcode{WC1E 7HX}
  }

\author{Leon Sch\"oppl}
\authornotemark[1]
\affiliation{%
 \institution{Munich Center for Mathematical Philosophy, LMU}
\streetaddress{Geschwister Scholl Platz 1}
  \city{Munich}
  \state{Bavaria}
  \country{Germany}}

\author{Ulrike Hahn}
\affiliation{%
\institution{Birkbeck College, University of London}
  \streetaddress{Malet Street}
  \city{London}
  \country{United Kingdom}
  \institution{and Munich Center for Mathematical Philosophy, LMU}
\streetaddress{Geschwister Scholl Platz 1}
  \city{Munich}
  \state{Bavaria}
  \country{Germany}
  }

\renewcommand{\shortauthors}{Assaad et al.}

\begin{abstract}
 In this paper, we introduce a new framework for modelling the exchange of multiple arguments across agents in a social network. To date, most modelling work concerned with opinion dynamics, testimony, or communication across social networks has involved only the simulated exchange of a single opinion or single claim. By contrast, real-world debate involves the provision of numerous individual arguments relevant to such an opinion. This may include arguments both for and against, and arguments varying in strength. This prompts the need for appropriate aggregation rules for combining diverse evidence as well as rules for communication. Here, we draw on the Bayesian framework to create an agent-based modelling environment that allows the study of belief dynamics across complex domains characterised by Bayesian Networks. Initial case studies illustrate the scope of the framework.
\end{abstract}


\received{14 Nov, 2023}


\maketitle

\section{Introduction}

Human societies are based on information exchange, deliberation, and negotiation. This means human societies rely fundamentally on argumentation. As a result, argumentation --broadly construed-- is a topic of active research interest across a wide range of disciplines. 

Some of these, such as research on argumentation \cite{walton2009argumentation} and research on persuasion \cite{maio2018psychology} have tended to focus on detailed characteristics of individual arguments. Others, such as research in computational social science  \cite{lazer2009computational} studying large-scale debates across online social platforms like Twitter or Facebook, have focussed in detail on the spread of arguments \cite{hofman2021integrating}--with the arguments themselves subjected to far more coarse-grained analysis in terms of keywords or sentiments (e.g., \cite{berger2012makes}). Research focussed on spread includes also agent-based modelling of belief- or opinion dynamics. Here, `arguments' have been highly stylised -represented only by numbers or elements of a vector \cite{mas2013differentiation}. Or they have been captured only implicitly by their effects, as in approaches that model opinion dynamics as a kind of `contagion'. 

This disconnect between research traditions focussed on individual arguments, and research traditions focussed on dynamics of spread has left a fundamental gap in the current understanding of how the exchange of arguments figures in human society. Moreover, this gap encompasses pressing theoretical and practical questions, for example concerning the impact of large online platforms on democratic debate, and with it, the health of democratic societies.  

Bridging that gap will, arguably, require bringing together tools and theories of research traditions that have focussed on individual arguments with those concerned with characteristics of spread. For example, both argumentation and persuasion research have historically focussed on dyads: communicating pairs exchanging reasons for claims in an ongoing exchange that involves competing and supporting arguments that combine in complex, often hierarchically nested ways. Researchers have sought to understand both argument `quality' and persuasive `success' in that dyadic frame of reference, developing both procedural rules for engagement \cite{van2003development,van2013fundamentals,van2015argumentation} and graphing techniques or ‘maps’ to aid argument evaluation and production \cite{gordon2007carneades}. This scales only partly to contexts with multiple, and possibly large numbers of, communicating agents (see also \cite{bonevac2003pragma,lewinski2014argumentative}) and the historical focus on dyads has left fundamental questions about argumentation and persuasion unaddressed. Conversely, the insights that might be gained from analysing the dynamics of the spread of arguments across large corpora of public debate are restricted by the level of content analysis applied to the arguments themselves. 

The research presented in this paper aims to help bridge this gap. Specifically, we introduce a new framework, NormAN --short for Normative Argument Exchange across Networks-- for agent-based modelling of argument exchange across social networks. This framework, we argue, combines important features of argumentation research on argument evaluation with agent-based modelling. Specifically, its goal is to incorporate multiple fundamental features of real-world communication: In discussing or debating a claim, individuals exchange arguments (individual reasons) for believing that claim to be true or false. Some of these may be arguments for, others arguments against, and some of these arguments may be better than others. And while some of this may be a matter of subjective evaluation, what arguments are available and how strong they are is also constrained by the topic at hand. Finally, communicative exchanges might take place in anything from small, tightly knit groups exchanging in-depth information, to large networks involving only fleeting exchange. This means understanding real-world arguments also requires understanding the impact of who agents communicate with and what they choose to exchange. 

No single model, let alone a single investigation, will be able to give equal focus to each of these fundamental features of real-world argumentation. As a result, NormAN is designed as a framework. In this paper, we set out this framework and introduce a basic model, NormAN version 1.0, that incorporates each of these core features. Specifically, the paper proceeds in three main parts. We first (\ref{section:Motivation}) briefly describe research across both sides of the `gap' in order to situate NormAN in the context of current research and motivate our specific design choices. We then introduce the framework and NormAN v. 1.0 in more detail (\ref{section:Introducing}). Finally, we describe two case studies to illustrate the features and benefits of the framework. In particular, we seek to show how this framework --though still highly stylised in nature-- affords both deeper understanding of longstanding questions, and opens up new avenues for research. 

\section{Motivation and Background}
\label{section:Motivation}
The goal of NormAN is to bring together the detail of research on argument evaluation and the simulation of multi-agent contexts in order to bridge the current gap and enrich both research traditions. We next provide background on the most important strands within these respective traditions. 

\subsection{Argument Quality and Dialectics}
\label{section:arg_dialectics}

Traditional research on argumentation has not used agent-based simulations. Rather, this highly interdisciplinary field has drawn on observation, formal analysis, and behavioural experiments.

\subsubsection{The Breadth of Argumentation Research}

 Philosophers have focussed on normative theories, that is, theories of 
how we \emph{should} behave. The traditional standard has been formal logic, but more recently, pragma-dialectical theories have focussed on the norms and conventions governing argumentative process (e.g., \cite{van2013fundamentals,van2015argumentation,walton1998new,walton2007informal}). Within psychology, `persuasion' has been a 
central topic of social psychological research (e.g., \cite{eagly1993psychology}). This  vast literature has identified many  moderating variables (e.g., speaker likeability, engagement, mode of presentation, fit with prior beliefs) that affect the degree to which persuasive communication will be effective. Developmental and education research has focussed on the way children’s argumentation skills develop and examined ways in which critical thinking and argument skills might be improved 
(e.g., \cite{felton2001development,kuhn2003development,von2008arguing}). Logicians and computer 
scientists have sought to devise argumentation frameworks for dealing with dialectical information, seeking 
to capture the structural relationships between multiple theses, rebuttals, and supporting arguments for use in computational argumentation system \cite{dung1995acceptability,prakken2001logics,rahwan2009argumentation}.

\subsubsection{The Central Role of Normative Concerns}
\label{section:indnorm}
The sheer breadth of disciplinary perspectives, research questions, and methods  makes for a bewildering array of literatures and findings on argumentation. Furthermore, many of these literatures have historically been largely or even wholly disconnected from one another. There is, however, a shared focal concern across most, if not all, argument research. This is the issue of argument quality or `what makes a good argument?', and, with that, the question of how good arguments can be distinguished from bad ones. This question is a normative, evaluative, question about what kinds of arguments should convince us, and which are the appropriate normative standards against which argument quality should be judged. Across fields and research interests, this question features both as an explicit topic of study and as an implicit concern.

It is of explicit interest within philosophy in research on human rationality and the epistemological question of
how we can arrive at secure knowledge of the world (\cite{rescher1977dialectics,dawid2015no,hartmann2021,eva2018bayesian,godden2018probabilistic}. In psychology,  cognitive psychologists study the quality of people’s argumentation (e.g., \cite{kuhn1991skills} as part of a long tradition of research on reasoning, judgment 
and decision-making (e.g., \cite{oaksford2009precis,stanovich2000individual,kahneman2011thinking,hahn2012rational}). And educational psychologists teaching or improving argument skills and critical thinking (e.g., \cite{van2003development}) must clarify their intended target. 

In other research, the question of argument quality is raised \emph{implicitly} by research goals and methodological constraints. For example, argument quality matters for logicians and computer scientists interested in argumentation as a tool for 
artificial 
intelligence systems  (e.g., \cite{jackson1986introduction,neapolitan1990probabilistic}), because, to work well,  such systems must adequately weigh and 
aggregate information. So how can argument quality be measured? What normative standards might be devised?

\subsubsection{Standards for Argument Quality}

 A wide range of tools, from different disciplines, has historically been applied to  the question of what makes a `good' argument: 
\begin{enumerate}
\item classical logic 
\item attempts to elucidate argument by mapping out structural relations between arguments:
\begin{itemize}
\item either informally by tagging them as ‘claims’, ‘warrants’, or ‘rebuttals’ (e.g., \cite{toulmin2003uses}) 
\item or in formal, computational frameworks (e.g., so-called ‘argumentation frameworks’, \cite{dung1995acceptability})
\end{itemize}
\item pragma-dialectical theories spelling out putative norms underlying argumentative discourse, such as 
a ‘right to reply’ or ‘burdens of proof’ \cite{van2004systematic}
\end{enumerate}
While all of these are useful and aid interesting research questions in different fields, they still miss much about argument quality.

Classical logic says nothing about most everyday informal arguments, other than that they are not logically valid \cite{toulmin2003uses,hamblin1970fallacies}, and, hence, it is too restrictive. \footnote{At the same time, it is too permissive in that it renders arguments strong that actually seem poor: For example, `A, therefore, B or not B', where  A is wholly irrelevant rather than providing a meaningful reason.} 
Likewise,  the quality of argument content cannot generally be reduced to procedural rules or to systems that map out support, attack and defeat relations. To illustrate: ``the book is in the library…no it’s not, because the moon is made of cheese'' 
involves an (intended) counter-argument, but is patently absurd \cite{hahn2020argument}. Simply noting that an argument is 
\emph{offered} as support or attack is itself a purely structural, syntactic evaluation. A content-based measure of argument strength is still needed in order to know whether intended ‘support’ or ‘defeat’ is successful. Likewise, pragma-dialectic notions such as `burden of proof' depend on the specific content of an argument in order to determine whether or not a burden of proof has actually been met \cite{hahn2007rationality}. 

This means that normative standards in addition to classical logic, procedural rules or merely syntactic renditions of the structural relations between arguments are necessary in order to capture argument content adequately. This has recommended a Bayesian approach to argumentation.

\subsubsection{Bayesian Argumentation}
\label{section:bayesarg}

The probability calculus is intensional  \cite{pearl1988}: the probabilities that attach to propositions are determined by their specific content, not (just) their logical form. The resulting ability of the probability calculus (or, where decisions and utilities are involved, Bayesian decision theory) to meaningfully capture normative questions about argument content is demonstrated by its application to the catalogue of so-called fallacies of argumentation. The fallacies are argument schemes such as `arguments from ignorance', `ad hominem arguments' etc. that have long posed a challenge for explanations of why exactly they are poor arguments  (see \cite{woodsirvine,hamblin1970fallacies}). One central difficulty encountered here was that not all instances of these schemes seem equally poor or fallacious, and a Bayesian treatment has elucidated those differences \cite{hahn2020argument}. 

The  Bayesian framework has also been applied to a broader set of schemes for everyday argument from the informal logic literature that, unlike the fallacies, are presumptively (but defeasibly) `good' arguments. Specifically, they provide reasonable, albeit defeasible, inferences for uncertain, ampliative reasoning (which 
sets them apart from logical schemes such as the classic set of syllogism or conditional reasoning schemes 
such as modus ponens, modus tollens etc.). The literature on informal argument previously catalogued 60+ such schemes \cite{walton2008argumentation} that identify recurring structures that occur with varying 
content (and hence varying strength) in everyday discourse.  As  \citep{hahn2016normative} seeks to show, the Bayesian framework can provide a normative basis for these schemes. It can thus further the long-standing goals of research on argument schemes, namely a computationally explicit treatment with guidance for lay reasoners on when particular instances of these schemes are weak or strong.

In the Bayesian framework, argument strength can be captured by considering the extent to which an argument or piece of evidence rationally changes one's beliefs. The 
posterior, $P(C|A)$, is affected by the likelihood (i.e., the sensitivity of the evidential test $P(A|C)$), and by the 
false positive rate (i.e., $P(A|notC)$) as captured in the likelihood ratio (i.e., $P(A|C)/P(A|notC)$). 

With the likelihood ratio, the Bayesian framework has a notion of informational relevance. This helps  
with the fallacies, given that fallacies are typically fallacies of relevance \cite{walton2004relevance}. It is also essential 
to capturing argument quality in general, and  elucidating 
the notion of relevance in a formally satisfactory, non-question-begging, way has been a long-standing challenge (see, \cite{sperber1986relevance,hahn2006bayesian}). 
Finally,  the Bayesian framework has a well-developed normative foundation that links to goals such as inaccuracy 
minimisation (on the link between `being Bayesian' and inaccuracy minimisation see \cite{pettigrew2016}, for discussion of normative foundations for argumentation more generally, see \cite{corner2013normative}).

Bayesian Argumentation has also been expanded to other features of argument (e.g., such as `argument cogency' or `degrees of justification' \cite{zenker2012bayesian,godden2018probabilistic}). At the same time, work by Hartmann and colleagues has extended the formal arsenal of Bayesian Argumentation in order to broaden the scope of possible inferences \cite{eva2018bayesian,eva2020learning} and has provided detailed treatments of scientific inference schemes (e.g., \cite{dawid2015no}) in a programme paralleling the treatment of everyday schemes. Specifically, Hartmann and colleagues have shown how forms of new ‘evidence’ not amenable to Bayesian conditionalization may be captured through the application of Kullback-Leibler divergence.

The body of work on Bayesian Argumentation arguably represents the state of the art with respect to measuring argument quality, in that both a quantitative measure and a well-developed normative basis is provided (see also \cite{nussbaum2011argumentation}). It is for this reason that we adopt the  Bayesian framework for NormAN.

While a Bayesian perspective on argument quality has arguably been productive, there are key features of argumentation --discussed next-- that have been neglected to date, not just by Bayesian argumentation, but by argumentation research as a whole.

\subsection{Beyond Dyads: Multi-Agent Models}

As noted above, most work on argumentation has, at best, concerned itself with dyads, that is, a proponent and an opponent engaged in direct exchange.
Public debate, however,  has many actors choosing when to contribute and when not, which arguments to repeat, which to ignore, which to address, and how. This fundamental feature of real-world argument has remained largely outside the view of argumentation research.

Even where argumentation research has concerned itself with large-scale debates, it has either attempted to assimilate these into dialogue-oriented models \cite{lewinski2014argumentative} or it has focussed exclusively on the arguments themselves (e.g., in argument mapping approaches to large-scale debates such as Kialo\footnote{See \hyperlink{https://www.kialo.com}{https://www.kialo.com}}; on such tools more generally see e.g., \cite{benetos2023digital}). This obscures all sense of the \emph{dynamics} of argument exchange in public debate and the many underlying decisions by participants that give rise to those dynamics. 
 
The dynamics of debate, however, have become a matter of research interest with the advent of online social media. With online social media came the ability to harvest and analyse large volumes of real-world debate. And the rise of computational social science has seen the analysis of online data from large platforms such as Twitter, Facebook or Reddit become a major topic of research \cite{cioffi2014introduction,lazer2009computational}. At the same time, putative negative impacts of large online platforms on misinformation, polarization, extremism, and a weakening of democratic norms \cite{lorenz2020behavioural,lewandowsky2020technology,lorenz2023systematic} has fuelled interest in belief and opinion formation across social networks.  
This has led to a wealth of modelling research to help understand how opinions spread across networks.

There remains, however, a significant gap:  the analysis of real world data from platforms such as Twitter has largely focussed on limited features of such data, focussing either on spread by analysing retweets \cite{java2007we,suh2010want,ten2014modelling,cha2010measuring} or analysing content in very restricted ways such as sentiment analysis \cite{hutto2014vader}, bags of words \cite{naveed2011bad,brady2017emotion,storey2022text} and/or topic modelling \cite{zhao2011comparing,corti2022social} (but see also more recently e.g., \cite{visser2020reason}). This is a far cry from the detailed analyses of content common within the research tradition concerned with argument quality outlined in the previous section. And, as the following sections will show,  models of belief or opinion-dynamics are arguably even more restrictive: At present, most ABMs do not involve the communication of reasons for claims. In other words, they do not capture argument at all.

\subsubsection{Models of Opinion Dynamics}
\label{section:opiniondynamics}

The modelling of opinion dynamics has seen multiple frameworks. Two of these import concepts from other disciplines: contagion and social physics models. Contagion-based models, in effect, treat the spread of opinions or behaviours as a kind of ``infection'' \cite{lopez2008diffusion,barash2011dynamics,centola2018behavior}. This allows models from epidemiology to be applied. Contagion based models have been used to examine everything from basic contagion dynamics \cite{barash2011dynamics,izquierdo2018mixing}, effects of network structure \cite{jackson2007relating,lopez2008diffusion}, the influence of word of mouth reports of events on elections \cite{moya2017agent},  extremism \cite{youngblood2020extremist}, to echo chambers and viral misinformation \cite{tornberg2018echo}. Methods have ranged from standard analytic models within epidemiology (see e.g., \cite{kiss2017mathematics}), through statistical models, to agent-based modelling.

The social physics approach draws on models developed within physics to capture opinion dynamics \cite{10.1103/revmodphys.81.591}. In particular, methods (e.g., mean field approximation) and models from statistical mechanics, such as diffusion models and the Ising model \cite{dorogovtsev2008critical}, have been used to model issues such as opinion polarization \cite{macy2003polarization} or the spread of misinformation \cite{budak2011limiting}.

Finally, one of the earliest and one of the most influential models of opinion dynamics was first put forward by statistician M. DeGroot \cite{degroot1974reaching}. The DeGroot model was proposed initially to shed light on how groups might use opinion pooling to reach a consensus judgement. It is based on repeated (weighted) belief averaging until beliefs converge. Iterated averaging also underlies the Lehrer-Wagner \cite{lehrer1981rational} model in philosophy that has been used extensively to develop notions of rational consensus, and the work of Hegselmann and Krause (e.g.,\cite{hegselmann2002opinion}), which we discuss further in the next section.

 With respect to motivating NormAN, we note two main observations about  models of belief- and opinion dynamics discussed so far:
First, they use an unanalysed aggregate --the opinion or belief in question, typically represented as a boolean or continuous variable-- without the provision of reasons;  this limits the research focus of such models to the population dynamics regarding that single quantity. This renders this body of work impossible to connect meaningfully to the research on argumentation described in section \ref{section:arg_dialectics} above. Second, there is no `ground truth' at stake in these models (but for an addition of `ground truth' to the DeGroot model see
Golub and Jackson \cite{golub2010naive}). Hence many questions about knowledge and accuracy, of either individual agents or the collective as a whole \cite{hahn2022collectives}, are necessarily outside the scope of these models.

\subsubsection{Agent Based Models in Social Epistemology}
\label{section:socepistemology}

Questions of how knowledge comes about and how it comes about specifically in social contexts are, however, the central concern of social epistemology.  

Considerable research within social epistemology has utilised the DeGroot model---either in the form of the Lehrer-Wagner (1981) or the Hegselman-Krause model \cite{hegselmann2002opinion,hegselmann2015opinion}. Hegselman-Krause added  the idea of weights reflecting differential `trust' in other members of the collective such that agents only  listen to others who are sufficiently `close' in their estimates (giving rise to so-called convergence threshold models \citep{hegselmann2015opinion}). Work using these models has focussed on understanding when networks do and do not converge (see for extensive analysis, \cite{krause2015positive}).

In order to connect better to the concerns of social epistemology, Hegselman and Krause \cite{hegselmann2006truth} later also added to their model the idea that a specific opinion value may be designated as `the truth' (for other extensions see \cite{douven2010extending}, including, outside of social epistemology, toward greater psychological realism \cite{9900065}; for a review, \cite{douven2019computational}). 

An influential further class of models in social epistemology is bandit models \citep{zollman2007communication,zollman2010epistemic}. These models use one- or multi-armed bandits \cite{slivkins2019introduction} to generate evidence about an underlying state of the world. That evidence may be observed directly by individual agents or indirectly through observation of other agents' states based on aggregates of that evidence, or received via communication. Used initially by economists to study social learning across networks and its effect on economic outcomes \cite{bala1998learning,bala2000strategic}, bandit-based models have been applied in social epistemology to questions such as  when (and when not!) communication is beneficial to the progress of science \cite{zollman2010epistemic}, the effects of misinformation \cite{o2018scientific}, and polarization \cite{o2018scientific}. Although they often involve Bayesian updating at the individual agent level, bandit models have also been combined with DeGroot-like averaging \cite{douven2019optimizing}. Conceptually, bandit models allow there to be a model ground truth, and the evidence dispensed by the bandit provides at least a very limited notion of `argument'.

A different model aimed at many of the same research questions is the model of testimony first proposed by Olsson and colleagues \cite{olsson2011simulation,OlssonVallinder2013Norms,olsson2013bayesian,angere2017publish}. The realisation that much of what humans believe to know stems from  the testimony of others \cite{coady1992testimony}, has fuelled research concerned with the conditions under which testimony is reliable and a meaningful guide to truth. A significant proportion of that work has drawn on probability theory to explicate those conditions in formal models \cite{olsson2005against,bovens2003bayesian,olsson2007reliability} including agent-based simulations. The Olsson (2011) model is such an agent-based model.

As it has inspired many of the features of the new framework presented in this paper, we outline it in some detail here. In the model, there is a single proposition (represented as a Boolean variable) at issue. Agents in the model each occupy a particular position in a social network. At each time step, there is a probability of acquiring a piece of evidence `from the world', and a probability of communication. Communication links are symmetrical, and communicating agents will affirm that proposition C, the target hypothesis under scrutiny in this artificial society, is true, whenever their degree of belief in C exceeds a threshold of assertion (say, p = .8). When belief drops below 1 minus the threshold of assertion, agents will assert not-C; on all other occasions they will remain silent. This is designed to capture the fact that real-world communication does not typically involve communication of point values, but rather involves the assertion of claims (`C is true'). The agents in the model are Bayesian, using Bayes' rule to revise both belief in the target hypothesis and the reliability of their sources (including their own inquiry into evidence from the world).\footnote{More precisely they are naive Bayesian agents in that they make the simplifying assumption that evidence is independent, \cite{ng2002discriminative}. For an examination of the consequences of this assumption see \cite{merdes2021formal,HahnPPS}.}

In the Olsson model, agents have a belief (in the claim at issue) and there is a ground truth. There is also a very simple type of `argument' or evidence which consists solely of assertion that the claim in question is true or false.

The Olsson model has been used, among other things, to study the impact of network topology on accuracy \cite{HahnHansenOlsson2018truth}, polarization \cite{olsson2013bayesian,olsson2020bayesian,pallavicini2021polarization}, the impacts of different strategies for estimating the reliability of testimonial sources \cite{hahn2018good,collins2018bi} and the dependence created through communication \cite{hahn2018communication}. This includes the role of communication in the context of science, specifically asking whether the overall progress of science is helped or hindered by frequent communication between scientists \cite{angere2017publish}.

This latter question has also been studied in so-called epistemic landscape models \cite{weisberg2009epistemic,pinto2018epistemic,grim2013scientific}. 
These models capture scientific exploration of a topic by agent-based probing of a fitness landscape: 
The boundaries of the landscape represent the boundaries of the 
topic; the coordinates of the landscape correspond to different approaches scientists could be bringing to its study and
the topography of the landscape represents the relative `significance' of the resultant scientific work. 

The recent argumentation-based ABM of \citep{BorgFreySeseljaStrasser019theory} represents yet a further attempt to study the same problem. In this model, agents seek to explore an underlying argument map setting out the relevant arguments on the issue. In effect, the `epistemic landscape' is the argument map. This argument map is formulated in the abstract argumentation framework of \cite{dung1995acceptability}  and agents exchange information about arguments they have encountered. 
This allows investigation of the impact of different argument selection and communication strategies by agents with respect to achieving full end-state knowledge of the relevant state of affairs. 

In both epistemic landscape models and the Borg et al. argumentation-based ABM, there is a `ground truth' of sorts implicit in the model. However,  the design of the underlying `landscape' (whether the fitness landscape or the argument map) is essentially arbitrary and unconstrained. To the extent that properties of that landscape matter to model behaviour and findings, results will remain somewhat difficult to interpret --in particular, with respect to how they generalise to real-world situations. At the same time, however, there is no explicit representation of agent beliefs regarding a particular target hypothesis, which separates these models fundamentally from models of opinion dynamics. 

To fully join dyadic models of argument with opinion- and belief dynamics, a modelling framework that distinguishes between arguments, aggregate beliefs or opinions, and the social network across which communication is happening is required (for related points see also \cite{grim2013scientific}). We return to these issues below.

\subsubsection{Multi-Agent Models of Argumentation}

Finally, the exchange of arguments between computational agents has been a focal point for research on multi-agent-systems (for introduction and overviews to multi-agent-systems see e.g., \cite{dorri2018multi,van2008multi}). Much of the modelling here has involved logical formalisms of one kind or another \cite{calegari2021logic,chesnevar2000logical}, though other argumentation frameworks such as Dung's abstract argumentation framework \cite{dung1991negations} and extensions thereof (e.g., \cite{bench2002value})  have also been used (see for an overview of relevant approaches \cite{rahwan2003argumentation,carrera2015systematic}). And there has been some (but comparatively limited) interest in the Bayesian framework (e.g., \cite{saha2004bayes,nielsen2007application,vreeswijk2004argumentation}). 

Both the tools used for capturing argument and some of the research questions asked have connections to traditional (dyad focussed) argumentation research as described in section \ref{section:arg_dialectics}. Not only are there researchers that have contributed to both communities, input has specifically been sought from non-computational argumentation researchers (see for illustration of this point e.g., \cite{rahwan2009argumentation}). However, the different focus of most research on argumentation for autonomous agents means that this body of research does not ultimately connect well to research on belief- or opinion dynamics, or to research concerned with the spread of arguments (at least at present). This stems from the fact that multi-agent-systems research typically has in mind practical applications for which collections of agents provide a potential computational solution. This makes a top-down approach to systems involving many agents the natural focus. The goal is to solve complex computing problems, and multi-agent systems, as a type of distributed artificial intelligence, provide a potential tool. By contrast,  most of the research discussed thus far is interested in the bottom-up question of dynamics and patterns that emerge naturally from groups of interacting agents.

\subsection{The Value of Normative Models}
\label{section:normative}

The preceding discussion should have made clear that there is a particular value to normative models in the context of argumentation research.
In the context of individual-focused, or (at best) dialectical research on argumentation the explicit and implicit normative focus is clear (section \ref{section:indnorm} above). Not only have normative issues been of direct, explicit, interest, but normative models have methodological value even in otherwise purely descriptive endeavours.  

Specifically, the new (to argumentation research) normative standard provided by Bayesian probability  not only addressed long-standing theoretical, philosophical concerns (e.g., about fallacies of argumentation), it also opened up novel paths for empirical, behavioural experimentation examining laypeople’s reasoning (e.g., \cite{corner2011psychological,bhatia2015discounting,corner2009evaluating,harris2012because,hornikx2018many}). And the specificity of research questions pursued in those studies goes considerably beyond what was possible with normatively limited frameworks such as the Toulmin framework (see also \cite{hahn2023argument} for further discussion of this point).

In the context of multi-agent models, the importance of normative concerns should also be clear. Again, there is considerable interest in normative accounts of collective discussion and information exchange in social epistemology. Likewise, there is considerable interest, for example, in improving online platforms, and for such endeavour an understanding of what is possible, in the best case, is important. Finally, normative models are important in making sense of descriptive data which all too often are simply assumed to reflect bias and irrationality whenever data patterns seem surprising (as illustrated, by the literature on shifts to extremity in group discussions or polarization, both examined in more detail below, \ref{section:case1} and \ref{section:case2}). 

At present, however, there is a large gap in that there are no normative models of argument exchange across collectives that would allow researchers to address issues such as accuracy and link up to classic, individual- and dyad-focussed research on argument.

Crucially, in order to achieve that, two components of the model need to have a normative grounding:
argument evaluation and aggregation  on the one hand, and, on the other, a grounding of the evidence distribution in a ground truth world model, against which beliefs can be compared and scored. 

Both the evaluation/aggregation rules used by agents and the distribution of (in principle) available evidence will affect belief dynamics. Consequently, making both of these aspects principled (and not `just so') seems of fundamental importance for connecting to real-world contexts in meaningful ways.

This gives rise to the following desiderata for an agent-based model of argument exchange. What is required is a model with (i) a ground truth world, (ii) evidence linked to that ground truth world, giving rise to a principled evidence distribution, and (iii) rational agents who form optimal beliefs given the evidence they have. Furthermore, such a model should be easy to use and extend. NormAN seeks to provide a general framework for creating just such models. 

\section{Introducing the NormAN Framework}
\label{section:Introducing}

\begin{figure}[h]

  \centering
  \includegraphics[width=0.8\linewidth]{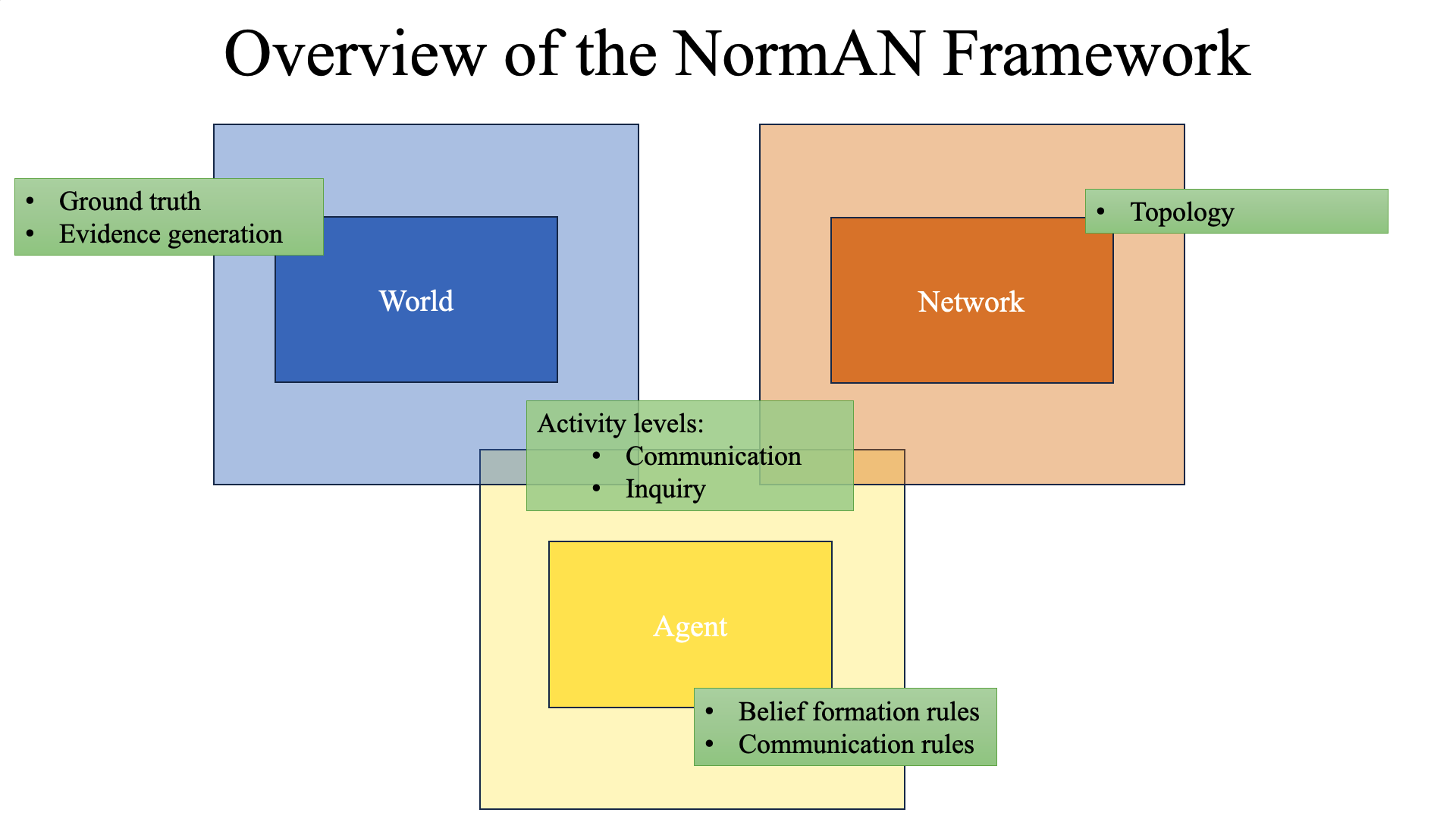}
  \caption{Illustration of the main components of NormAN. A model within the NormAN framework specifies (1) a ground truth world, (2) a social network, and (3) individual agents communicating across that network. Each of these three components has core specifications. In addition, the square in the middle ``activity levels'' refers to aspects that could variously be conceived of as properties of world, network or agent. The assignment of model parameters in the current version of NormAN (1.0) to these various model aspects is described in the text.}
  \Description{This is an overview of NormAN. It shows the three core components "world" (blue box), "network" (red box), and "agent" (yellow box) along with the core features/specifications of these components (green boxes).}
  \label{fig:overview}
\end{figure}

The core conceptual components of the NormAN framework are illustrated in Figure \ref{fig:overview}. It comprises the three core elements that make up a NormAN model: a ground truth’ `world', individual `agents', and the social `network' across which these agents communicate. The ground truth world determines the true state of the claim (hypothesis) at issue in the discussion, along with the evidence for it that could be discovered in principle. Agents receive evidence about that world (through inquiry) and may communicate that evidence to others as arguments and receive it in turn.\footnote{The framework of Bayesian argumentation elides the difference between evidence and arguments in as much as it models arguments with the same basic machinery used to model other evidence (though substantive differences may, of course, arise as a result of that formalisation). For clarity, it can be helpful to restrict the term `evidence' for information received in the model `from the world' (whether through initial assignment or subsequent inquiry), and `argument' for communication of that evidence.} Agents aggregate all evidence/arguments that they have encountered to form a degree of belief in the claim at issue. Communication, finally, takes place across a social network (including a possible `null network' in which no communication takes place for comparison). 
 
NormAN sets out a general framework in as much as each of these core components is inherently modifiable: users can modify the `world', the evidence received, aggregation rules, communication rules, and network topology. Moreover, NormAN is designed this way because it is our contention that all three components are essential in coming to a deep understanding of argumentation, communication, opinion dynamics and deliberation. As a consequence of this foundational assumption, even the initial release of NormAN (version 1.0) already has sufficient flexibility to allow users to define models selected from a broad space of models varying world, agent and network characteristics. 

Moreover, NormAN is freely available (see \ref{section:online}) so as to allow users to readily adapt and extend the code and create new models by modifying the core components. 

As outlined above (see \ref{section:normative}),  the absence of agent-based models capturing argument exchange over a ground truth world is central to the current `gap'.  A key problem in developing NormAN was thus how to build a ground truth world. Extant models such as the  Olsson (2011) model or  bandit models (e.g., \cite{zollman2007communication} )
utilise a simple binomial process to this end. The modeller stipulates a particular `hypothesis' (say `1') to be true, and a binomial process with probability $p$ --representing the accuracy of the evidence source\footnote{Assuming $p(E|H) = p(E|notH)$, i.e.,  sensitivity and specificity of the `test' are set to be equal, see also \cite{hahn2018good} for discussion of the extent to which this does and does not limit generalisability of model results.}-- produces a stream of 0s and 1s as `evidence' which can then form the basis of Bayesian updating on the part of the agents. 

Rich argument exchange over a ground truth world requires extending this kind of approach in appropriate ways. This means not just generating an evidence distribution that is plausible vis a vis the real world, but also generating evidence in such a way as to allow agents to form probabilistically coherent beliefs (`be Bayesian') \emph{at least in principle}. While one's analytic interests need by no means be limited to optimal agents (ours are not), the mere possibility of implementing a rational agent exhibiting normatively correct reasoning (or as close as possible to that) means one cannot simply generate wholly unconstrained, arbitrary combinations of  evidence, because doing so may potentially generate evidence combinations over which no rational belief is possible.    

To satisfy the dual demands of a plausible world and meaningful evidence distribution, NormAN adopts Bayesian Belief Networks (BNs) as a tool for generating the world. BNs are graphical representations of multi-variable relationships \cite{pearl1988,pearl2000causality,korb2010bayesian,scutari2021bayesian}. They are widely used across many disciplines (ranging from computer science, through statistics, philosophy, psychology and sociology, among others) for both theoretical analysis and practical software and engineering applications (e.g., \cite{kammouh2020probabilistic}). Specifically, BNs summarise the way variables do and, more importantly, do not influence one another in a graphical representation that simplifies Bayesian calculations. BNs thus have a normative (Bayesian) foundation and they connect to extant work on argumentation (see section \ref{section:indnorm} above) including Bayesian models of argument generation \cite{zukerman1998bayesian,zukerman1999exploratory,jitnah2000towards,keppens2019explainable,timmer2015explaining}. Furthermore, their use for statistical analysis \cite{salini2009bayesian} and decision-support systems \cite{fenton2018risk} mean that there exist repositories of BNs (e.g., the bnlearn repository, \hyperlink{https://www.bnlearn.com/bnrepository/}{https://www.bnlearn.com/bnrepository/})  that putatively capture real-world structure within the respective application domain. 

NormAN allows users to select a BN and use it to generate a suitable ground truth world through a simple trick. A variable in the network is selected as the target hypothesis or claim at issue; its value is set for the purposes of one or more model runs to represent the 'true state of the world' regarding that claim. The resultant probabilities for the remaining variables (given that true state) are then used to stochastically generate a body of evidence that is available, in principle, in that ground truth world (for fuller details see below). Agents may subsequently receive evidence from that world and exchange what evidence they have received via communication. Conceptually, this generating procedure encompasses the simple binomial processes used in past models as a special case.

Finally, while the use of this generating procedure is an integral part of the appeal or value of NormAN (at least to us), it should be noted that the framework is general enough to allow incorporation of other 'world representations' beyond BNs. Agent-based models defined over arbitrary argument graphs (such as  \cite{BorgFreySeseljaStrasser019theory}, see Section \ref{section:socepistemology}), for example, can readily be captured as a type of NormAN model that uses an argument graph instead of a BN as an underlying world, and in which agents' belief aggregation is disabled.

The key feature of the basic NormAN agents (as implemented in version 1.0)  is that they optimally aggregate evidence via Bayes' rule. To do so, they too, draw on a BN, which in the basic version of the model is a  veridical model of `the world', that is, in essence, a matching (subjective) BN `in the head' (future extensions of the model involve relaxing that requirement of veridical match between world and model in the head and also the subjective models of other agents). Crucially, agents must also communicate and NormAN version 1.0 implements a variety of different rules and constraints on what and when agents communicate. This includes both rules motivated by past work (e.g., \cite{mas2013differentiation}) and initial suggestions for other communication rules in a multi-argument selection context. 

NormAN also allows users to vary the structure of the communication network across which the argument exchange takes place. In the current version, this includes selecting from a range of network types, as well as basic network parameters such as network size and link density (the number of connections between agents). 

Finally, the framework allows modellers to determine the relative balance between evidence `from the world' (both initial and as a result of ongoing inquiry) and the amount of communication. This feature derives from the Olsson (2011) model and is of importance because it has been shown to affect a variety of outcomes from individual and collective accuracy \cite{angere2017publish,hahn2018communication} to polarization \cite{hahnpolarisation,HahnPPS}.

We describe the framework and its current implementation in more detail next. For a full technical description of NormAN 1.0 following the ODD protocol \cite{grimm2010odd}, see the Appendix (Section \ref{section: ODDfull}).


\subsection{Main Components}

The core parameters of NormAN are shown in Table 1. 
We describe the world, the agents, and the networks in turn. 

\begin{table}[H]
  \begin{tabular}{cccl}
    \toprule
    Entity & Variable & Value range/Type & Description\\
    \midrule
    World & \texttt{causal-structure} & Bayesian network & Determines evidence nodes \\
     &&&  and their causal relation  to hypothesis node.  \\
     
     & \texttt{hypothesis}  & Variable & The hypothesis proposition (truth values: \textit{true} or \textit{false}).\\

    & \texttt{hypothesis-probability}  & $0 - 1$  & Probability that hypothesis is true.\\
    & \texttt{evidence-list} & List & Stores truth values of evidence nodes (values: \textit{true}/\textit{false}).\\

   Agents & \texttt{agent-evidence-list} & List  \small (dynamic)\normalsize  & Stores truth values of evidence nodes agents encountered.\\
   & \texttt{agent-belief} & 0 - 1  \small (dynamic)\normalsize  & Belief in the hypothesis. \\
    & \texttt{initial-belief} & 0 - 1  \small (dynamic)\normalsize  & Unconditional prior belief in hypothesis. \\
      & \texttt{update-list} & List \small (dynamic)\normalsize  & Stores  impact.\\
     & \texttt{\texttt{recency-list}} & List \small (dynamic)\normalsize & Governs communication via the recency rule. \\
     
     & \texttt{chattiness} & $0 - 1$ & Probability of communication. \\
     & \texttt{curiosity} & $0 - 1$ & Probability of inquiry (evidence gathering). \\
     & \texttt{conviction-threshold} & $0 - 1$ & Conviction in claim  required to join debate. \\
      & \texttt{max-draws} & Integer &  Determines number of inquiries agents can perform.\\
      & \texttt{share} & Chooser & Determines agents'  communication rule. Values: \\
      
      & &  & \textit{random}, \textit{impact}, \textit{recent}.\\

    Network & \texttt{number-of-agents} & 1 - 1000 & Determines the size of the network.\\
     & \texttt{social-network} & Chooser & Network type: \textit{null}, \textit{complete}, \textit{small-world}, \textit{wheel}.\\
        
  \bottomrule
    \label{tab:CoreParams}
\end{tabular}
\caption{The Core Parameters of NormAN governing world, agents, and social network (for a complete table, cf. Appendix Fig \ref{fig:table-full}).}

\end{table}

\subsubsection{The World}
\label{section:the_world}

\begin{figure}[h]
    \centering
    \includegraphics[width=0.5 \linewidth]{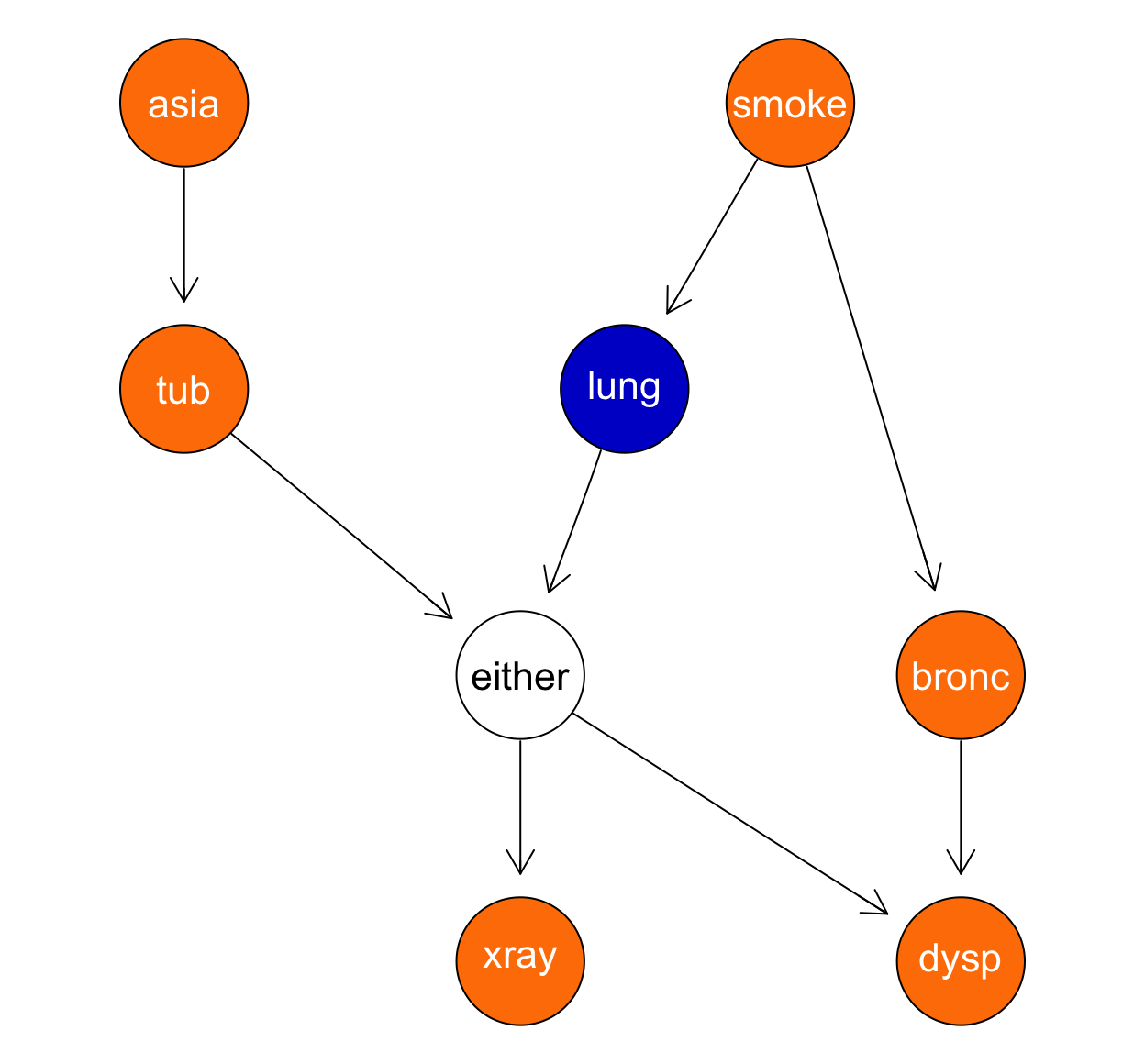}
    \caption{The original `Asia' lung cancer network \cite{Lauritzen1988}. The Asia BN model was accessed via the bnlearn Bayesian Network Repository (\url{https://www.bnlearn.com/bnrepository/discrete-small.html}); it is also one of the exemplar BNs used in the bnlearn package documentation (\url{https://www.bnlearn.com/documentation/bnlearn-manual.pdf}). This BN was constructed using a hypothetical case of qualitative medical knowledge to illustrate the utility of Bayes' rule for expert systems. The target hypothesis is ‘Lung’ (whether or not lung cancer is true, shown as the blue node), and there are seven observable evidence nodes (shown as the orange nodes): Asia (recent visit to Asia); smoking; tuberculosis; bronchitis; dyspnoea (shortness of breath); and, x-ray (chest x-ray). The likelihood of lung cancer is increased when smoking, x-ray, bronchitis and dyspnoea are set to true. Combinations of evidence lead to interactions. The ‘either’ node is a deterministic node that is used in this network to represent the fact that both tuberculosis and lung cancer can result in positive x-ray.  Network properties: Number of nodes = 8, Number of arcs = 8, Number of parameters = 18, Average Markov blanket size: 2.5, Average degree = 2 and Maximum in-degree = 2.}
    
    \label{fig:asianet}
\end{figure}

The world model consists of a BN comprising a set of nodes (variables) and the probabilistic/causal links between them. In general, Bayesian networks  consist of a so-called directed acyclic graph and a matching probability distribution that specifies the conditional probabilities of each variable in accordance with the graph.\footnote{$B = \langle G,P \rangle$ is a directed acyclical graph $G = \langle V,E \rangle$, with a set of nodes (variables) $V$ and edges $E$, and a joint probability distribution $P$ over $V$ such that $G$ satisfies the parental Markov condition together with $P$.} As an example, consider the well-known lung cancer/asia network \cite{lauritzen1988local}, as seen in Fig. \ref{fig:asianet}: a hypothetical network from the medical field that models the causal (probabilistic) relationships between a patient's habits (smoking, visiting Asia) and their symptoms (dyspnoea, positive X-ray).

In such a BN, the modeller identifies a variable as the hypothesis variable (\texttt{hypothesis}), or \texttt{H} for short, (e.g., `lung cancer') and chooses a subset of the other nodes as evidence nodes ($\texttt{E}_1, \texttt{E}_2, \ldots, \texttt{E}_n$). In NormAN 1.0, hypothesis nodes and evidence nodes must be two-valued, that is, they are either \textit{true} or \textit{false}. The model assigns such a truth value to the hypothesis (manually or probabilistically). The following procedure then determines the values of the evidence nodes. The marginal conditional probability of the evidence is calculated; and on initialisation, this chance stochastically determines the truth value of each piece of evidence. For example, if it is true that the patient has lung cancer, and $P(bronchitis|lungcancer)=0.2$, then there is a 20\% chance that the value of the variable \texttt{bronchitis} is \textit{true}. Since the evidence nodes are two-valued, this procedure yields a chain of evidence in the form of $\neg E_1, E_2, E_3, \ldots, \neg E_n$ (where $\neg E_i$ denotes `$\texttt{E}_i$ is false' and $ E_i$ denotes `$\texttt{E}_i$ is true'). This list of indexed truth values is stored in \texttt{evidence-list}.

Crucially, this evidence assignment determines what evidence, on a given run, counts as evidence for or against the hypothesis. While the structure of the BN determines the evidential impact of a piece of evidence (e.g., the degree to which the presence of smoking [`smoke'] increases belief in `lung cancer'), it is the actual value assigned to that variable on a given run which determines the evidence as for or against \emph{in this world}: if the value of `smoke' is initialised to false, it provides evidence against the hypothesis lung cancer as knowledge of that fact will lower degree of belief in `lung cancer' being true.

\begin{figure}[H]
    \centering
    \includegraphics[width=0.8 \linewidth]{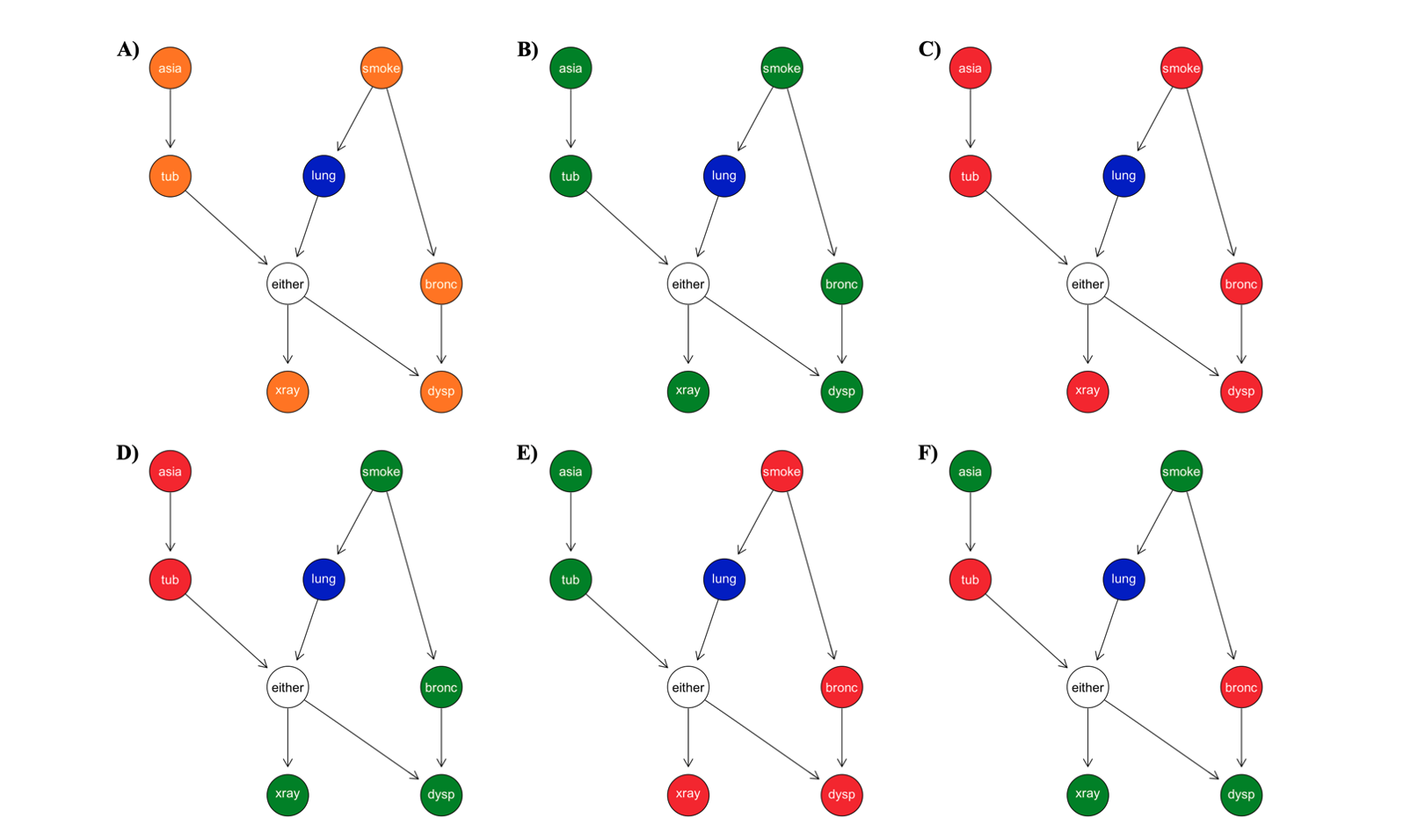}
    \caption{Different instantiations of the `world' defined by the  `Asia' lung cancer network \cite{Lauritzen1988}. On a given model run, the base net (a), can give rise to different `worlds' with varying arguments `for' (green) and `against' (red), depending on the stochastic initialisation.}.
    \label{fig:asianet_manyinitialisations}
\end{figure}

This means also that many possible combinations of evidence for and against will be generated by a single BN world model, see Fig \ref{fig:asianet_manyinitialisations}. And in the user interface (UI) NormAN users can determine not only which BN they wish to work with, but also whether or not the evidence is re-initialised on a given run. 

\subsubsection{The Agents}

Each agent is characterised by (a) their degree of belief in the hypothesis (variable \texttt{agent-belief}), (b) their representation of the causal structure of the world, and (c) a list of evidence they have already encountered.  We go through each feature in turn. First, each agent assigns a degree of belief to the hypothesis (variable \texttt{agent-belief}). Second, they use a BN that connects the evidence to the hypothesis as their representation of the world to compute said belief. Third, they store the truth values of the evidence they have already encountered (variable \texttt{agent-evidence-list}). These three aspects are related in a dynamic, straightforward way. Suppose an agent $A$ stores the following list of evidence at time $t$: $\texttt{agent-evidence-list}_{A}^{t} =\{E_1, \neg E_3\}$. In that case, they will use their Bayesian network to compute $\texttt{agent-belief}_A^{t}=P(H|E_1, \neg E_3)$ by using Bayesian conditionalization. Whenever agents encounter a new piece of evidence (e.g., $E_2$), they update their degree of belief (e.g., $\texttt{agent-belief}_A^{t+1} = P(H|E_1, E_2, \neg E_3)$). When the agent's \texttt{agent-evidence-list} is empty, that is, when they have not yet encountered any evidence, their \texttt{agent-belief} is simply the base rate (marginal probability) of the hypothesis node in their BN. This value is stored in the agent-variable \texttt{initial-belief} as their agnostic, pre-evidence belief in the hypothesis.

In the first version of NormAN presented here, we assume that each agent's BN simply corresponds to the world model's network: that is, we assume that agents represent the world correctly (on relaxing this assumption, see \ref{section:future} below).  This homogeneity of worldviews entails that whenever two agents have access to the same evidence, they also have the same degree of belief in the hypothesis. This assumption can be interpreted as fulfilling the uniqueness standard of rationality, that is, the claim that for any body of evidence $E$ and proposition $P$, $E$ justifies at most one doxastic attitude toward $P$ \cite{feldman2011reasonable, white2019epistemic}. This homogeneity also means that disagreements are entirely the result of asymmetric information. Heterogeneity in beliefs arises because  agents may have access to different sets of evidence.

\subsubsection{The Social Network}

\begin{figure}

         \centering
         \includegraphics[width=0.3\textwidth]{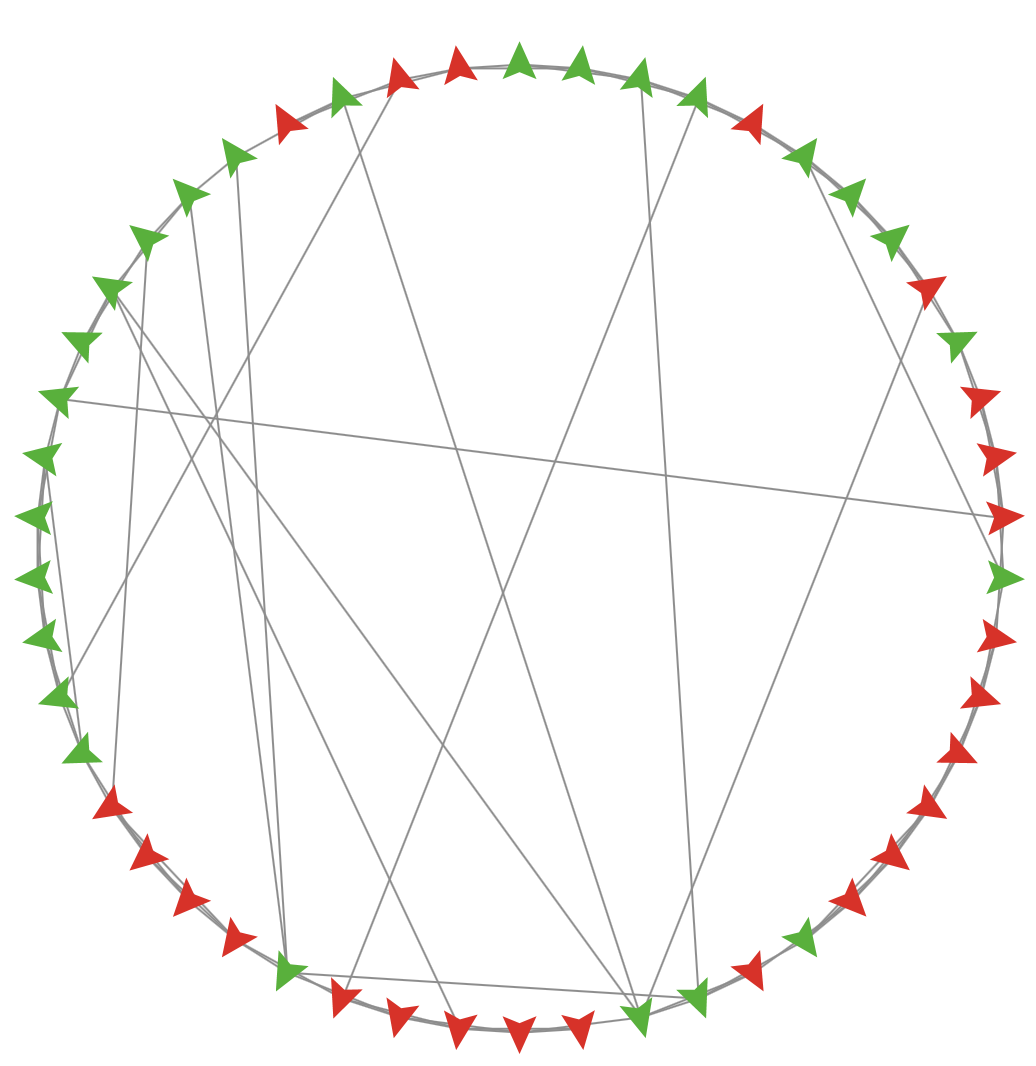}
         \includegraphics[width=0.3\textwidth]{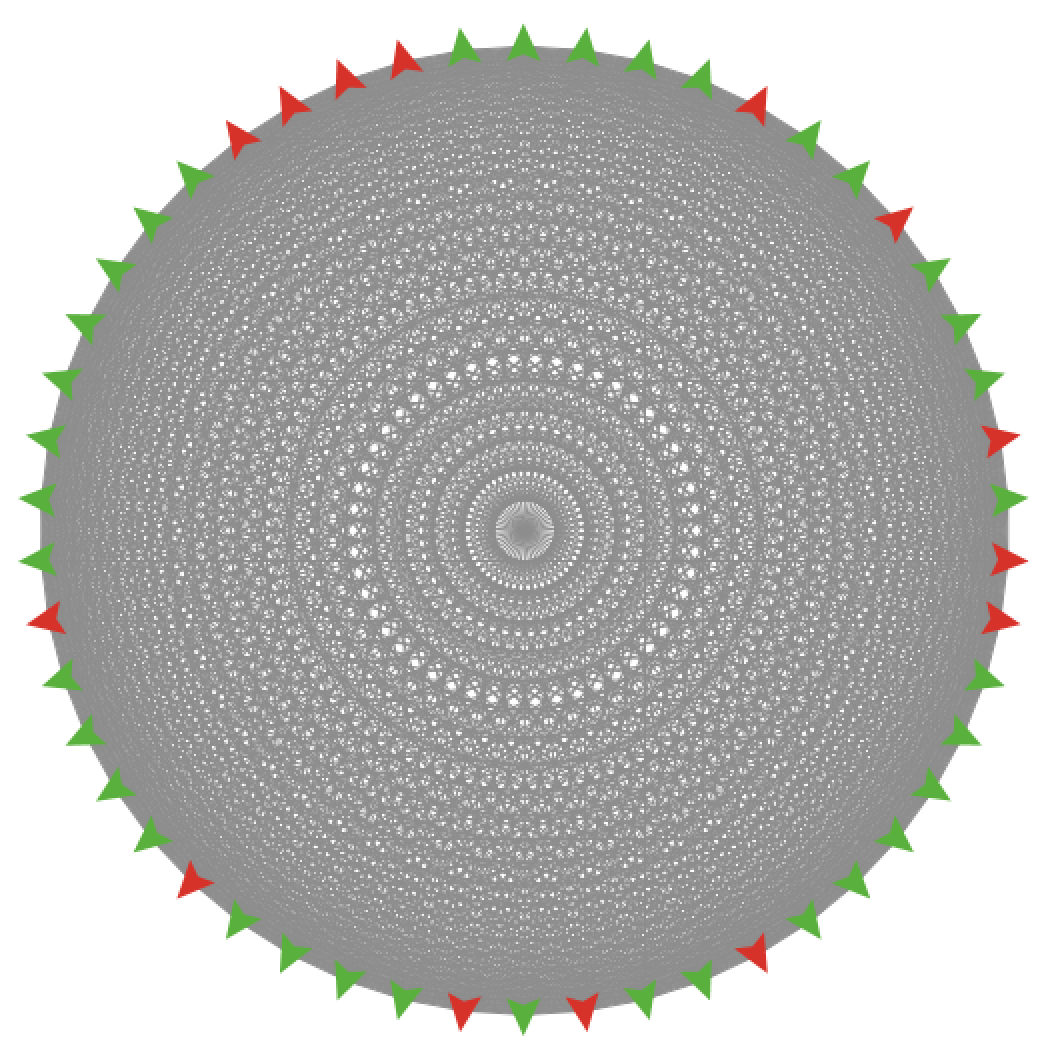}
     
         \caption{Two groups of 50 agents connected in a social network:  a `small-world' network on the left, and a complete network on the right. Green triangles represent agents who currently support the hypothesis, and red those who do not (cf. Section \ref{section: Process}). Both network types are used in the case studies (sections \ref{shifttoextremity}, \ref{polarizationversusconvergence}). Parameters of the small-world network: \texttt{rewiring-probability=0.2}, \texttt{k}=2.}
         \label{networks}
\end{figure}

The model places `\texttt{number-of-agents}' agents on a grid and then specifies who is connected to whom via undirected communication links. Agents can only communicate with their link neighbours. NormAN provides a number of different network structures that the user can select before initialisation (via the chooser variable \texttt{social-network}), such as the complete network, a `wheel' (cf. \cite{zollman2010epistemic, frey2020robustness}) and small-world networks (also known as Watts- Strogatz networks \cite{watts1998collective, wilensky2005netlogo}). The latter are a type of network structure found in many social and biological networks. They are characterised by comparatively short paths between nodes in the network (`five degrees of separation') and comparatively high clustering, although the density of connections is relatively low (see Fig. \ref{networks} for a visualisation). 

\subsection{Process Overview}
\label{section: Process}

Deliberation unfolds dynamically, in discrete time steps. At each step, agents follow this protocol: 

\begin{itemize}

    \item[(1)] Collect evidence: agents may collect a new piece of evidence from the world.
    \item[(2)] Communication: agents may share one piece of evidence they have already encountered with their link neighbours.
\end{itemize}

Collecting evidence facilitates the flow of information into the network, and communication facilitates the flow of information through the network. This subsection explains when and how agents perform each activity (a detailed description and a flowchart of the protocol can be found in section \ref{section:process}, Fig. \ref{Agent-Update-Flowchart}). 

To collect evidence (or `inquire'), agents randomly select an item from the world model's \texttt{evidence-list} that they have not yet encountered. They add this truth value to their personal \texttt{agent-evidence-list}.\footnote{As an example, suppose that at time $t$, agent $A$ stores the truth values $\texttt{agent-evidence-list}_{A}^{t}=\{E_1, \neg E_3\}$. Through inquiry, they may find that $E_2$ is indeed true, thus extending their list to $\texttt{agent-evidence-list}_{A}^{t+1}=\{E_1, E_2, \neg E_3\}$.} Inquiry is therefore modelled by `drawing' from the world's  \texttt{evidence-list}. Learning any new piece of evidence (be it via inquiry or communication) is modelled as learning that a certain piece of evidence is \textit{true} or \textit{false}. Two agent variables govern inquiry. First, agents have a fixed maximum number of inquiries (the variable \texttt{max-draws} determines how many pieces they may each collect during one simulation). Second, agents will not necessarily inquire every round. Rather, their chance of inquiry is determined by a \texttt{curiosity} parameter. Hence, agents only collect evidence if they are `curious', and if they still have `draws'.

Next, in each round, agents may communicate and receive evidence through communication. In  NormAN 1.0, communication is modelled via a simple transmission mechanism: the communicating agent $A$ chooses which piece of evidence to transmit to their link neighbours. Each link neighbour, e.g., $B$, then either adds this evidence to their  \texttt{agent-evidence-list}, and computes a new \texttt{agent-belief}, or ignores this evidence if they have already heard it.\footnote{For instance, if $A$'s list is $\texttt{agent-evidence-list}^{i}=\{E_1,\neg E_3\}$, and $B$'s \texttt{agent-evidence-list} is $\texttt{agent-evidence-list}_B^{i}=\{ \neg E_3\}$, then $A$'s sharing $E_1$ will enrich agent $B$'s list to $\texttt{agent-evidence-list}_B^{i+1}=\{E_1,\neg E_3\}$. Had $A$ chosen $\neg E_3$, $B$'s list would have remained unchanged.} In NormAN, agents recognise distinct pieces of evidence and never `double count' pieces of evidence.   

Although this mechanism of informational `passing the parcel' is simple in that it avoids the complexity of testimony, it can be used to capture distinct, complex styles of communication.  In NormAN 1.0, three sharing rules are examined:

\begin{enumerate}
    \item Random: Agents share a random piece of evidence from their \texttt{agent-evidence-list}.
    \item Recency: Agents share the piece of evidence they most recently encountered. 
    \item Impact: Agents share the piece of evidence that they hold to be the best piece of evidence in favour of their current position.
\end{enumerate}

Since the random rule is self-explanatory, we briefly explain how the recency and impact rules work (for a detailed, technical explanation, see Section \ref{section:process}). Under the recency rule (loosely inspired by Maes and Flache's model of bi-polarization \cite{mas2013differentiation}), agents are most likely to share the last piece of evidence they heard. This is implemented by each agent's \texttt{recency-list}, which keeps track of the order of receipt.\footnote{Importantly, even if agents receive a piece of evidence they have already encountered, this piece is `popped' to the top of the \texttt{recency-list}.} With a high probability of $x$ the agents share the last element of this list, but with a probability of $1-x$ they share another random element from the list.\footnote{In the base model, $x=0.9$.}

The impact-sharing rule provides a very basic implementation of the idea that speakers seek to communicate what they consider to be (most) relevant. This means sharing what they consider to be their best---strongest---piece of evidence. In our simple impact rule, this is the piece of evidence which most convinces agents of their current position. In NormAN, agents track the magnitude of the belief update that evidence triggers, that is, its `impact'. To measure this, for each evidence $E_i$, agents store the update magnitude $P(H|E_i)-P^{initial}(H)$, where $P^{initial}(H)$ marks the agent's prior, pre-evidence belief (\texttt{initial-belief}). That is, the impact of a piece of evidence is measured by how far it moved (or would move) an agent's belief away from their agnostic prior. Each agent has an $\texttt{update-list}$ recording the impact of each piece of received evidence. If an agent currently supports the hypothesis, they share the evidence with the highest update value (and they share the evidence with the lowest, i.e., largest negative impact if they currently oppose it). NormAN models this `support'  as follows: if the agent's \texttt{agent-belief}>\texttt{initial-belief}, they support the hypothesis, and they oppose it if \texttt{agent-belief}<\texttt{initial-belief} (cf. Fig. \ref{networks}).  Hence, an agent's position is measured relative to their pre-evidence, agnostic prior.

Communication is regulated by a conviction threshold, a percentage value that serves as a cut-off point for when an agent's belief departs sufficiently from their agnostic, pre-evidence prior (\texttt{initial-belief}) for them to jump into the discussion. This threshold is set by the global variable \texttt{conviction-threshold}, which determines a percentage by which the agent's conviction needs to exceed their initial, pre-evidence belief.\footnote{Specifically, it defines a lower bound and an upper bound for agent beliefs. The lower bound is computed as ($\texttt{initial-belief} - \texttt{initial-belief} \times \texttt{conviction-threshold}$). The upper bound is computed as ($\texttt{initial-belief} + (1 - \texttt{initial-belief}) \times \texttt{conviction-threshold}$).} If an agent's \texttt{agent-belief} does not exceed the threshold (above or below), they will not share. Note that if $\texttt{conviction-threshold}$ is set to $0$, the sharing condition is trivially met in most cases: agents will share whenever their \texttt{agent-belief} $\not =$ \texttt{initial-belief}.\footnote{As an example, if \texttt{conviction-threshold} $=0$, \texttt{initial-belief} $=0.3$, and agents use the impact sharing rule, they will share pieces of evidence `against' $H$ if their current belief is below 0.3 (and vice versa for \texttt{agent-belief}$>0.3$).}
    
One last agent-variable co-determines the frequency of communication:  agents' chattiness $\in [0,1]$, that is, the chance that they will communicate with their link neighbours on each round (determined by the global variable \texttt{chattiness}). If an agent passes the conviction threshold and is chatty, they will send an argument (item from to their \texttt{agent-evidence-list}) to their link neighbours. 

To summarise, in each time step, agents may first collect new evidence from the world (if they are curious and still have `draws'). Then, if they cross the threshold and are chatty, they share one of their pieces of evidence with their neighbours (according to the sharing rule chosen by the model user). Whenever they learn of a new piece of evidence, they compute their new belief in the hypothesis.

\subsection{Implementation and usage}

In order to make the NormAN framework accessible to researchers from a broad range of backgrounds we chose to implement it in NetLogo \cite{Wilensky:1999,wilensky2015introduction}. Designed initially to teach programming to beginners, NetLogo is an accessible, well-documented, platform that has been and continues to be widely used in agent-based modelling research \cite{gunaratne2021nl4py}, including specifically for research on opinion dynamics \cite{lorenz2017modeling,wang2022multiagent}, belief dynamics \cite{hahn2018good,hahn2018communication,hahnpolarisation}, and social epistemology \cite{weisberg2009epistemic,pinto2018epistemic}.

Its benefits lie in the fact that much of the machinery required for setting up an ABM and running simulations with it is in-built, leading to very compact code: the initial version of NormAN (version 1.0) has only 500 lines of code (excluding the BN definitions). 

Moreover, Netlogo has extensions for both R \cite{thiele2012agent} and Python \cite{jaxa2018pynetlogo}, that allow two-way interactions with the extensive scientific computing resources of those platforms. For our initial version of NormAN we chose the R extension (a version with Python is planned in future). Specifically, NormAN draws on the R-package bnlearn \cite{scutari2009learning,scutari2021bayesian} to handle all Bayesian belief updating over BNs. NormAN 1.0 was developed using NetLogo version 6.2.1, which efficiently implements the R extension (developed by \cite{thiele2010netlogo}). 

The combination of R (bnlearn) and NetLogo makes for a very flexible modelling environment: to characterize the world BN, the modeller can load whole $R$ files into the NetLogo model, or simply copy and paste the lines of R code into the indicated section of the NetLogo code. One can also use one of eight preset BN's (see \ref{section:BNworlds}). The NetLogo interface handles the rest: sliders, inputs and switches determine the size and shape of the social network, the agent variables such as sharing styles, as well as the specification of which BN nodes ought to count as the evidence and hypothesis. 

\subsubsection{Running and Evaluating Simulations}
With respect to running and evaluating simulations, the use of the R extension means that users have two routes for controlling `experiments' and model explorations: Netlogo's built-in BehaviorSpace \cite{tisue2004netlogo} and directly through R \cite{thiele2012agent}.

While the use of a high-level language such as NetLogo does come at a performance cost, we found simulations with NormAN version 1.0 to not only (easily) be efficient enough for practical purposes in the range of network sizes we think modellers will most likely wish to explore in detail (up to around 100). It is also possible to run larger networks. We have, albeit slowly, run networks with 100,000 agents in the NetLogo User Interface (on a 2020 MacBook Pro with 2 GHz Quad-Core Intel Core i5, and 16 GB of RAM). 

It is thus possible, even in the current implementation, to check how findings scale and whether `more is different' for the target phenomenon of interest \cite{anderson1972more}.  Furthermore, much of the processing time for large networks involves the construction of the social network (in particular, the small world network), suggesting paths for scalable future versions \cite{railsback2017improving}. 

While we consider the balance between accessibility and performance to be a suitable one with respect to our current goals (see also \cite{burbach2020netlogo,railsback2017improving}),  re-implementing the NormAN framework not just with other extensions \cite{salecker2019nlrx,gunaratne2021nl4py}, but also within other platforms and languages is a goal for future work. 

\section{Initial Case Studies}

In the third main part of this paper, we seek to demonstrate the utility of NormAN with two case studies. These have been chosen to illustrate the value of its features and demonstrate the reasons for our basic design choices.  In particular, they serve to underscore the claims of section \ref{section:normative} above, that a suitable model of argument exchange needs normative grounding both with respect to the aggregation of evidence/arguments by individual agents and with respect to the ground truth world. The case studies have been chosen also to illustrate how NormAN, as a normative model, may contribute to extant theoretical concerns across a range of disciplines. 

\subsection{Case Study 1: Shift to Extremity}
\label{shifttoextremity}

The so-called `shift to extremity' \cite{stoner1968risky} is the original `polarization' phenomenon. Although the term `polarization' has now become associated with belief or opinion \emph{divergence} within a population, the term was first used to describe the phenomenon whereby deliberating groups tended to shift further in the direction of an initial opinion over the course of deliberation (for a review see e.g., \cite{isenberg1986group}). This shift to extremity attracted considerable research spanning six decades to date and has proved highly reliable (it has been observed with lab-based studies, mock juries, deliberative polling \cite{myers1976group} and citizen debates \cite{lindell2017drives} and with tasks as diverse as risk acceptance, probability judgment, policy choice and negotiation \cite{lamm1988review}), though it is not observed in every group discussion.  This interest has been fuelled not just by the phenomenon's practical relevance, but also the fact that it (at least initially) seemed counter-intuitive and in need of explanation: after all, one might expect a deliberation to surface both arguments for and against a claim. This made it seem surprising that beliefs might shift regularly in one particular direction. 

Multiple lines of explanation were pursued in the literature, such as the idea that the shift reflects social comparison processes \cite{sanders1977social} or social identity considerations \cite{abrams1990knowing}: people may become comfortable expressing positions they initially feared others might view as extreme or they may, as a matter of identity,  seek to adopt attitudes stereotypical of the group. A third account,  by contrast, attributed the shift to the arguments that surfaced within a debate. Burnstein and Vinokur's `argumentative theory' proposed that group members lean toward an initial position because they have more (or stronger) arguments in favour of that position and more (or stronger) arguments in favour of that position will consequently be available for exchange in the deliberation \cite{burnstein1977persuasive}. Experimental research subsequently sought to distinguish these competing (but ultimately not mutually exclusive) accounts \cite{lamm1988review,sanders1977social,vinokur1978depolarization}. 

The argumentative theory was also supported through simulations in an agent-based model by \cite{mas2013differentiation}. In this model, Maes and Flache implement the substantive assumptions of the persuasive argumentation theory and combine them with the modelling of homophily in order to understand bi-polarization or belief divergence (an issue we turn to in our next case study). In effect, their model of the latter combines the shift to extremity afforded by persuasive argumentation with homophily-based segregation to explain divergence. In their model, agents have a numerically valued opinion (drawn from the interval -1 to 1)  representing their stance on the issue in question. Additionally, there is a set of arguments that address that issue. The valence of an argument is expressed numerically (pro = 1, con = -1), and all arguments carry equal weight. An agent's current stance is based on the average value of the arguments they are currently considering,  and arguments for communication are selected randomly from the agent's current relevance set --a subset of the encountered arguments determined by recency. Resultant findings support the argumentative theory in as much as the positions of agents within the homophily-driven clusters become more extreme.

One limitation of the model, however, is that both the generation and evaluation of arguments lack a principled basis. And the initial distribution of arguments in the population is essentially `just so'.

Most recently, it has been pointed out that both of these concerns may be addressed by adopting a Bayesian perspective on argument \cite{HahnPPS}, as described in section \ref{section:bayesarg} above. From that perspective, multiple interlocking components give rise to the shift to extremity: Group members' pre-deliberation beliefs are based on a sample of the arguments available `in the world'. The available population of arguments for and against the claim is likely to contain stronger arguments in one direction than the other. Pre-deliberation beliefs based on samples will, on average, reflect those characteristics. By the same token, combining samples via group discussion is more likely to see individual members add stronger arguments supporting the initial direction, which in turn will shift the group mean.

The core component that the expected  distribution of arguments available \emph{in the world} is skewed follows from the Bayesian conceptualisation of argument --outlined in section \ref{section:bayesarg}-- whereby an argument or piece of evidence is strong or diagnostic to the extent that it is much more likely to be found if the hypothesis is true than if it is false (expressed by the likelihood ratio $P(e|H)/P(e|notH)$). This translates into an expected distribution of arguments by virtue of the fact that how likely a piece of evidence is  (its so-called marginal probability) is determined by `total probability': $P(e) = P(e|H)\times P(H) + P(e|notH)\times P(notH)$. In other words, the probability of a piece of evidence is determined by the probability of obtaining it if the underlying hypothesis is true, weighted by the probability of the hypothesis, plus the probability of a false positive, weighted by the probability of the hypothesis being false. This means that (all other things equal),  strong evidence in favour of a hypothesis is more likely than equally strong evidence against if the hypothesis is true \cite{HahnPPS}. 

In short, the Bayesian framework helps make explicit the fundamental point that --at least for claims about issues of fact-- what arguments are available in a domain is determined by the underlying structure of the world. And that evidence distribution, in turn, will impact agents' subsequent beliefs. The shift to extremity is simply an initially counter-intuitive consequence of that fundamental point. We should expect a group with prior evidence/arguments to lean, on average, in a particular direction and expect that exchanging that evidence will likely lead to a further shift in that direction.

\begin{figure}
  \centering
  \includegraphics [width=8cm]{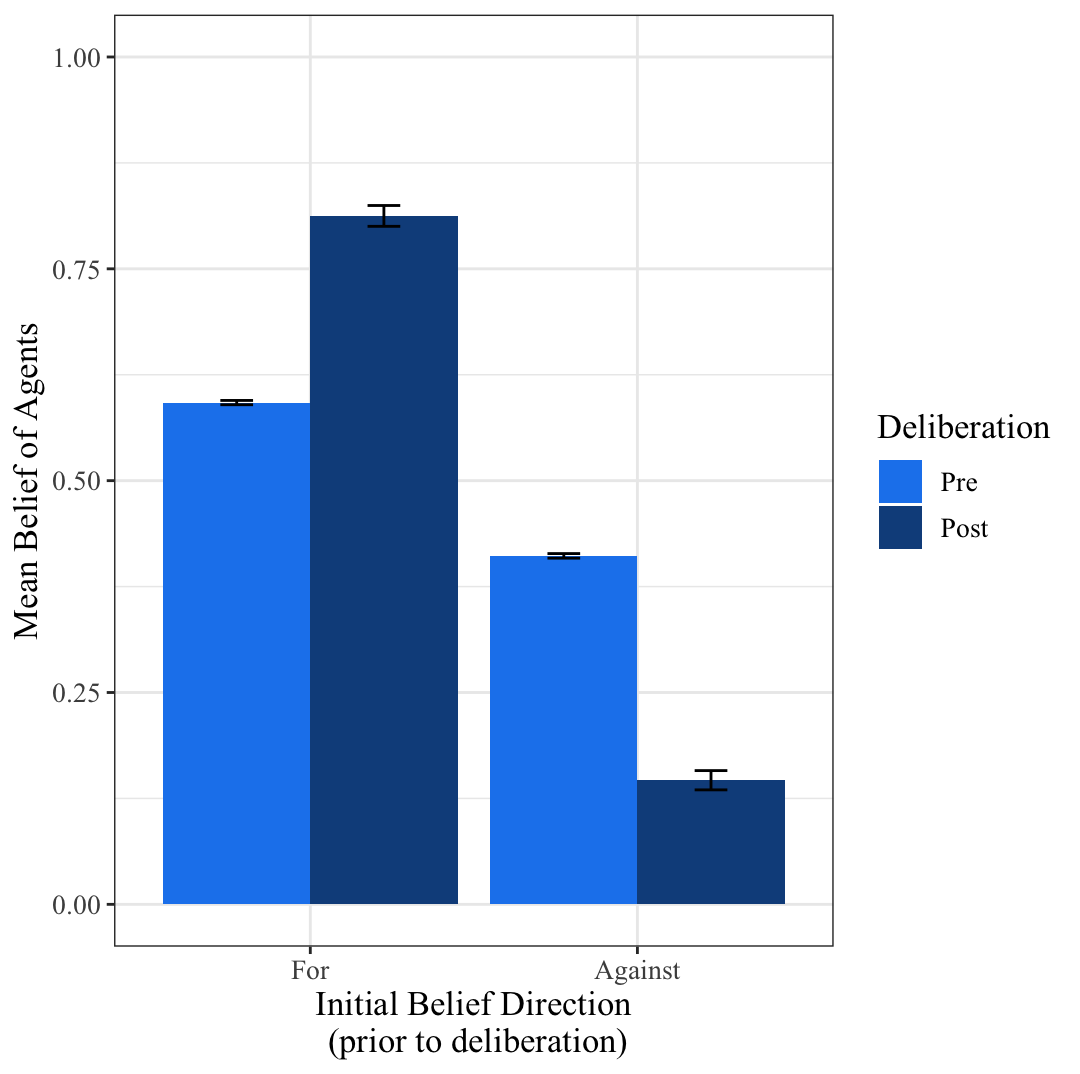}
  \caption{Results from simulation using the `Big Net' network. Shown are the mean beliefs of agents in the target hypothesis and standard error bars, across all 800 experiment runs, at the pre (0) and post-deliberation phase (in this experiment after 25 exchanges/steps). Groups are split by the agents' mean initial direction of belief for a given run.}
  \Description{This is the Shift to Extremity}
  \label{fig:shift}
\end{figure}

\begin{figure}[h]
  \centering
  \includegraphics[width=\linewidth]{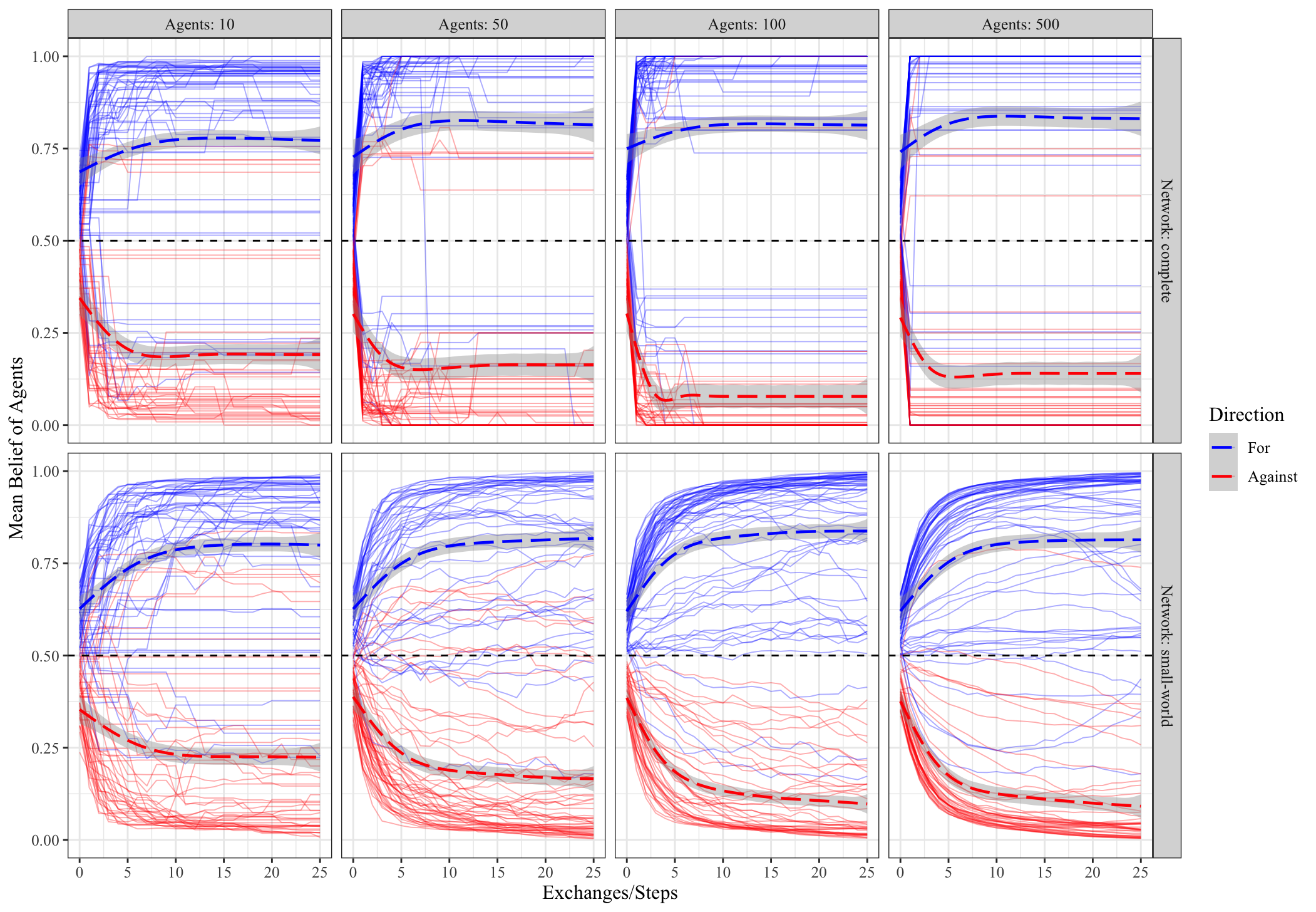}
  \caption{The Figure shows the trajectories of mean group beliefs across time (thin lines) and the average of those means (dashed lines) with standard error (grey shaded area), split by whether the group initially leaned 'for' (blue) or 'against' (red). The top row shows the simulation results for complete networks of various sizes (10,50,100, and 500 agents). The bottom row shows the same dynamics in small-world networks with corresponding numbers of agents. Note that the `pre-evidence' neutral point in these simulations is .5 (base rate of hypothesis), as indicated by the black dotted line.}
  \Description{This is the Shift to Extremity}
  \label{fig:shift_runs}
\end{figure}

In our first case study, we demonstrate these theoretical points through simple simulations with the NormAN framework. To this end, we simply ran NormAN using one of our default networks --the `Big Net' network (see Fig. \ref{fig:bignet} in Appendix \ref{section:BNworlds} below). This stylised network is useful for making the current point because it is symmetrical by design: the structure contains three arguments that, when true, are evidence for the claim, and three that, when true, are equally strong evidence against. On average, instantiations of this world will not, however, lead to equal numbers of arguments for and against (though this will sometimes occur) for the reasons just outlined. As a result, the shift to extremity is readily observed despite the balanced nature of this stylised network. To show this, we simulated many model runs, graphing the mean belief at 2 points in time: once after the initial argument-draw by agents, and once at the end of the run. The former represents the pre-communication, and hence the pre-deliberation state, the latter the end of deliberation (for full details see Appendix \ref{section:case1})
Figure \ref{fig:shift} shows the respective pre- and post-deliberation means split by whether the group's initial belief (given their initial access to evidence prior to deliberation) leans 'for' or 'against' the claim in question. The second figure, Figure \ref{fig:shift_runs} shows the same measurements but split by individual runs of the simulation. It plots the belief dynamics across time for the individual runs, split again by whether the group's initial belief (given their initial access to evidence prior to deliberation) leans `for' or `against' the claim in question (i.e.,  the target hypothesis). As can be seen, the shift does not always happen but happens most of the time.

The experimental literature on the shift typically considers fairly small groups, under conditions where members typically all hear one another. In network terms, participants constitute a complete network. These conditions are captured by the top left panel Figure \ref{fig:shift_runs}. But we can also explore how increasing the group size influences the dynamics. To this end, the four columns of row 1 in Fig. \ref{fig:shift_runs} show group sizes of 10, 50, 100 and 500 agents respectively. These show (for a constant pool of available arguments) a sharpening of the shift as a function of group size. This reflects the fact that the available evidence enters the group discussion more quickly.  The bottom row of Fig. \ref{fig:shift_runs} shows the same information for a  small world network \cite{watts1998collective}. One can see a dampening of the shift, due to the slower diffusion of arguments among agents.  Finally, to demonstrate that these findings are not a unique feature of the particular BN `world' selected, Fig. \ref{fig:shift_runs_asia} in the Appendix section \ref{section:case1} shows the same results for a different network included with NormAN version 1.0, the `Asia network' of Fig. \ref{fig:asianet} above. The Appendix also contains the full simulation details, model parameters, and further supplementary figures that elucidate model behaviour.

These simple simulations illustrate the different components of the `shift to extremity' that a Bayesian perspective helps unify and make explicit. In so doing, they also illustrate how additional insight and refinement of understanding becomes available as a result of moving from the argumentative theories' initial, purely verbal, formulation, through Maes and Flache's agent-based model \cite{mas2013differentiation} to a normative framework. Most importantly, however, these simple simulations highlight the important point that evidence or argument distribution matters fundamentally to understanding model behaviour. By the same token, it matters fundamentally to the understanding of the kind of real-world behaviours these models are trying to elucidate. 

\subsection{Case Study 2: Polarization versus Convergence}
\label{polarizationversusconvergence}

The goal of consensus as unanimous agreement is one of the key motivations for deliberation in early theories of deliberative democracy \cite{landemore2015deliberation}. 
Conversely, polarization as (extreme) belief divergence is seen as a threat to contemporary democracies worldwide \cite{sunstein2018republic}. 
Under which conditions can we expect a group to converge on consensus---and correct consensus at that? And under which conditions does polarization emerge? Computational models of deliberation have identified conditions undermining consensus, that is, conditions that lead to non-convergence. While such models, to date, have yielded a wealth of insight, in particular formal insight (e.g., \cite{krause2015positive}), there are key features of the most popular paradigms  that significantly limit or distort insight into polarization as a real-world phenomenon.

As discussed in section \ref{section:opiniondynamics} above,  most opinion dynamic or social-influence models revolve around the notion that individuals communicate their opinions and influence is typically implemented as opinion averaging \cite{hegselmann2002opinion,french1956formal,friedkin2011social,deffuant2005individual}. They thus abstract away entirely from supporting arguments themselves. As a result, these models have several consequences that are strongly at odds with what is empirically observed. For one, they typically exhibit an inevitable drive to convergence \cite{abelson1964mathematical,lorenz2006consensus,krause2015positive} which has meant that other factors preventing convergence and giving rise to polarization must additionally be imposed (e.g., \cite{o2018scientific}). As Maes and Flache (2013) note, many of these factors can be understood as negative influence of one form or another \cite{baldassarri2007dynamics,macy2003polarization,mason2007situating,olsson2020bayesian} giving rise to two competing forms of influence---positive influence from similar, like-minded agents, and negative influence from dissimilar agents. One aim of the Maes and Flache (2013) model introduced in the previous section is to demonstrate how positive influence alone (in the form of persuasive argumentation coupled with homophily) can give rise to divergence. At the same time, models based on opinion averaging also fail to capture the  shift to extremity. Averaging implies that averaging agents will not adjust their opinions when they interact with others with whom they already agree \cite{mas2013differentiation}. This follows necessarily from the fact that the average of two near-identical values will not only be similar to these values but will be less extreme than the more extreme of the two. It is thus difficult to generate the empirically observed dynamics whereby deliberation leads to views more extreme than those of any of the participants prior to the interaction. Yet this possibility has been observed widely in the empirical research on the shift to extremity discussed above. 

A model that is rich enough to include individual arguments thus seems essential to fully understanding real-world divergence of beliefs. In this context, the dynamics of NormAN help clarify a fundamental mechanism producing non-convergence: namely, the incomplete sharing of information on a social network. Whenever two agents start with the same prior degree of belief and also interpret pieces of evidence in the same way, the acquisition of different sets of evidence will drive their posterior beliefs apart. Hence, in our model, whenever deliberation does not result in a state where all agents have access to the same evidence, then final beliefs may be scattered as well. This is a natural consequence of the uniqueness assumption implemented in NormAN 1.0: for each set of evidence, there is only one permissible `doxastic state', that is, only one possible degree of belief. Other models identify a different cause of non-convergence: differences in the interpretation of evidence. Cook and Lewandowski \cite{cook2016rational} highlight how Bayesian agents entertaining different BNs (in particular, different conditional probability distributions) will exhibit belief divergence when receiving the same evidence (an experimental finding dating back to \cite{lord1979biased}). Indeed, relaxing the uniqueness requirement and allowing agents to have different conceptions of the world's causal structure will make perfect convergence via deliberation (as characterized here by the exchange of arguments/evidence) rather hard to attain: even if all agents have access to the same evidence, they may still disagree.

\begin{figure}

         \centering
         \includegraphics[width=0.48\textwidth]{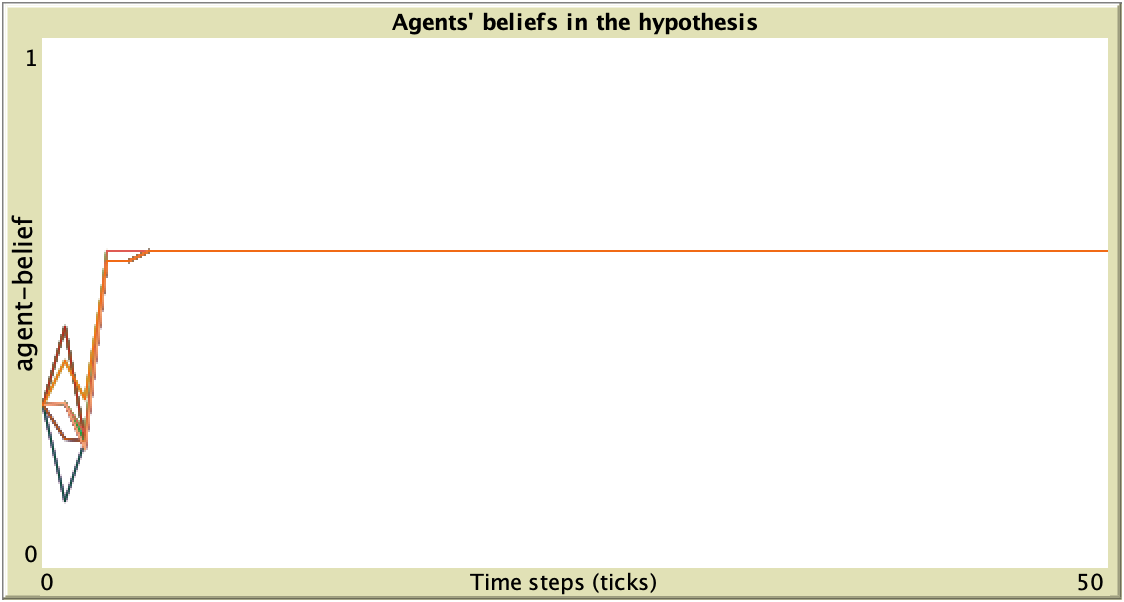}
         \includegraphics[width=0.48\textwidth]{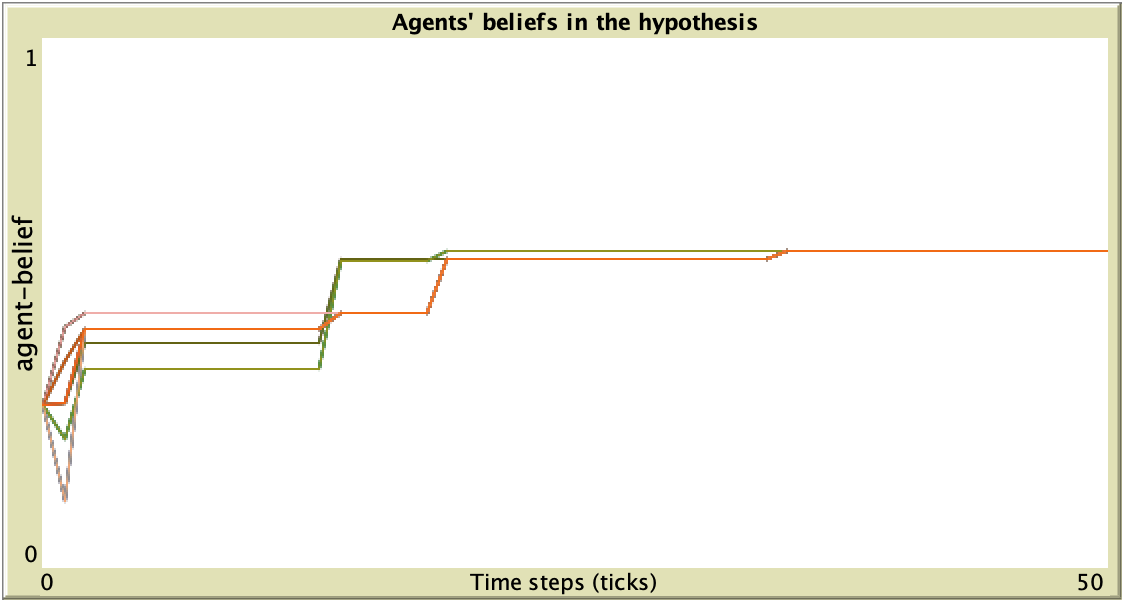}
        
     
         \centering
         \includegraphics[width=0.48\textwidth]{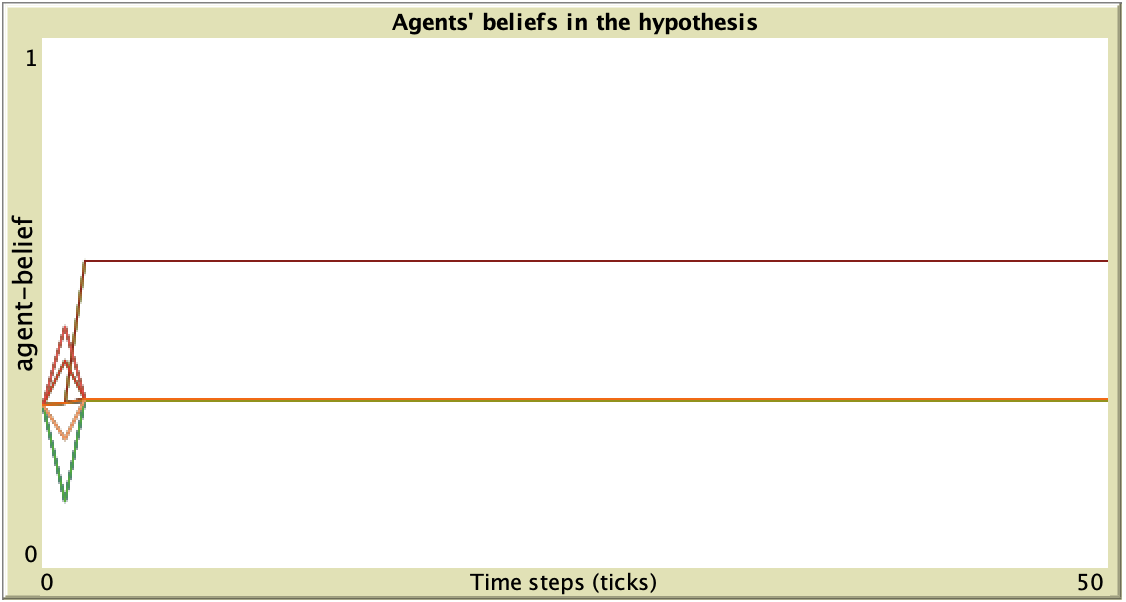}
         \caption{The beliefs of a population over time, for three sharing rules: random (top left), recency (top right) and impact (below). Each function tracks the state of one agent's \texttt{agent-belief} at each time step. Initial conditions: \texttt{causal-structure}= Vole (see the Appendix, section \ref{section:BNworlds} for an explanation), \texttt{chattiness}=0.5, \texttt{conviction-threshold}=0, \texttt{curiosity}=0, \texttt{initial-draws}=1, \texttt{max-draws}=1, \texttt{social-network}=complete, \texttt{number-of-agents}=50. The prior \texttt{initial-belief}$\approx 0.3$.}
         \label{fig:beliefscomplete}
\end{figure}

When can we expect agents to end up with identical sets of evidence, and therefore in consensus? In this brief case study, we use the NormAN framework to exemplify how a common style of communication can undermine consensus: impact-driven sharing (as explained in section \ref{section: Process}). Instead of sharing all their evidence, agents only share the piece of evidence that most convinces them of their current position, that is, the argument or piece of evidence that they find most impactful. Crucially, this type of incomplete, selective sharing need not be an irrational or bad-faith communication strategy. Using the impact rule simply means curating what one shares, leading to communication of what one considers one’s best, most truth-conducive piece of evidence. Given that both communicating and processing arguments is costly, such a strategy may be pragmatically appropriate in many contexts. It is intuitive that a communication style between agents who do not (or cannot) share their entire evidence may lead to non-convergence: even if no piece of evidence is completely unknown to the group, the impact rule, or any incomplete sharing rule (i.e., rules where agents communicate only a subset of their evidence), will make an end state of fully shared information less likely. Instead, each agent’s access to evidence is filtered by their social environment: If their neighbours believe that the hypothesis is true, they are more likely to receive confirmatory pieces of evidence through communication (and vice versa).

Figure \ref{fig:beliefscomplete} shows three exemplifying model runs of a population of 50 agents connected in a complete network (the world model uses the `Vole' network, explained in the Appendix, Section \ref{section:BNworlds}). The graphs track the agents' \texttt{agent-beliefs} in the hypothesis during a very simple deliberation process: initially, each agent has one piece of evidence (i.e., their \texttt{agent-evidence-list} contains the truth value of one randomly chosen evidence node). Subfigure 1 illustrates the evolution of beliefs resulting from deliberation following the `random' sharing rule. Convergence is virtually guaranteed as agents will share all their evidence eventually (the random rule is a `complete' sharing rule). The recency rule (Subfigure 2) creates a similar dynamic: although convergence takes longer to achieve, agents do, eventually, form a consensus. Subfigure 3 illustrates non-convergence as a consequence of the `incomplete', selective impact sharing. This non-convergence is easily explained:  meaningful communication between two agents is interrupted when the sender has communicated what they consider to be their best evidence (the sender may repeat themselves, but the receiving agent, having already incorporated the information, will now ignore this communication with respect to their belief in the hypothesis). It is only resumed when the sender learns even stronger information in favour of their position or when they change their mind (i.e., cross the conviction threshold). However, once agents find themselves in a group of like-minded neighbours, they are unlikely to receive further evidence that changes their minds. Consequently, communication ceases and divergent beliefs stabilise.

\begin{figure}

         \centering
         \includegraphics[width=0.9\textwidth]{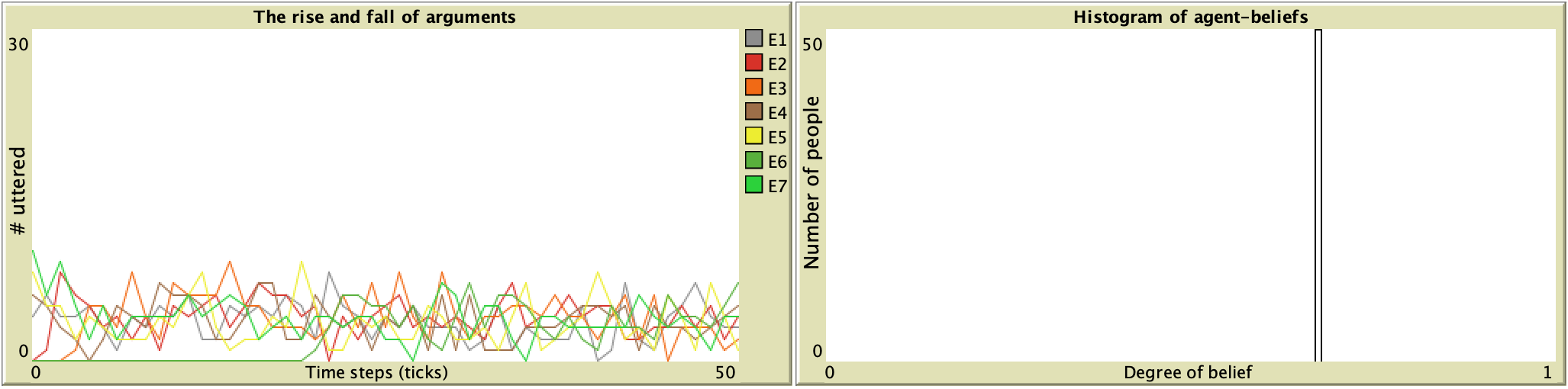}
         
         \includegraphics[width=0.9\textwidth]{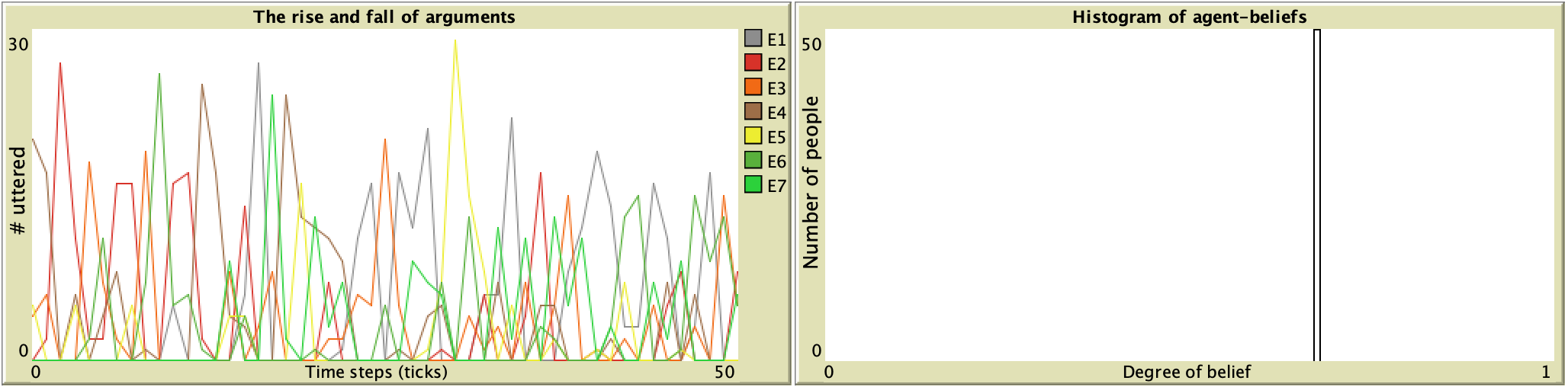}

     
         \centering
         \includegraphics[width=0.9\textwidth]{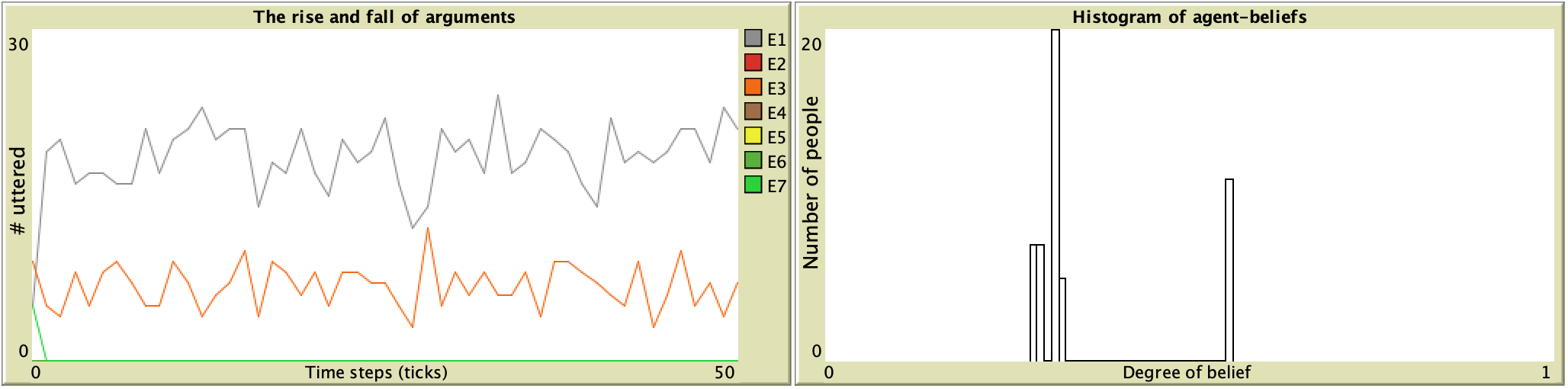}
         \caption{Different deliberations tracked by (left) the number of transmissions of each argument per round and (right) the degrees of belief of the agents. As for the tracking of arguments (left), if the argument $S$ was shared by 20 agents at time step $10$, the red line marks $\#uttered=20$ on the y-axis. On the right-hand side, we see a histogram of the agents' \texttt{agent-beliefs} at time step 50 (the beliefs have stabilized at that point). The top figure shows a run driven by the random rule. The middle figure shows a run controlled by the recency rule. The bottom figure shows a run controlled by the impact rule. Initial conditions: \texttt{causal-structure}= Vole (see the Appendix, section \ref{section:BNworlds} for an explanation), \texttt{chattiness}=0.5, \texttt{conviction-threshold}=0, \texttt{curiosity}=0, \texttt{initial-draws}=1, \texttt{max-draws}=1, \texttt{social-network}=complete, \texttt{number-of-agents}=50. The prior \texttt{initial-belief}$\approx 0.3$.}
         \label{fig:arguments}
\end{figure}

This non-convergence result contrasts with models of homophily and echo chambers where agents themselves (possibly unknowingly) select sub-optimal information environments. The present simulations with  NormAN reveal that divergence may arise also as a function of good faith communication strategies whereby agents seek simply to convey what they consider to be their `best evidence', without homophily or any other attempt to filter their (input) sources, and also without any form of negative influence. These sample runs also demonstrate a further important point. Paradigmatic models of polarization typically showcase only one type of non-convergence: wholly driving apart two subgroups, possibly to the extremes (e.g.,  0, that is, certainty that the central hypothesis is false and 1, certainty of truth) \cite{olsson2013bayesian,pallavicini2021polarization, bramson2017understanding}.  Realistic models of deliberation that connect meaningfully to real-world data, however, plausibly ought to reconstruct non-convergence as a scattering of intermediate beliefs across the unit interval as well.

As the model runs of NormAN show, scattering of beliefs can arise as the result of incomplete sharing in a social network. In the simulations (as, arguably, in the real world), exactly how much the group’s beliefs diverge, will depend partly on the nature of the available evidence: situations where there are strong pieces of evidence both in favour and against the hypothesis will prove more divisive than more homogenous distributions. Furthermore, using scarcer, more realistic network structures (as opposed to the complete networks in the above sample run) will also exacerbate divergence effects: using a small-world network slows down belief convergence induced by the complete sharing rules (recency and random), and exacerbates the scattering effect of the impact rule (cf. Fig. \ref{fig:smallworlddynamics} in the Appendix.)

The case study simulations show how different communication rules will give rise to different patterns. Needless to say, in the real world these factors may additionally interact with other factors that have been shown to give rise to polarization such as homophily or differences in trust. A richer underlying model of communication that involves argument exchange thus opens up new avenues for understanding the complex interplay of factors likely at work in real-world settings.

A corollary of the fact that communication rules are so central is that it highlights the need to understand in more depth \emph{what has been communicated}. Different belief distributions come about because agents end up being exposed to different subsets of the evidence. To analyze the dynamics of argument exchange, NormAN allows the modeller to track the sharing frequency of particular pieces of evidence. Fig. \ref{fig:arguments} shows three model runs (with the same initial conditions as in Fig. \ref{fig:beliefscomplete}), with the left panels tracking the number of transmissions of each piece of evidence per time step (on the right we see the final distributions of the agent's beliefs). Both the first (top) Subfigure, showing a random run, and the second (middle), showing a run driven by the recency rule, end in convergence. Note, however, the differences in the frequency of argument sharing: While the frequencies of arguments sent remain roughly the same when sharing is random (each argument is similarly popular), the frequency of arguments governed by the recency rule rises and falls in waves. Finally, in the Subfigure below, which shows a run governed by the impact rule, we can see that two groups of disagreeing agents (one with \texttt{agent-belief}<\texttt{initial-belief}, the other with \texttt{agent-belief}>\texttt{initial-belief}) each select what they consider to be their `best' evidence and stabilize on sharing those arguments repeatedly. Therefore, less impactful evidence is not communicated and, as can be seen on the right-hand side, the agents' beliefs do not converge. 

Crucially, this illustrates also that NormAN can be used to study not only belief dynamics but also \emph{argument dynamics}. However, studying the dynamics of what arguments are exchanged is not just a matter of supplementary analysis. It deserves to be seen as a research topic in its own right. As noted in  \ref{section:opiniondynamics} above, computational social science has seen large amounts of research  devoted to aspects of how particular messages spread across real-world social networks. That analysis has, however, remained coarse-grained, arguably, in good part because of a lack of models to help formulate theory-driven questions and hypotheses. We would hope that NormAN simulations could provide new impetus here.

\subsection{Implications of the Case Studies and Future Directions}

Both case studies illustrate how the NormAN framework may contribute to extant research. While these case studies are intended to be of interest in their own right, we think also that they illustrate the breadth of disciplines and research questions that the NormAN framework could contribute to. Case Study 1 on the shift to extremity helps further illuminate an empirical phenomenon that has exercised, in particular, psychologists and political scientists. Case Study 2 on polarization contributes to a research topic that has attracted interest across a broad range of disciplines, ranging from psychology (e.g., \cite{fasce2023science,brown2022social}, communication studies \cite{moore2020prime,kubin2021role,lee2014social}, sociology (e.g., \cite{mas2013differentiation}), epistemology (e.g., \cite{olsson2013bayesian}), political science (e.g., \cite{baldassarri2007dynamics,fiorina2008political}), economics (e.g., \cite{fang2021does}), as well as complex systems researchers examining models inspired by perspectives as diverse as epidemiology (e.g., \cite{vasconcelos2019consensus}), and physics (e.g., \cite{galam2005local}). 

Furthermore, our case studies underscore key points that have shaped the design of the framework. First, they serve to highlight why one cannot simply study belief dynamics with representations that involve only a single numerical opinion. As models of polarization have shown, this does not generalise well. In particular, the difficulty of obtaining anything other than convergence in the majority of extant models of opinion dynamics illustrates that point. Enriching models with arguments or reasons in support of beliefs is thus essential for a more realistic understanding.

Second, doing so highlights how the distribution of arguments that are, in principle, available to agents---and from which the arguments they personally have available are sampled---strongly influences belief dynamics. This opens the door for a deeper understanding of extant theories, alternative mechanisms for known phenomena, and novel predictions that can be brought to real-world data. 

Third,  the fact that `the world' matters to observable belief dynamics furthermore makes it important that the distributional assumptions about underlying arguments or evidence are sufficiently grounded. The Bayesian framework helps with this problem because BN models of domains can be learned from real-world data \cite{heckerman2008tutorial}. 

Fourth,  the rules by which agents evaluate arguments and form beliefs also clearly matter. Hence it is important to supply agents with principled argument evaluation and aggregation rules. This is not to claim that humans are perfect Bayesians. They may approximate Bayesian norms more or less faithfully in some contexts, however \cite{peterson1967man,chater2006probabilistic,chater2010bayesian}. The analytic value of considering `rational' or optimal aggregation, though, does not ultimately rest on the precise extent of that approximation. Rather, consideration of optimal rules aids the identification of deep structural challenges that cognitive agents face and the attribution of `bias' or irrationality is meaningful only against the foil of the performance of an ideal agent \cite{hahn2014does}.

Fifth, even the simple simulations of our case studies highlight the fundamental impact of agents' communication rules, that is, what evidence they choose to communicate and why. This makes clear how much of the final model behaviour depends on a key component that is itself without a (clear) normative basis. This reflects a more general gap in normatively oriented research: normative accounts of \emph{where evidence actually comes} from are, arguably, under-developed in this and other contexts \cite{meder2022makes}. 

In fact, early work in social epistemology emphasised how simulations might aid the discovery of appropriate norms for communication \cite{olsson2011simulation}. Beyond the question of the frequency or intensity of communication within epistemic collectives (e.g., \cite{angere2017publish,zollman2010epistemic,BorgFreySeseljaStrasser019theory,hahn2018communication,hahn2022collectives} very little progress has been made with respect to this question. Arguably, this is because extant models have been too restrictive to support much in the way of variation in communication rules. Even the simple BNs explored in this paper, however, are rich enough to allow one to formulate key elements of the factors that plausibly influence communication in real life, such as speaker confidence (influencing whether or not to speak at all), own belief and perceptions of argument strength which feed into the pragmatic rules or maxims governing cooperative exchange \cite{grice1969utterer,levinson1983pragmatics}, as well as deviations in non-cooperative exchange.

Sixth and last, communication rules shape the rise and fall of arguments. Incorporating arguments into the model opens a new world of exploring argument dynamics alongside belief dynamics. Exploring how argument dynamics are shaped by communication rules should open up new avenues for linking argumentation research to work in computational social science. For one, examining real-world patterns of argument spread across online social media and comparing this with models could inform the identification of underlying (real-world) communication rules.

We conclude with some indication of future directions. 

\subsubsection{Future Directions} 
\label{section:future}
Crucially, NormAN is conceptualised not as a specific model, but as a framework in which to develop concrete models by adjusting key components: world, agent belief formation, communication rules,  and network characteristics.

As argued extensively above, the fact that our framework has normative grounding is (to us) an essential requirement. That said, however, it is entirely possible to strip away any such interpretation and simply treat the Bayesian belief revision implemented for our agents as purely descriptive, that is, as `just another update rule'. From that perspective, Bayesian belief revision is simply another weighted averaging rule \cite{jones2011bayesian}. This makes it an interesting research question how it compares, on both the individual and collective level, to other popular rules in models of opinion dynamics. 

Second, our initial case studies highlight just how much model behaviour (and, by the same token, real-world belief- and opinion dynamics) are shaped by agents' communication rules. This makes studying the impact of communication rules a central topic for future research. Accordingly, future versions of NormAN should implement richer notions of communication: for example, rules that include a model of the listener (e.g., that listener's knowledge states \cite{levinson1983pragmatics}). This would also enable agents to include strategic considerations in their communications \cite{roth2007strategic,matt2008game,rahwan2009arggames}.

Third, as mentioned above, future NormAN models should allow agents to possess subjective models of the world (BNs) that differ from the ground truth world, and that differ across agents. 

Fourth,  richer agent models should incorporate notions of trust in order to connect with the rich modelling literature on testimony  (e.g., \cite{olsson2011simulation,bovens2003bayesian,shafto2012epistemic}) and with bounded confidence models of opinion dynamics (e.g., \cite{hegselmann2002opinion}). 

With respect to `the world', future research should involve systematic exploration of the impact of different worlds, including richer models involving many more variables. 

Last, but not at all least,  there should be a much deeper exploration of the impact of network topology than the current version allows. In particular,  it will be important to study not just other types of networks (e.g., preferential attachment networks \cite{barabasi1999emergence}), and their impact on argument dynamics. This should also include dynamic networks, in which agents can change who they communicate with \cite{sekara2016fundamental}; this not only affords greater realism, but it will specifically allow the study of the epistemological and argumentative impacts of homophily \cite{mcpherson2001birds}.  Finally, this should include hierarchical networks \cite{ravasz2003hierarchical}.

\section{Conclusions}

In this paper, we argued that there is currently a significant gap in the research literature. On the one hand, traditional research on argumentation does not connect well with the many real-world contexts that involve more than two agents or competing perspectives. On the other, the wealth of research trying to understand belief and opinion dynamics across social networks is limited by the fact that it has not considered, or been able to consider properly, individual arguments as the actual information exchanged that drives those dynamics. In order to bridge that gap, agent-based models involving argument exchange are required. 

We argued further that a normative, Bayesian perspective provides a privileged way to build such a model. We have sought to outline why normative models, more generally, are relevant not just for research concerned with how we ought to behave, but also for descriptively oriented research concerns. More specifically, we have detailed how, within the argumentation literature, the Bayesian framework allowed one to capture the content of arguments with sufficient detail to advance long-standing research questions. We have detailed also how the Bayesian framework allows one to capture belief dynamics and evidence/argument aggregation. We have shown a novel application of the Bayesian framework: namely how Bayesian Belief Networks, can be employed to create a ground truth world and evidence distribution for agent-based simulations.  

These aspects are embodied in NormAN, a new framework for the study of argument exchange across social networks. We have sought to illustrate with case studies different ways in which NormAN models might benefit extant research. It is hoped that NormAN will help bridge the current `gap' and support new research across the breadth of research on argumentation, opinion dynamics, and communication discussed in this paper.

\begin{acks}
The research reported in this paper was supported by the UK's Arts and Humanities Research Council grant AH/V003380/1, and the Deutsche Forschungsgemeinschaft (DFG, German Research Foundation) project number 455912038. L.A. was supported by a Konrad-Adenauer Stiftung scholarship. Special thanks go to Michael Maes, Davide Grossi for many helpful discussions and Borut Trpin for feedback on an initial draft of this manuscript.

\subsection{CRediT statement}
Conceptualization: L.A., R.F., U.H., A.J., and L.S.
Data curation: K.P.
Formal analysis: K.P.
Funding acquisition: U.H.
Investigation: K.P.
Methodology: L.A., R.F., U.H., A.J., K.P., and L.S.
Project administration: U.H.
Resources: K.P.
Software: L.A., A.J., and L.S.
Validation: L.A., U.H., K.P., and L.S.
Visualization: R.F., U.H., and K.P.
Writing - original draft: L.A. and U.H.
Writing - review \& editing: L.A., R.F., U.H., A.J., K.P., and L.S.
\end{acks}

\bibliographystyle{ACM-Reference-Format}

\bibliography{Tidy}


\appendix

\small  

\begin{table}[H]
  \label{tab:FullParams}
  \begin{tabular}{cccl}
    \toprule
    Entity & Variable & Value range/Type & Description\\
    \midrule
    World & \texttt{causal-structure} & Bayesian network & Determines relation between evidence and hypothesis, \\
    
   &&& used by agents and world model to compute probabilities.  \\

     & \texttt{hypothesis}  & Variable & The hypothesis proposition (truth values: \textit{true} or \textit{false}).\\
     
    & \texttt{hypothesis-probability}  & $0 - 1$ & Probability that hypothesis node is true.\\
    
    & \texttt{evidence-propositions}  & List & The evidence propositions (truth values: \textit{true} or \textit{false}).\\
     
& \texttt{evidence-probabilities-list}  & List & The marginal probabilities of the individual pieces of evidence, \\
& & & given the truth/falsity of the hypothesis.\\

    & \texttt{evidence-list} & List & Stores truth values of evidence nodes.\\
    & \texttt{optimal-posterior} & $0 - 1$ & Stores the probability of the hypothesis conditional on the truth  \\
    & & & values of all pieces of evidence (i.e., \texttt{evidence-list}).\\

   Agents & \texttt{agent-evidence-list} & List & Stores truth values of evidence encountered by agents.\\
   
   & \texttt{agent-belief} & 0 - 1 & Belief in the hypothesis. \\
      & \texttt{update-list} & List & Stores update magnitude on evidence reception.\\
       & \texttt{\texttt{recency-list}} & List & Basis for the recency sharing rule.\\
     & \texttt{chattiness} & $0 - 1$  & Probability of communication. \\
     & \texttt{curiosity}  & $0 - 1$ & Probability of inquiry (evidence gathering). \\
     & \texttt{conviction-threshold}  & $0 - 1$ & Co-determines conviction in claim agent requires to join debate. \\
     & \texttt{initial-belief}  & $0 - 1$ & Stores the unconditional prior belief in the hypothesis. \\
     
      & \texttt{max-draws} & Integer & Determines the maximal number of evidence collections. \\
      & \texttt{initial-draws} & Integer & Number of evidence collections performed before the model run. \\
      & \texttt{share} & Chooser & Determines agents'  communication rule: \\
      
      & &  & Selective, Random, Recent.\\

    Network & \texttt{number-of-agents} & 1 - 1000 & Determines the size of the network.\\
     & \texttt{social-network} & Chooser & Selects network type: Wheel, Complete, Small-world, Null.\\
       & \texttt{rewiring-probability} & $0 - 1$ & Co-determines small-world network.\\
        & \texttt{k} & $0 - 10$ & Co-determines small-world network. \\
 
 Simulation & & &\\
 
 \tiny (variables in the user interface) \normalsize &\texttt{ticks} & Integer & Counts discrete time steps (NetLogo primitive).\\
 \normalsize &\texttt{setup} & Button & Initializes a new simulation run.\\
 &\texttt{go} & Button & Starts a model run.\\

      & \texttt{reset-world-?} & On/Off & Determines if world model is reset when \texttt{setup} is pressed.\\
     & \texttt{reset-social-network-?}  & On/Off & Determines if the social network is reset when \texttt{setup} is pressed.\\
       & \small \texttt{reset-agents-initial-evidence-?} \normalsize  & On/Off & Determines if the initial evidence \\
       & & & of agents is reset when \texttt{setup} is pressed.\\
      
       & \texttt{max-ticks} & Integer & Determines the length of the model run. \\
        & \texttt{stop-at-max-ticks-?} & On/Off & Determines if simulation stops when \texttt{ticks}= \texttt{max-ticks}. \\
        & \texttt{stop-at-full-information-?} & On/Off & Determines if simulation stops when all agents know all \\
        &&&  evidence \small (i.e., \texttt{agent-evidence-list}= \texttt{evidence-list}). \normalsize 
        \\
        
       & \texttt{show-me-?} & On/Off & Optionally prints the agents' protocols in NetLogo's monitor. \\
        & \texttt{plotting-type} & Chooser &  Values: \{$uttered$, $sent-to$, $received-as-novel$\}. \\
        & & & Determines the type of argument graph plotted in the interface.\\
       
        & \texttt{approximation} & Chooser & Values: \{\textit{seed}, \textit{repeater}, \textit{received-as-novel}\}. \\
        & & & \small  Type of approximation used when performing any \textit{cpquery()}.\normalsize \\

        & \texttt{seed} & Integer & Input for \textit{set-seed()} used in the seed approximation. \\
        & \texttt{repeater} & Integer & Determines the repetitions used by the $repeater$ approximation. \\
        & \texttt{evidence-nodes} & Output \tiny (List) \small & Outputs the chosen evidence node from the Bayesian network. \\
        & \texttt{hypothesis-node} & Output & Outputs the chosen hypothesis node from the Bayesian network. \\
 & \texttt{causal-structure} & Chooser & Takes as values the preset BNs and \textit{custom}. \\
        & \texttt{path-to-custom-DAG} & Input & File path to bnlearn file (\texttt{causal-structure}= $custom$). \\
          & \texttt{evidence-nodes-custom-DAG} & Input  \tiny (List) \small & Determines evidence nodes in customized BN. \\
            & \texttt{hypothesis-node-custom-DAG} & Input  \tiny (List) \small & Determines hypothesis node in customized BN. \\
            
            & \small \texttt{custom-evidence-and-hypothesis-?} \normalsize  & On/Off & Activates custom evidence and hypothesis nodes \\

            & & & (overriding presets). \\

  \bottomrule
  
\end{tabular}
\caption{Full list of model parameters implemented in NormAN version 1.0.}
\label{fig:table-full}
\end{table}
\normalsize 
\newpage 

\section{ODD Protocol (Full)}
\label{section: ODDfull}
This section follows `The ODD protocol: a review and first update' \cite{grimm2010odd}.

\subsection{Purpose}
\label{section:purpose}

The purpose of the NormAN 1.0 model is to introduce a flexible framework for modelling argument exchange across social networks. NormAN 1.0 is an agent-based model in which agents discuss a central hypothesis by exchanging distinct pieces of evidence (i.e., argumentation). It consists of a world model that includes the hypothesis (the claim at issue), evidence, and the causal structure of the world. It consists further of agents within a social network. And it consists of rules that govern the deliberation dynamics. NormAN was developed to meet the following requirements: 

\begin{itemize}
    \item Capture a ground truth world
    \item Represent evidence as linked to the ground truth world, i.e., a principled evidence distribution
    \item Agents who are rational with respect to belief formation
    \item Facilitation of easy extension and usage
\end{itemize}

The questions that can be addressed with the NormAN framework are diverse and include issues such as polarization and the emergence of consensus, argument dynamics across social networks, the truth-tracking potential of deliberative bodies and many more.

\subsection{Entities, State Variables and Scales}
\label{section:entities}

The model features three different kinds of entities: agents, the world model, which specifies the deliberation `environment', and the social network, which specifies the social environment. A list of the specific state variables can be found in Table \ref{fig:table-full}. We divide this section into three subsections: the world, the agents, and the networks.

\subsubsection{The World}
The world model consists of a Bayesian network. Formally, a Bayesian Network (BN) $B = \langle G,P \rangle$ is a directed acyclic graph $G = \langle V,E \rangle$, with a set of nodes (variables) $V$ and edges $E$, and a joint probability distribution $P$ over $V$ such that $G$ satisfies the parental Markov condition together with $P$. Each node $\in V$ has a set of possible truth values. (In the networks we deal with in this paper, most nodes are Boolean. So their truth values are \textit{true} and \textit{false}.) NormAN 1.0 provides a number of BNs (some well-known, others novel) as presets. They are described in detail in Section \ref{section:BNworlds} of this Appendix.

In this network, the modeller identifies one variable as the \texttt{hypothesis}, or \texttt{H} for short (e.g., `lung cancer' in the well-known `asia' network), and chooses a subset of the other nodes as evidence nodes $\texttt{E}_1, \texttt{E}_2, \ldots, \texttt{E}_n$. In NormAN 1.0, the evidence nodes and the hypothesis node must be Boolean---although other nodes in the BN  may be many-valued. NormAN assigns a truth value to the hypothesis (manually or probabilistically). This value is used to determine the values of the evidence nodes: The marginal conditional probability of each evidence (conditional on the value of the hypothesis node) is calculated; and on initialisation, this chance stochastically determines the truth value of the evidence. For example, if it is true that the patient has lung cancer, and $P(bronchitis|lungcancer)=0.2$, then there is a 20\% chance that the value of $bronchitis$ is \textit{true}. If the evidence nodes are Boolean, this procedure yields a chain of evidence in the form of $\neg E_1, E_2, E_3, \ldots, \neg E_n$ (where $E_i$ denotes `$\texttt{E}_i$ is true', and   $\neg E_i$ denotes `$\texttt{E}_i$' is false). Whenever an agent encounters a piece of evidence, they receive information about its truth value. Therefore, in the remainder of this model description, we will speak of agents receiving pieces of evidence such as $\neg E_3$.

These features are captured by the following variables: \texttt{causal-structure} determines the chosen BN, \texttt{evidence-list} stores the list of the evidence nodes' truth values. The two are mediated by \texttt{evidence-probabilities-list}, which stores the marginal probabilities of the evidence given the actual value of the \texttt{hypothesis} (which is stochastically determined by \texttt{hypothesis-probability}). Finally, the world model computes and stores the so-called `optimal posterior', the posterior degree of belief an agent would compute if they maintained \texttt{causal-structure} and knew all the evidence (formally, this is the case for an agent whenever their \texttt{agent-evidence-list} is the same list as \texttt{evidence-list}). In model runs of NormAN, the convergence of \texttt{agent-belief} variables to \texttt{optimal-posterior} is a common result of certain sharing rules.

\subsubsection{The Agents} Each agent is characterised by (i) their degree of belief in the hypothesis (variable \texttt{agent-belief}), (ii) their representation of the causal structure of the world, and (iii) a list of evidence they have already encountered (variable \texttt{agent-evidence-list}). First, each agent assigns a degree of belief to the hypothesis (\texttt{agent-belief}). Second, they use a Bayesian network that connects the evidence and the hypothesis and acts as their representation of the world. Third, they store the truth values of all encountered evidence in their \texttt{agent-evidence-list}. These three aspects are related in a dynamic, straightforward way. Suppose an agent $A$ stores the following chain of evidence at time $t$: $\texttt{agent-evidence-list}_A^{t} =\{E_1, \neg E_3\}$.\footnote{Technically, this is implemented in the NormAN code by a personalised list of values as long as the list of evidence of the world model. The agent's \texttt{agent-evidence-list} stores the truth value of evidence they have already encountered, leaving empty list items for evidence they have not yet encountered. For example, if the world's evidence list is $\{E_1, E_2,\neg E_3, E_4\}$, agent $A$'s \texttt{agent-evidence-list} list could be $\{E_1,-,\neg E_3,-\}$.} In that case, they will use their Bayesian network to compute $\texttt{agent-belief}_A^{t}=P(H|E_1, \neg E_3)$. Whenever agents encounter a new piece of evidence (e.g., $E_2$), they update their degree of belief (e.g., $\texttt{agent-belief}_A^{t+1} =P(H|E_1, E_2, \neg E_3)$). We call this updating 
 procedure of \texttt{agent-belief} through conditionalization on the agents' \texttt{agent-evidence-list} `\texttt{COMPUTE-POSTERIOR}'. When the agent's \texttt{agent-evidence-list} is empty, that is, when they have not yet encountered any evidence, their \texttt{agent-belief} is simply the base rate (marginal probability) of the hypothesis node in their BN. This prior value is stored in an agent-variable \texttt{initial-belief}.

In the first version of NormAN presented here, we assume that each agent's Bayesian network simply corresponds to the world model's network, \texttt{causal-structure}: that is, we assume that agents represent the world correctly. Furthermore, three agent variables characterize agents in their deliberation procedure: their \texttt{chattiness}, their \texttt{curiosity} and their \texttt{conviction-threshold}. The nature and function of these variables are explained in the next subsection (\ref{section:process}), for they govern the model's dynamics.

\subsubsection{The Social Network}

The model places \texttt{number-of-agents} (slider in the interface) agents on a grid and then specifies who is connected to whom via undirected communication links. Agents can only communicate with their link neighbours. NormAN provides four different network structures that the modeller can select before initialization:  complete network, `wheel' (cf. \cite{zollman2010epistemic, frey2020robustness}), a `null' network (i.e., all agents are disconnected) and small-world networks (also known as Watts-Strogatz networks \cite{watts1998collective}, implemented in NetLogo using \cite{wilensky2005netlogo}).  Which network is used is determined by the variable \texttt{social-network} (chooser in the interface). The specific form of the Watts-Strogatz networks is (stochastically) determined by two variables: \texttt{rewiring-probability} and \texttt{k}, which are determined by the modeller.

\newpage 

\begin{figure}[H]
    \centering
    \includegraphics[width=0.8\linewidth]{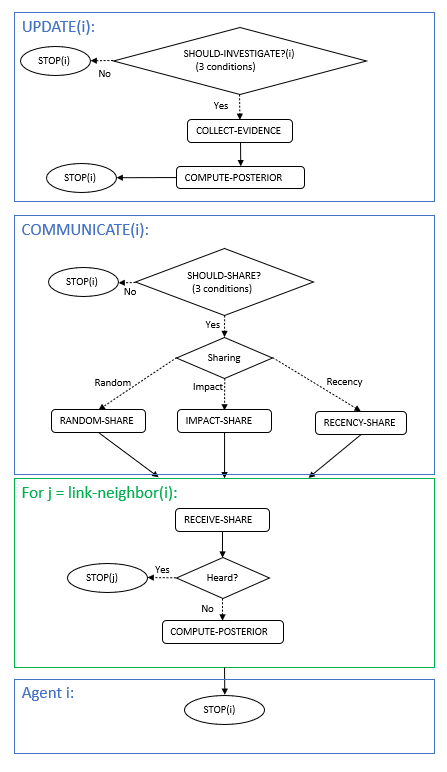}
    \caption{Flowchart for the update performed by each  agent $i$ and induced updates for out-link-neighbours $j$ called by $i$ during sharing.}
    \label{Agent-Update-Flowchart}
\end{figure}

\subsection{Process Overview and Scheduling}
\label{section:process}

The model evolves in discrete time steps. In each step, agents have the opportunity to inquire, that is, to collect evidence and to communicate/receive communication from their link neighbours. Collecting evidence facilitates the flow of information into the network, and communication facilitates the flow of information through the network. Whenever an agent learns a new piece of evidence, they re-compute their belief (\texttt{COMPUTE-POSTERIOR}). The dynamics of the model are entirely determined by the behaviour of the agents. Simply put, each simulation step has the following schedule:

\begin{enumerate}
    \item Collect evidence: The agents collect a new piece of evidence.
    \item Communication: The agents share a piece of evidence they have already encountered with their link neighbours.
\end{enumerate}

However, agents only collect and only communicate if certain conditions are met; and communication may be determined by different rules. This subsection explains the agents' schedule in detail.

First, in each round, agents may inquire by collecting evidence. To do so, they add the truth value of one random evidence node from \texttt{evidence-list}, which they have not yet encountered, to their \texttt{agent-evidence-list}.\footnote{As an example, suppose that at time $t$, agent $A$ stores the truth values $\texttt{agent-evidence-list}_A^{t}=\{E_1, \neg E_3\}$. Through inquiry, they may find that $E_2$ is indeed true, thus extending their list to $\texttt{agent-evidence-list}_A^{t+1}=\{E_1, E_2, \neg E_3\}$.} Inquiry is therefore modelled by `drawing' from the world model's \texttt{evidence-list}. Two agent variables govern inquiry. First, agents have a fixed maximum number of inquiries, \texttt{max-draws}: they can only collect a limited amount of times. Hence, if  \texttt{max-draws} is smaller than the number of evidence, agents in NormAN have limited access to the evidence: if they do not communicate with their peers, they will not learn the entirety of the evidence on their own.  Second,  agents have a \texttt{curiosity} score (between 0 and 1), which determines the probability they will inquire in any given round. The values of \texttt{curiosity} and \texttt{max-draws} are currently determined globally (i.e. they are the same for each agent). Agents only draw from the \texttt{evidence-list} if they are curious (stochastically determined), and if they still have draws. Meeting these two conditions is noted as the condition `\texttt{SHOULD-INVESTIGATE-?}' in the flowchart (\ref{Agent-Update-Flowchart}). We call the triggered collection procedure \texttt{COLLECT-EVIDENCE}.

Second, in each round, agents may communicate and receive evidence via communication. In this first version of NormAN, communication is modelled via a simple transmission (`passing the parcel') mechanism: the communicating agent $A$ chooses which piece of evidence (i.e., which truth value) to transmit to their link neighbours. If the recipient has not yet heard this piece of evidence, they add it to their \texttt{agent-evidence-list} and compute a new \texttt{agent-belief}. If they have heard it, their \texttt{agent-evidence-list} list remains unchanged. Hence, NormAN's agents recognize indexed, distinct pieces of evidence and never `double count'. \footnote{For instance, if $A$'s evidence list is $\texttt{agent-evidence-list}_A^{i}=\{E_1,\neg E_3\}$, and $B$'s evidence list is $\texttt{agent-evidence-list}_B^{i}=\{ \neg E_3\}$, then $A$'s sharing $E_1$ will enrich agent $B$'s evidence list to $\texttt{agent-evidence-list}_B^{i+1}=\{E_1,\neg E_3\}$. Had $A$ chosen $\neg E_3$, $B$'s list would have remained unchanged; hence agents recognize evidence, and never `double count' a piece. }  

Although this mechanism of `passing parcels' is simple in that it avoids the complexity of testimony, it can still be used to capture distinct, complex styles of communication. This is because it is non-trivial what evidence the agent chooses to transmit. In this first version of NormAN, three sharing rules are implemented:

\begin{itemize}
  
    \item \texttt{RANDOM-SHARE}: Communicating agents share a random piece of evidence from their \texttt{agent-evidence-list}.
    \item \texttt{RECENCT-SHARE}: Communicating agents (likely) share the piece of evidence they most recently encountered. 
    \item \texttt{IMPACT-SHARE}: Communicating agents share the piece of evidence that they hold to be the best piece of evidence in favour of their current positions.
\end{itemize}

Although \texttt{RANDOM-SHARE} needs no further elaboration, \texttt{RECENCT-SHARE} and \texttt{IMPACT-SHARE} do. The recency rule (loosely inspired by Maes and Flache's model of bi-polarization \cite{mas2013differentiation}) follows this procedure: agents use an ordered \texttt{\texttt{recency-list}} that functions as a stack. It stores the order in which evidence has last been encountered, pushing novel pieces of evidence and popping previously heard ones. For instance, agent $A$ might entertain the recency list $\texttt{\texttt{recency-list}}_A^{t}=\{E_3, E_4, E_1\}$. If they now hear a new piece of evidence, e.g., $E_5$, they extend this list to $\texttt{\texttt{recency-list}}_A^{t+1}=\{E_3, E_4, E_1, E_5\}$. If they have already accepted a piece of evidence, encountering it again will bring that piece to the top of the list. For instance, if $A$ encounters $E_3$ while entertaining $\texttt{\texttt{recency-list}}_A^{t}=\{E_3, E_4, E_1\}$, their new recency list will be $\texttt{\texttt{recency-list}}_A^{t+1}=\{E_4, E_1, E_3\}$. The order of the list is crucial because agents will preferably share what they last encountered (i.e., the last list item). In NormAN 1.0, this tendency is encoded stochastically: with a probability of $x$ (in the base model set to $x=0.9$), communicating agents will choose to share the last item of their \texttt{\texttt{recency-list}}. With a probability of $1-x$, they will share a random other piece from their list. 

The impact-sharing rule captures the following rationale: agents share what they currently hold to be their best piece of evidence. By best, we mean the piece of evidence that most supports the agent's current position on the hypothesis --- the strongest piece. To this end,  NormAN agents track the magnitude of the belief update---the impact--- that each piece of evidence induced. Hence, for each piece of evidence $E_i$ received, agents store the following update magnitude $\in [-1, 1]$: $P(H|E_i)-\texttt{initial-belief}$ (where \texttt{initial-belief} is the agent's prior belief before receiving any evidence). This impact is stored in each agent's own \texttt{update-list}. If the agent believes the hypothesis to a higher degree than their \texttt{initial-belief} they choose to communicate the piece of evidence with the highest update value (and they choose the lowest if they believe the hypothesis to a lower degree). That is, agents will share evidence for the direction in which their own \texttt{agent-belief} has been moved over the course of the simulation.

Which sharing rule is used is determined by a global variable \texttt{share} which takes one of the values  $\{\textit{random}, \textit{recent}, \textit{impact}\}$ (determined by the modeller in the interface). The value fixes every agent's sharing rule (there are no mixed populations). Whenever an agent shares a piece of evidence, they transmit it to the entirety of their link-neighbours, as determined by the social network.\\

\noindent Agents communicate (using one of the above rules) given three conditions are met (\texttt{SHOULD-SHARE?} in the flowchart \ref{Agent-Update-Flowchart}):
\begin{enumerate}
  \item Trivially, agents may only share a piece of evidence if they have previously encountered one (i.e., \texttt{agent-evidence-list} must not be empty).
    \item The global value \texttt{conviction-threshold} $\in [0,1]$ determines when an \texttt{agent-belief} is sufficiently distinct from their \texttt{initial-belief} that they choose to communicate.  The conviction threshold is a percentage value that serves as a cut-off point for when an agent’s belief departs sufficiently from the agnostic (pre-evidence) prior (\texttt{initial-belief}) for them to jump into the discussion. Specifically, passing the threshold requires an agent's belief to fulfil the following conditions: either $\texttt{agent-belief} < (\texttt{initial-belief} - \texttt{initial-belief} \cdot \texttt{conviction-threshold})$ (i.e., their belief passes below the lower bound) or  $\texttt{agent-belief} > (\texttt{initial-belief}  + (1 - \texttt{initial-belief} ) * \texttt{conviction-threshold})$ (i.e., their belief surpasses the upper bound). When either of these conditions is met, we deem an \texttt{agent-belief} to pass the threshold. Note that if  $\texttt{conviction-threshold}$ is set to $0$, this condition is trivially met in most cases (specifically, whenever \texttt{agent-belief} $\not =$ \texttt{initial-belief}).
    \item The global variable \texttt{chattiness} $\in [0,1]$ encodes any agent's chance that they will communicate with their link-neighbours on a given round. 
\end{enumerate}

The precise protocol is summarized in the following pseudo-code.


\subsection*{Pseudo-Code}
\medskip
Typesetting for data types:
\begin{itemize}
    \item \texttt{variables} and \texttt{lists}
    \item \textit{values} 
    \item \textit{\textbf{Boolean values}} (true/false)
    \item \texttt{PROCEDURES} 
    \item \textbf{control flow and logic operators}
    \item Agent-index $i$ for variables, e.g. \texttt{shared-piece}($i$) is omitted when there is no ambiguity.
\end{itemize}
\medskip
Special Primitives:
\begin{itemize}
    \item \texttt{RANDOM-FLOAT}($input$): function that outputs a random real number smaller than $input$ (number).
    \item \texttt{ENQUEUE}(\textit{value};\texttt{list}): puts \textit{value} at end (last position) of \texttt{list}. 
    \item \texttt{DEQUEUE}(\texttt{list}): outputs last item from \texttt{list}.
    \item \texttt{REMOVE}(\textit{value};\texttt{list}): removes all occurrences of \textit{value} from \texttt{list}.
    \item \texttt{INSERT}(\textit{value}; \texttt{list}; $index$): inserts \textit{value} to \texttt{list} at position $index$.
    \item \texttt{POSITION}(\textit{value}; \texttt{list}) outputs first such $index$ of \textit{value} in \texttt{list}.
    \item \texttt{POSTERIOR}(\textit{prior}; \texttt{evidence})) outputs the posterior probability $P(h|e_1,...,e_n)$ for \textit{prior} (probability value) $h$ and \texttt{evidence} (list) $e_1,...,e_n$.
\end{itemize}

\bigskip

After initialising the model (specifying parameter settings for agents and loading in the selected DAG), the main procedure (\texttt{GO}) runs until the stopping condition (when every agent has seen each piece of evidence) is satisfied. The general procedure looks like this:

\begin{itemize}
 \item[]------------------------------------------------------
    \item[]\textbf{NormAN}: \texttt{GO}
    \item[] \textbf{while} STOPPING-CONDITION = \emph{false}
    \begin{itemize}
    \item[] \textbf{for} $i\in$ [0; number-of-agents]
        \begin{itemize}
           \item[] UPDATE-AGENT($i$) 
        \end{itemize}
    \item[] \textbf{end for}  
    \item[] \textbf{for} $i\in$ [0; number-of-agents]
        \begin{itemize}
           \item[] COMMUNICATE($i$) 
        \end{itemize}
    \item[] \textbf{end for}  
    \end{itemize}
    \item[] \textbf{end while}
    \item[] \textbf{End NormAN}: \texttt{GO}
     \item[]------------------------------------------------------
\end{itemize}

\bigskip

\newpage

The corresponding simple flow diagram is:

\begin{figure}[H]
    \centering
    \includegraphics[width=0.6\linewidth]{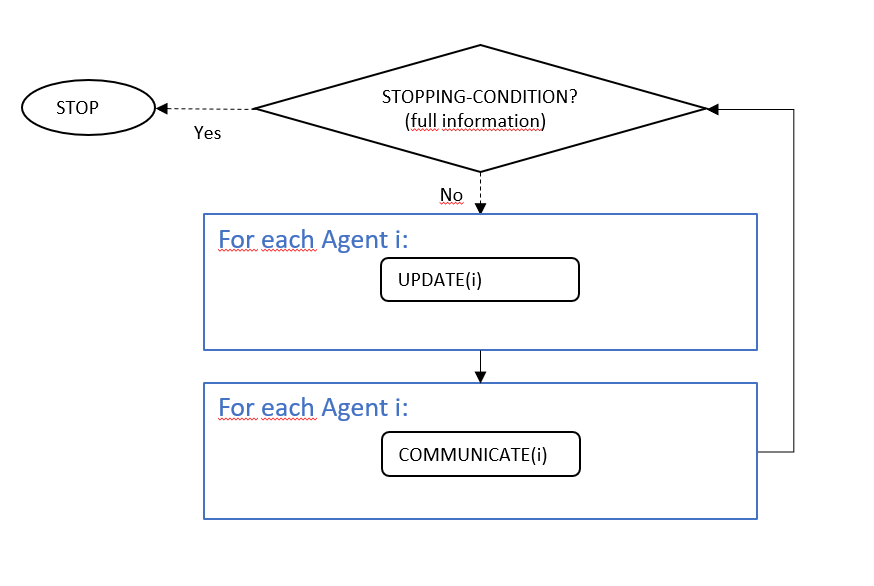}
    \caption{Flowchart for the global GO-Procedure.}
    \label{fig:enter-label}
\end{figure}

\bigskip

We next break down the Updating procedure for each agent (explained in pseudocode and visualized as a flow chart in Fig. \ref{Agent-Update-Flowchart} above):

\bigskip

\begin{itemize}
   \item[]------------------------------------------------------
    \item[]\textbf{Algorithm 0.I}: \texttt{UPDATE-AGENT}($i$) 
    \begin{itemize}
        \item[] \texttt{SHOULD-INQUIRE?}($i$)
        \item[] \textbf{if} \texttt{SHOULD-INQUIRE?}($i$) = \textit{\textbf{true}}, \textbf{then}
        \begin{itemize} 
              \item[] \texttt{COLLECT-EVIDENCE}($i$)
              \item[] \texttt{COMPUTE-POSTERIOR}($i$)
        \end{itemize}
        \item[]\textbf{end if}
    \end{itemize} 
    \item[] \textbf{End Algorithm 0.I}: \texttt{UPDATE-AGENT}($i$)
   \item[]------------------------------------------------------
\end{itemize}

\bigskip

\begin{itemize}
 \item[]------------------------------------------------------
    \item[]\textbf{Algorithm 0.II}: \texttt{COMMUNICATE}($i$) 
     \item[] \textbf{if} \texttt{SHOULD-SHARE?}($i$) \textbf{then}  
        \begin{itemize} 
            \item[] \textbf{if} \texttt{sharing} = \emph{random} \textbf{then} 
            \begin{itemize}
                \item[] \texttt{shared-piece}($i$) $\leftarrow$ \texttt{RANDOM-SHARE}($i$)
            \end{itemize}
            \item[] \textbf{else if} \texttt{sharing} = \emph{impact} \textbf{then} 
            \begin{itemize}
                \item[] \texttt{shared-piece}($i$) $\leftarrow$ \texttt{IMPACT-SHARE}($i$)
            \end{itemize}
            \item[] \textbf{else if} \texttt{sharing} = \emph{recency} \textbf{then} 
              \begin{itemize}
                \item[] \texttt{shared-piece}($i$) $\leftarrow$ \texttt{RECENCY-SHARE}($i$)
            \end{itemize}
            \item[]\textbf{end if}
            \item[] \textbf{for} $j\in$ \texttt{outlink-neighbors}
            \begin{itemize}
                \item[] \texttt{RECEIVE-SHARE}($j$) 
                \item[] \textbf{if} \texttt{shared-piece}($i$) $\not\in$ \texttt{agent-evidence-list}($j$) \textbf{then}
                    \item[] ~~ ~~ ~~ ~~ ~~ ~~ ~~ ~~ ~~ ~~ ~~ ~~ \texttt{COMPUTE-POSTERIOR}($j$) 
                \item[] \textbf{end if} 
            \end{itemize}
            \item[] \textbf{end for} 
        \end{itemize}
        \item[]\textbf{end if}
        \item[] \textbf{End Algorithm 0.II}: \texttt{COMMUNICATE}($i$)
   \item[]------------------------------------------------------
\end{itemize}

\bigskip

\begin{itemize}
\item[]------------------------------------------------------
    \item[]\textbf{Algorithm 1}: \texttt{SHOULD-INQUIRE?}($i$) 
    \begin{itemize}
        \item[] \textbf{if} \texttt{draws} > 0 \textbf{and} \texttt{curiosity} > \texttt{RANDOM-FLOAT}(1) \textbf{and} |\texttt{agent-evidence-list}| < |\texttt{evidence-list}| \textbf{then} 
        \begin{itemize} 
              \item[] \textbf{output} \textit{\textbf{true}} 
        \end{itemize}
        \item[] \textbf{else}
         \begin{itemize} 
              \item[] \textbf{output} \textit{\textbf{false}} 
        \end{itemize}
        \item[] \textbf{end if}
    \end{itemize} 
    \item[] \textbf{End Algorithm 1}: \texttt{SHOULD-INQUIRE?}($i$)
   \item[]------------------------------------------------------
\end{itemize}

\bigskip

\begin{itemize}
\item[]------------------------------------------------------
    \item[]\textbf{Algorithm 2}: \texttt{COLLECT-EVIDENCE}($i$)
    \begin{itemize}
        \item[] $e\leftarrow$ \texttt{RANDOM-SELECT}(1, \texttt{evidence-list})
        \item[] \textbf{if} $e\not\in$ \texttt{agent-evidence-list} \textbf{then} 
        \begin{itemize}
            \item[] \texttt{agent-evidence-list} $\leftarrow$ \texttt{INSERT}($e$; \texttt{agent-evidence-list}; \texttt{POSITION}($e$; \texttt{evidence-list}))
        \end{itemize}   
    \end{itemize} 
    \item[] \textbf{End Algorithm 2}: \texttt{COLLECT-EVIDENCE}($i$)
    \item[]------------------------------------------------------
\end{itemize}

\bigskip

\begin{itemize}
\item[]------------------------------------------------------
    \item[]\textbf{Algorithm 3}: \texttt{COMPUTE-POSTERIOR}($i$)
    \begin{itemize}
        \item[] \texttt{singular} $\leftarrow$ \texttt{POSTERIOR}(\texttt{initial-belief}; $e$)
        \item[] \texttt{agent-belief} $\leftarrow$ \texttt{POSTERIOR}(\texttt{agent-belief};\texttt{agent-evidence-list}) 
        \item[] \texttt{update} $\leftarrow$ \texttt{singular} $-$ \texttt{initial-belief} 
        \item[] \texttt{update-list} $\leftarrow$ \texttt{ENQUEUE}(\texttt{update}; \texttt{update-list})
    \end{itemize}
    \item[]\textbf{End Algorithm 3}: \texttt{COMPUTE-POSTERIOR}($i$)
\item[]------------------------------------------------------
\end{itemize} 

\bigskip

\begin{itemize}
\item[]------------------------------------------------------
    \item[] \textbf{Algorithm 4}: \texttt{SHOULD-SHARE?}($i$) 
    \item[] |\texttt{agent-evidence-list}| > 0 \textbf{and} \texttt{chattiness} > \texttt{RANDOM-FLOAT}(1) \textbf{and} \texttt{agent-belief} $\in$ \texttt{conviction} \textbf{then}
        \begin{itemize} 
              \item[] \textbf{output} \textit{\textbf{true}} 
        \end{itemize}
        \item[] \textbf{else}
         \begin{itemize} 
              \item[] \textbf{output} \textit{\textbf{false}} 
        \end{itemize}
        \item[] \textbf{end if}
    \item[] \textbf{End Algorithm 4}: \texttt{SHOULD-SHARE?}($i$)
   \item[]------------------------------------------------------
\end{itemize}

\bigskip

\begin{itemize}
\item[]------------------------------------------------------
    \item[]\textbf{Algorithm 5a}: \texttt{RANDOM-SHARE}($i$)
              \begin{itemize}
                  \item[] \texttt{shared-piece} $\leftarrow$ RANDOM-SELECT(1,\texttt{agent-evidence-list})
              \end{itemize}
    \item[]\textbf{End Algorithm 5a}: \texttt{RANDOM-SHARE}($i$)
    \item[]------------------------------------------------------
\end{itemize} 

\bigskip

\begin{itemize}
\item[]------------------------------------------------------
\item[] \textbf{Algorithm 5b}: \texttt{IMPACT-SHARE}($i$)
      \begin{itemize}
                  \item[] \textbf{if} \texttt{agent-belief} > \texttt{initial-belief} \textbf{then}
                  \begin{itemize}
                      \item[] \texttt{shared-piece} $\leftarrow$ $\max$ \texttt{update-list}
                  \end{itemize}
                  \textbf{else}
                  \begin{itemize}
                      \item[] \texttt{shared-piece} $\leftarrow$ $\min$ \texttt{update-list}
                  \end{itemize}
                  \item[] \textbf{end if}
              \end{itemize}
        \item[]\textbf{End Algorithm 5b}: \texttt{IMPACT-SHARE}($i$)
        \item[]------------------------------------------------------
\end{itemize}

\bigskip

\begin{itemize}
\item[]------------------------------------------------------
\item[] \textbf{Algorithm 5c}: \texttt{RECENCY-SHARE}($i$)
       \begin{itemize}
                  \item[] \texttt{shared-piece} $\leftarrow$ \texttt{DEQUEUE}(\texttt{recency-list})
          \end{itemize}
        \item[]\textbf{End Algorithm 5c}: \texttt{RECENCY-SHARE}($i$)
        \item[]------------------------------------------------------
\end{itemize}

\bigskip

\begin{itemize}
\item[]------------------------------------------------------
    \item[]\textbf{Algorithm 6}: \texttt{RECEIVE-SHARE}($j$)
    \begin{itemize}
        \item[] \texttt{recency-list}($j$) $\leftarrow$ \texttt{REMOVE}(\texttt{recency-list}($j$); \texttt{shared-piece}($i$)) 
        \item[] \texttt{recency-list}($j$) $\leftarrow$ \texttt{ENQUEUE}(\texttt{shared-piece}($i$); \texttt{recency-list}($j$))
        \item[] \texttt{agent-evidence-list}($j$) $\leftarrow$ \texttt{INSERT}(\texttt{agent-evidence-list}($j$); \texttt{shared-piece}($i$); \texttt{POSITION}(\texttt{shared-piece}($i$); \texttt{evidence}))
    \end{itemize}
    \item[]\textbf{End Algorithm 6}: \texttt{RECEIVE-SHARE}($j$)
   \item[]------------------------------------------------------
\end{itemize} 

\newpage

\newpage

\subsection{Design Concepts}
\label{section:designconcepts}
\subsubsection{Basic Principles}

The NormAN 1.0 model introduces a new framework for modelling the exchange of multiple arguments across agents in a social network: Normative Argument Exchange across Networks. The model is composed of two submodels (elaborated in section `Sub-Models' \ref{section:submodels}). The way agents update their beliefs with respect to new evidence is handled by a purely Bayesian framework. This encompasses the revisions of the agents' \texttt{agent-belief} and \texttt{update-list}. Similarly, the values of the global variables \texttt{evidence-list} and \texttt{optimal-posterior} are computed based on Bayesian conditionalization. The values of these variables then co-determine the agents' behaviour. 

All other factors co-determining agent behaviour, that is, inquiry and sharing, are handled by a second submodel, which encompasses the \texttt{social-network} and the agents' sharing and inquiring behaviour (e.g. \texttt{max-draws} and the \texttt{share} variable). These features fall outside the purely Bayesian framework. Their design has either been taken from (or inspired by) the literature or constitutes an initial suggestion on our part for the purposes of exploration in NormAN. Among the former (derived from the literature) we count the network structures embedded via \texttt{social-network} (the Watts-Strogatz network \cite{watts1998collective}, the wheel network \cite{zollman2007communication}), and the recency sharing rule embedded in \texttt{share} (loosely inspired by Maes and Flache's model of bi-polarization \cite{mas2013differentiation}).

\subsubsection{Emergence}

The behaviour of NormAN is determined by changes in the agents' doxastic states, that is, their \texttt{agent-beliefs} and their \texttt{agent-evidence-list}. As the case studies \ref{section:case1}  and \ref{section:case2} show, this means that the main emergence of interest is the trajectory of the group's beliefs. Since NormAN is a flexible framework, this paper does seek to demarcate the set of possibly emergent phenomena; rather, its case studies showcase that known social phenomena can be captured by the model. In particular, the case studies capture the shift to extremity, belief divergence, polarization and convergence to consensus (all measured via \texttt{agent-beliefs}). Additionally, NormAN provides insights into the relative sharing frequencies of arguments (individual pieces of evidence drawn from \texttt{evidence-list}) across the social network.

\subsubsection{Sensing}

Agents have no direct access to the truth value of the hypothesis node and only limited access to \texttt{evidence-list}. Agents fill in their own \texttt{agent-evidence-list} as they encounter evidence through inquiry or communication.  

Agents have full access to their own variables determining their doxastic states, that is \texttt{agent-evidence-list}, \texttt{agent-belief}, \texttt{update-list} and \texttt{\texttt{recency-list}}, which form the basis for the agents' communications. However, agents have no insights into their link-neighbours' variables. Hence, in NormAN 1.0, the content of communication is exclusively governed by the agent variables of the sender, and sharing is not tailored to the recipients' current beliefs or evidence lists (i.e., no strategic communication).

\subsubsection{Interaction}

Agents interact by sharing pieces of evidence which they store in their \texttt{agent-evidence-list} lists, as described in the process overview (\ref{section:process}).

\subsubsection{Stochasticity}

Many processes and variables are subject to stochasticity. In the world model, the truth of the hypothesis is determined stochastically via \texttt{hypothesis-probability}, and the truth values of each piece of evidence (stored in \texttt{evidence-list}) are stochastically determined by their marginal probabilities (determined by the Bayesian network and stored in \texttt{evidence-probabilities-list}). For the social network, the creation and form of the  Watts-Strogatz social network are subject to stochasticity (\cite{watts1998collective}). Lastly, the agent's behaviour and the deliberation dynamics are also subject to randomness: \texttt{curiosity} determines the chance that an agent will inquire. The variable \texttt{chattiness} determines the probability of sharing a piece of evidence. 

\subsection{Initialisation}

The initialisation process consists of three steps: (1) The world model, (2) the social network and (3) the agents are initialised. Each of these facets is determined by the modeller's/user's
choice of parameters and input data (cf. the complete table of model parameters, Fig. \ref{fig:table-full}).
\begin{itemize}
    \item[(1)]  For the world model, a \texttt{causal-structure},  the Bayesian network, is chosen (see section \ref{section:BNworlds} for a detailed description). Next, the modeller sets which of the propositions (nodes) therein constitutes the hypothesis under discussion and which nodes are considered evidence (for the pre-defined networks included in NormAN v. 1.0, this has already been pre-set, but can be modified with toggles in the UI). The truth values of the hypothesis and the evidence nodes are determined according to the procedure explained above. The resulting values are stored in the global \texttt{evidence-list}.
    \item[(2)] To create the population, NormAN generates \texttt{number-of-agents} agents ($\in [1, 1000]$).  The social network, consisting of a graph of undirected edges between agents, is created. The shape of this network (i.e., complete, wheel, small-world or null) is determined by the modeller's choice of \texttt{social-network}. 
    
    \item[(3)]  Each agent is then initialized. If not otherwise specified, agents start with empty \texttt{agent-evidence-list} lists. In this case, their \texttt{agent-belief} is determined by calculating the unconditional prior degree of belief in the hypothesis node as determined by the Bayesian network that characterises both the world and the agents' perceptions (\texttt{causal-structure}). The value of this computation is stored by the agents' \texttt{initial-belief}. Their \texttt{agent-evidence-list} lists are initialized as empty. However, using the global variable \texttt{initial-draws}, the user can initialize agents with a number of pieces of evidence, even before the procedure starts. In that case, the agent inquires \texttt{initial-draws} times and computes their \texttt{agent-belief} and \texttt{agent-evidence-list} list as specified above. Hence, if \texttt{initial-draws} is 0, agents start with empty \texttt{evidence-lists} and therefore identical priors, whereas in the case of \texttt{initial-draws} $> 0$ they may start with different evidence lists and different beliefs. Note that the agents' \texttt{initial-belief} is always computed as the unconditional prior degree of belief in the hypothesis. Hence, \texttt{initial-belief} does not hinge on the value of \texttt{initial-draws}.
\end{itemize}

An important note for users of the model: The switches in the NetLogo model's user interface determine which parts of the model are reinitialised when the model is set up again, i.e. whenever \texttt{setup} is pressed. The switch \texttt{reset-world-?} resets the truth values of the evidence nodes, while \texttt{reset-social-network-?} resets all agents and the social network that connects them. Finally, \texttt{reset-agents-initial-evidence-?} resets the set of evidence that each agent starts with upon initialisation (only relevant if \texttt{initial-draws}>0).

\subsection{Input data}
\label{section:inputdata}


 The world model's \texttt{causal-structure} is determined by a Bayesian network implemented in the R package bnlearn (more on the connections between NetLogo and R follows in the next subsection below). One key feature of NormAN is its ability to accommodate a broad range of causal structures in the form of BNs. Well-known BNs implemented in R can be found in online repositories (such as the \hyperlink{https://www.bnlearn.com/bnrepository/}{bnlearn repository} from which we retrieved the 'Asia' network, Fig. \ref{fig:asianet} above). Preset networks available in NormAN 1.0 are described further in section \ref{section:BNworlds} below.

\subsection{Submodels}
\label{section:submodels}

NormAN uses two main submodels (as anticipated in Section \ref{section:designconcepts}). First, NormAN is implemented as a NetLogo model (Version 6.2.1, \cite{Wilensky:1999}). NetLogo is a programming language and integrated development environment for agent-based modelling, and as such, NetLogo stores and governs the agents and their behaviour, the social network, the world, and all other parameters determining the model's dynamics (cf. Table \ref{fig:table-full},  Section \ref{section:process}).

As a second submodel, NormAN draws on the R-package bnlearn \cite{bnlearn} to handle all Bayesian belief updating and calculation of marginal, objective evidence probabilities. This submodel consists of a Bayesian network (\texttt{causal-structure}), i.e., the network structure consisting of nodes and edges and its corresponding conditional probability distribution, written as bnlearn R code (embedded in NetLogo, an extension created by \cite{thiele2010netlogo}). Whenever an entity computes a probability, it uses an R query to compute a conditional probability. As an example: to compute their \texttt{agent-belief} based on their \texttt{evidence-list}$=\{E_1, \neg E_3\}$, an agent uses the query  \textit{cpquery(bn, event = (H == 'yes'), evidence = (E1 == 'yes' \&  E3 == 'no'))}. This R query takes as its input the BN specified in \texttt{causal-structure} (called $bn$ in the code) and the truth values from the agent's \texttt{agent-evidence-list}. Hence, bnlearn's $cpquery()$ estimates the conditional probability of an event given evidence using the method specified in the method argument. All Bayesian belief computations are handled by such $cpquery()$ commands, and therefore use the bnlearn submodel. 

A note for users: for technical reasons,  the probabilities returned by $cpquery()$ are approximate estimates. Hence, two identical queries (e.g., by two agents entertaining the same \texttt{agent-evidence-list} list) may lead to slightly different values. This is naturally at odds with NormAN 1.0's `uniqueness' assumption, i.e., that for each set of evidence, there is a unique rational degree of belief. To circumvent the divergent effect of estimation errors, our model implements two different methods: R's $set. seed()$ function or a `repeater'. The modeller can select by choosing a value for the global interface variable \texttt{approximation} (either $seed$ or $repeater$). The $seed$ function uses the R $set.seed()$ function to guarantee that the same random values are produced each time the same code is run. A global \texttt{seed} is chosen for every $cpquery()$, such that every identical query outputs the same probability. Of course, this also means that each query produces the same rounding error. The second method, `repeater', does not fix a seed but rather calculates $cpquery()$ a fixed number  of times (determined by \texttt{repeater}) and returns the mean result.

\section{World Models (Bayes' Nets) Included in NormAN 1.0}
 \label{section:BNworlds}

 We collected a variety of BNs for inclusion. They represent: hypothetical examples of a simple reasoning problem (`Wet Grass’); dependencies between statistical probabilities and base rates (`Sally Clark’); legal idioms and complex (circumstantial) evidence structures (`Vole’); and, medical diagnoses (`Asia’). The full (often rather large) conditional probability tables for these models can be found here: \url{https://osf.io/treaw/?view_only=7d840891096b4572925420945a52c97e}

The `Wet Grass' \cite{Jensen1996} BN model, accessed via `agena.ai.modeller' software (version 9336, www.agena.ai), is a widely used example  BN. The target hypothesis is `Sprinkler', and the event in question is whether or not the sprinkler was left on. The alternative cause (which cannot be directly observed or conditioned on) is `Rain'. These two potential causes of (possibly) wet grass in Holmes's garden (evidence node 'Holmes') are independent and not mutually exclusive. The second evidence node is the  wet grass in Watson's garden (evidence node `Watson'), which only has one possible cause (`Rain'). This model is commonly used to illustrate a common effect node, termed `collider' (here `Holmes'). Conditioning on the collider, i.e., observing the wet grass, creates dependence between the two causes. When `Rain' is made more likely by conditioning on `Watson', i.e., when also observing that Watson's grass is wet, `Sprinkler' becomes less likely (and vice versa). This effect is called `explaining away'.

Network properties: Number of nodes = 8, Number of arcs = 3, Number of parameters = 8, and Average Markov blanket size: 2.

\begin{figure}[H]
  \includegraphics[width=.5\linewidth]{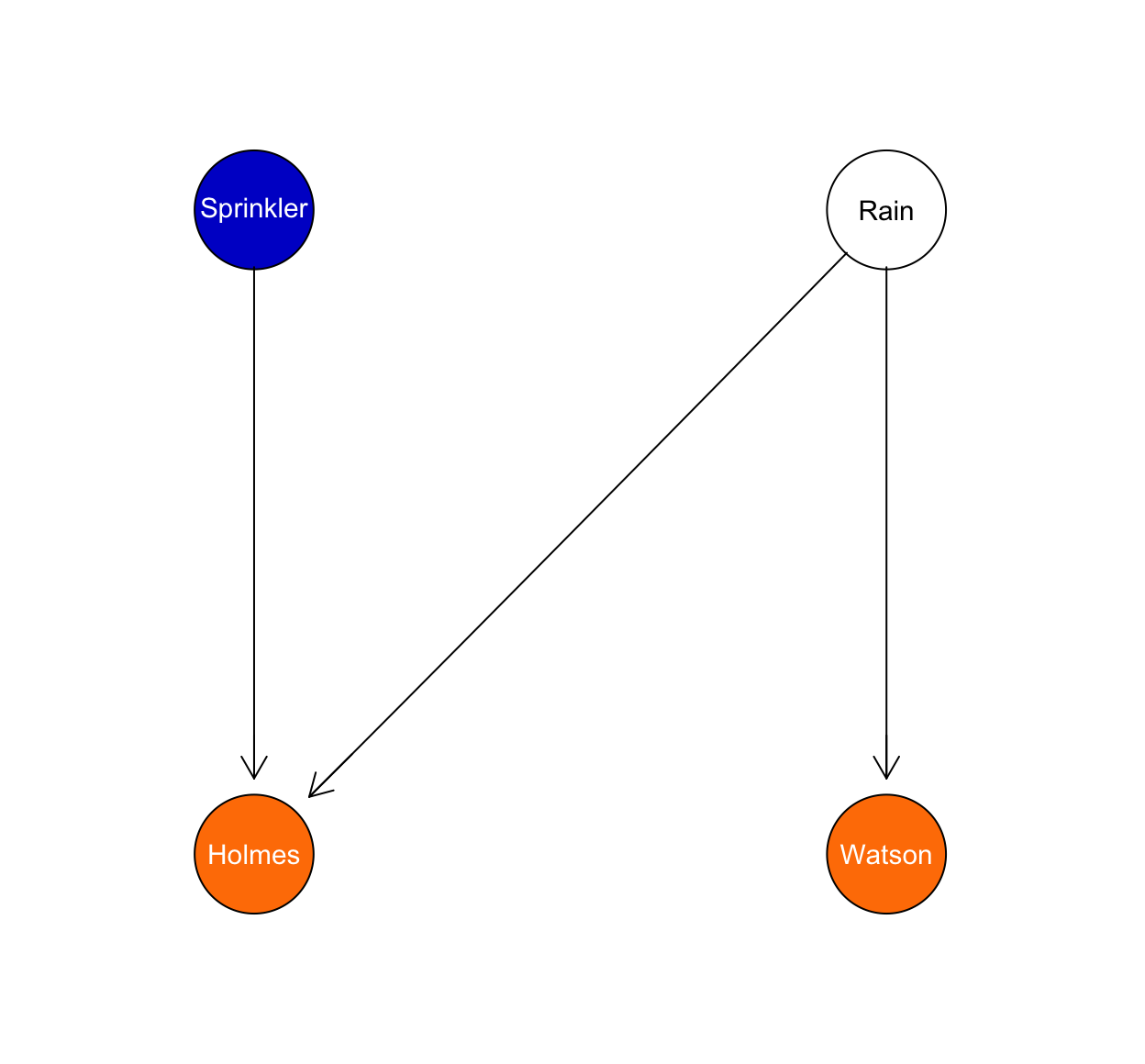}
  \caption{`Wet Grass' \cite{Jensen1996} Bayesian network.}
  \Description{`Wet Grass' BN network }
    \label{fig:wetgrass}
\end{figure}

 The `Sally Clark' \cite{fenton2018risk,fenton2014} BN model was accessed via `agena.ai.modeller' software (version 9336, www.agena.ai). This model is based on a real-world case of wrongful conviction, due to a flawed statistical analysis of the facts in question. The prosecution argued that the infinitesimal odds of two infants both dying of Sudden Infant Death Syndrome (SIDS) was proof of guilt. This  critically ignored (a) the dependence between the two deaths as the infants were related (and therefore both could have an underlying condition) and (b) the base rate for parental murder of an infant (as opposed to the base rate for SIDS). It was an important but clear flaw in the original case that the alternative base rate was not considered; and it was not recognised that the likelihood of two related infant deaths caused by SIDS was much greater than the likelihood of two related infant deaths caused by parental murder (for further discussion see \cite{fenton2014}). The BN model represents a more appropriate reconstruction  of the real-world facts of the case. The target hypothesis is `Guilt’ (whether or not Sally Clark was guilty of murder), and the four observable evidence nodes are whether or not Child A and/or Child B had bruising and whether or not Child A and/or Child B had signs of an underlying disease. This BN illustrates how to correct the identified errors from the original analysis. Firstly, the structure of the BN reflects the dependence between the cause of death of Child A and Child B. Secondly, the conditional probabilities of the cause nodes reflect the base rate of murder for Child A, and for Child B the probability of murder is closely aligned to the likelihood of murder for Child A (i.e., this dependence is reflected in structure and parameterisation). Finally, the `Findings' node is a deterministic node (states only 1 or 0), meaning its values are derived from the states of its parent nodes. In this case, the Findings node represents the three possible conclusions relevant to a guilty finding, either both infants were murdered, one infant was murdered, or neither was murdered. The target hypothesis `Guilt’ is also a deterministic node and combines the first two likelihoods of the ‘Findings’ node as both of these states are probative of a guilty verdict. Note: the states of both cause nodes were changed from `SIDS' and `Murder' to `no' and `yes' respectively. 

Network properties: Number of nodes = 8, Number of arcs = 8, Number of parameters = 22, and Average Markov blanket size: 2.0.

\begin{figure}[H]
  \centering
  \includegraphics[width=.5\linewidth]{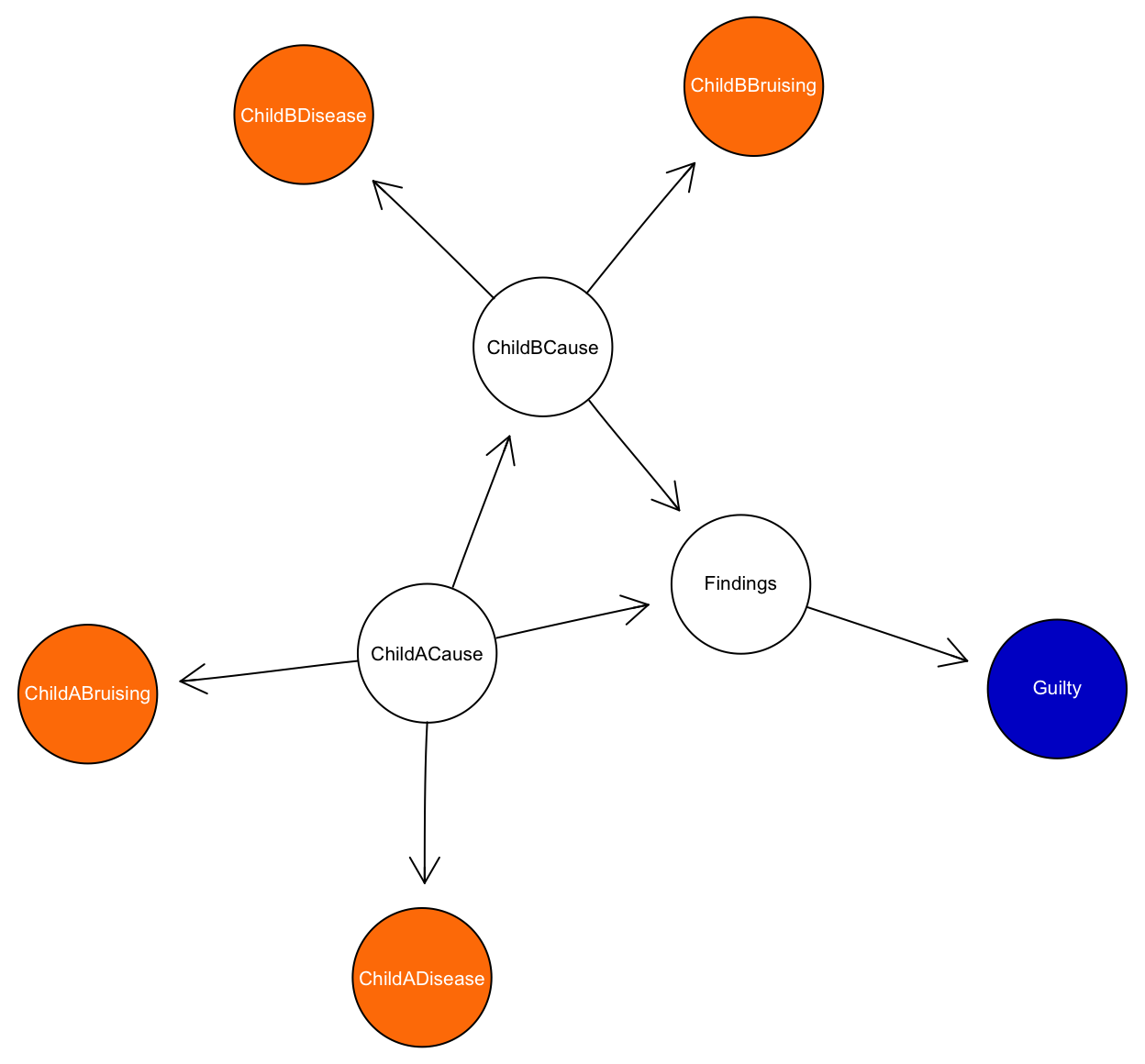}
  \caption{`Sally Clark' \cite{fenton2018risk} Bayesian network.}
  \Description{`Sally Clark' BN network.}
    \label{fig:sallyclark}
\end{figure}

The `Vole' BN model \cite{fenton2018risk} was accessed via `agena.ai.modeller' software (version 9336, www.agena.ai). This model is based on a fictional case: the Agatha Christie play `Witness for the Prosecution’, where the character Vole is accused of killing French. It is  based on Lagnado’s BN model of that case (see \cite{Lagnado2011}), revised by \cite{fenton2013general} to incorporate and illustrate how idioms (a small number of basic causal structures) can be used to present complex legal arguments (e.g., a murder case) in a BN. Idioms included in this model are evidence idioms (denoted `E’), evidence-accuracy idioms (denoted `A’), idioms that deal with motive (denoted `M’) and opportunity (whether Vole was present), an alibi evidence idiom (in this case E2 and A2) and explaining away idioms (in this case the constraint node). The target hypothesis is `H0’ (whether or not Vole was guilty) and there are seven observable evidence nodes (all denoted `E’): E1 = Maid testifies that Vole was present; E2 = Vole alibi (testifies that he was not present); E3 = Blood matches French; E4 = Blood matches Vole; E5 = Vole shows scar; E6 = Letters as evidence of lover; and, E7 = Romaine testifies that Vole admitted guilty. In terms of the construction of the CPTs, a hypothesis of 'guilty' is supported by E1, E3 and E7 being set to true, and  `not guilty'  is supported by E2, E4, E5 and E6 being set to true. Finally, the auxiliary node is another example of a deterministic node and is used here to enforce the constraint that H2 and H4 are mutually exclusive and one of them must be true. For a further discussion of the model and the use of idioms see \cite{fenton2013general}.

Network properties: Number of nodes = 22, Number of arcs = 25, Number of parameters = 63, and Average Markov blanket size: 2.82.

\begin{figure}[H]
  \centering
  \includegraphics[width=.5\linewidth]{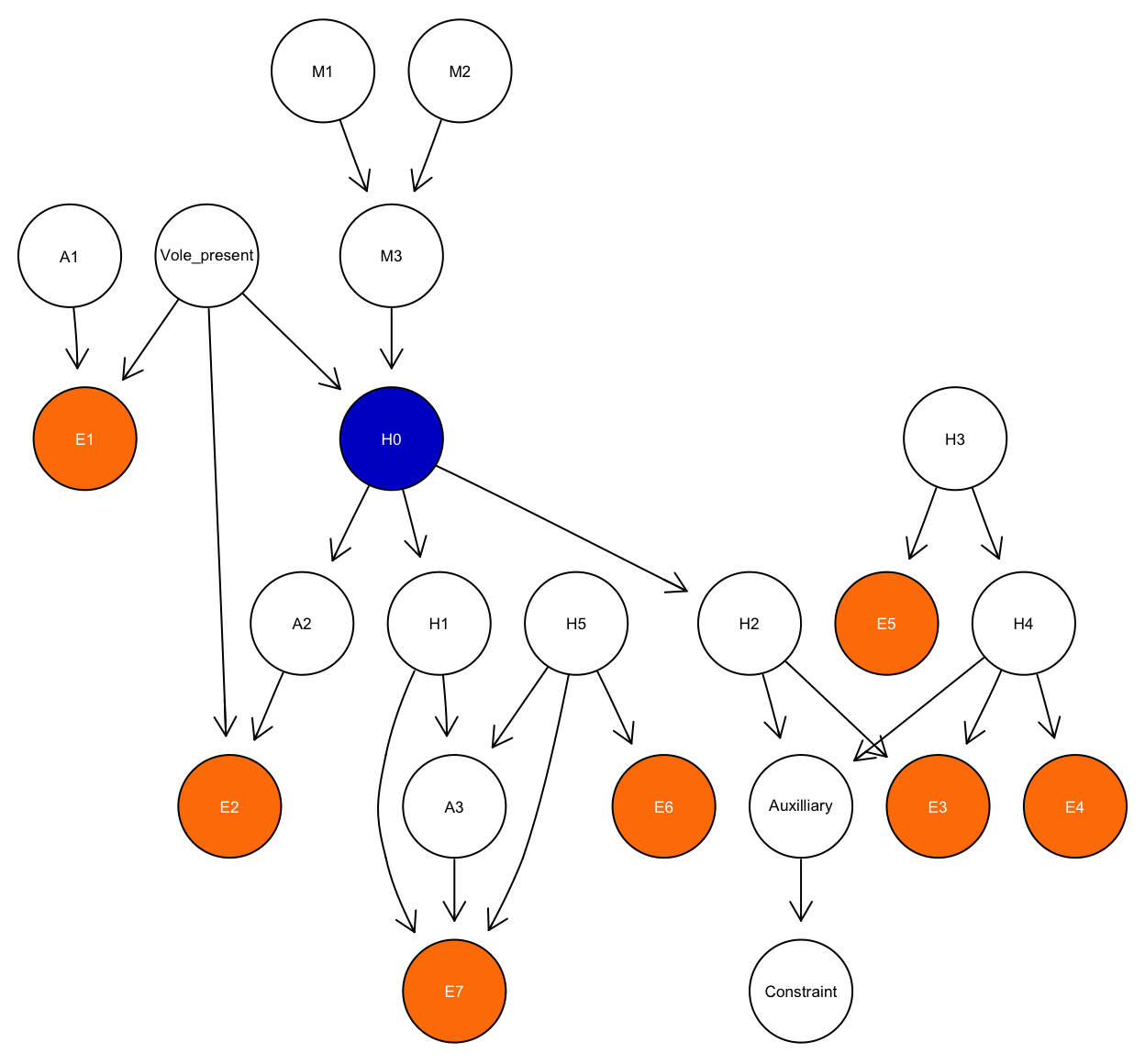}
  \caption{`Vole' \cite{fenton2018risk, fenton2013general} Bayesian network.}
  \Description{`Vole' Bayesian network.}
    \label{fig:vole}
\end{figure}

The `alarm' network is a Bayesian network designed to provide an alarm message system for patient monitoring. It originates in Beinlich et al.'s ``The ALARM Monitoring System: A Case Study with Two Probabilistic Inference Techniques for Belief Networks'' \cite{beinlich1989alarm}. Its bnlearn instantiation is freely available in the bnlearn Network Repository \url{https://www.bnlearn.com/bnrepository/discrete-medium.html#alarm}.

Network properties: Number of nodes = 37, Number of arcs = 46, Number of parameters = 509, Average Markov blanket size = 3.51, Average degree = 2.49, and Maximum in-degree = 4.

\begin{figure}[H]
  \centering
  \includegraphics[width=.5\linewidth]{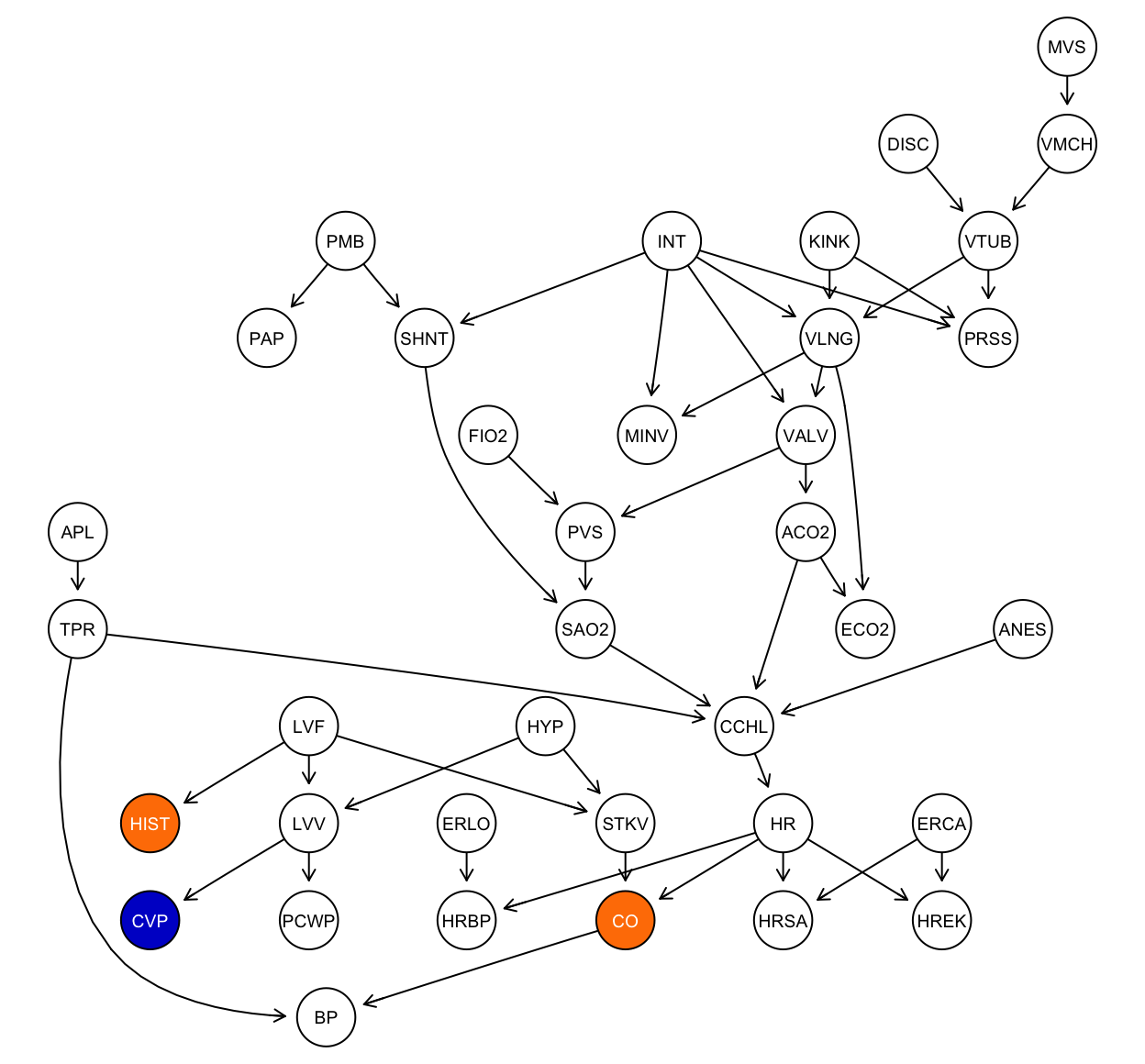}
  \caption{`Alarm' \cite{Beinlich1989} Bayesian network.}
  \Description{`Alarm' BN network.}
  \label{fig:alarm}
\end{figure}

We also included a hand-crafted network that we found useful for comparison purposes. The 'Big Net' is a generic network that represents the following causal structure: A Boolean hypothesis node \texttt{A} has three direct (Boolean) descendants (\texttt{B}, \texttt{C}, \texttt{D}), where the truth of \texttt{B} is fairly diagnostic evidence for \texttt{A}, the truth or falsity of \texttt{C} is non-diagnostic for \texttt{A}, and the truth of \texttt{D} is evidence against \texttt{A}. Each of these evidence nodes has three children (the children of \texttt{B} are \texttt{one}, \texttt{two}, \texttt{three}, the children of \texttt{C} are \texttt{four}, \texttt{five}, \texttt{six} and the children of \texttt{D} are \texttt{seven}, \texttt{eight}, \texttt{nine}). Each piece has a slightly different diagnostic value in relation to its parents. 

Network properties: Number of nodes = 13, Number of arcs = 12. The conditional probability distribution is: 
\small 

 \begin{itemize}
     \item[]$P(\texttt{A})=0.5$
     \item[] $P(\texttt{B}|\texttt{A})=0.9$; $P(\texttt{B}|\neg \texttt{A})=0.1$
     \item[] $P(\texttt{C}|\texttt{A})=0.5$; $P(\texttt{C}|\neg \texttt{A})=0.5$
     \item[] $P(\texttt{D}|\texttt{A})=0.1$; $P(\texttt{D}|\neg \texttt{A})=0.9$
     \item[] $P(\texttt{one}|\texttt{B})=0.9$; $P(\texttt{one}|\neg \texttt{B})=0.1$
     \item[] $P(\texttt{two}|\texttt{B})=0.8$; $P(\texttt{two}|\neg \texttt{B})=0.2$
     \item[] $P(\texttt{three}|\texttt{B})=0.7$; $P(\texttt{three}|\neg \texttt{B})=0.3$
     \item[] $P(\texttt{four}|\texttt{C})=0.9$; $P(\texttt{four}|\neg \texttt{C})=0.1$
     \item[] $P(\texttt{five}|\texttt{C})=0.8$; $P(\texttt{five}|\neg \texttt{C})=0.2$
     \item[] $P(\texttt{six}|\texttt{C})=0.7$; $P(\texttt{six}|\neg \texttt{C})=0.3$
     \item[] $P(\texttt{seven}|\texttt{D})=0.9$; $P(\texttt{seven}|\neg \texttt{D})=0.1$
     \item[] $P(\texttt{eight}|\texttt{D})=0.8$; $P(\texttt{eight}|\neg \texttt{D})=0.2$
     \item[] $P(\texttt{nine}|\texttt{D})=0.7$; $P(\texttt{nine}|\neg \texttt{D})=0.3$
 \end{itemize}
 
 \normalsize

\begin{figure}[H]
  \centering
  \includegraphics[width=.5\linewidth]{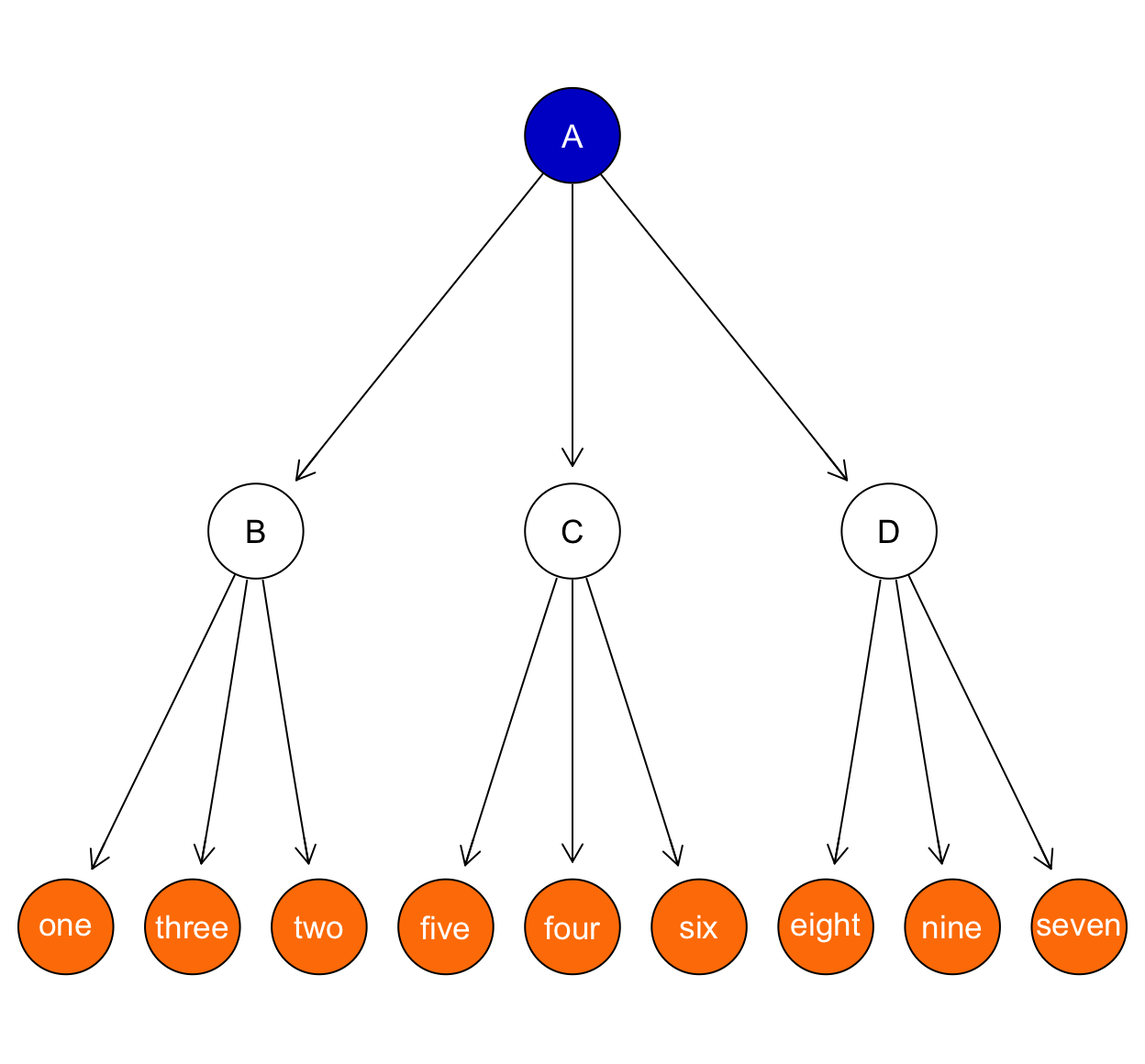}
  \caption{`Big Net' Bayesian network.}
  \Description{`Big Net' BN network.}
  \label{fig:bignet}
\end{figure}

Lastly, we included a smaller hand-crafted network, which follows the same rationale as `Big Net', called `Small Net'.  `Small Net' is a generic network that represents the following causal structure: A Boolean hypothesis node \texttt{V} has two direct (Boolean) descendants (\texttt{CS}, \texttt{VT}), where the truth of \texttt{CS} and \texttt{VT} are both rather diagnostic evidence in for \texttt{V}. Both of these evidence nodes have two children: the children of \texttt{CS} are \texttt{I} and \texttt{M};  both \texttt{I} and \texttt{M} are evidence in favour of \texttt{CS}. \texttt{VT} has two children:  \texttt{RS} and \texttt{WHO}. Whereas \texttt{WHO} (when true) is supportive of \texttt{VT}, \texttt{RS} is evidence against \texttt{VT}. We select the leaf nodes as evidence in our simulations. 

Network properties: Number of nodes = 7, Number of arcs = 6. The conditional probability distribution is: 

\begin{itemize}
     \item[]$P(\texttt{V})=0.5$
     \item[] $P(\texttt{CS}|\texttt{V})=0.9$; $P(\texttt{CS}|\neg \texttt{V})=0.1$
     \item[] $P(\texttt{VT}|\texttt{V})=0.8$; $P(\texttt{VT}|\neg \texttt{V})=0.2$

     \item[] $P(\texttt{I}|\texttt{CS})=0.7$; $P(\texttt{I}|\neg \texttt{CS})=0.3$

    \item[] $P(\texttt{M}|\texttt{CS})=0.8$; $P(\texttt{M}|\neg \texttt{CS})=0.2$

    \item[] $P(\texttt{RS}|\texttt{VT})=0.2$; $P(\texttt{RS}|\neg \texttt{VT})=0.8$
    \item[] $P(\texttt{WHO}|\texttt{VT})=0.8$; $P(\texttt{WHO}|\neg \texttt{VT})=0.2$
      
\end{itemize}

\begin{figure}[H]
  \centering
  \includegraphics[width=.5\linewidth]{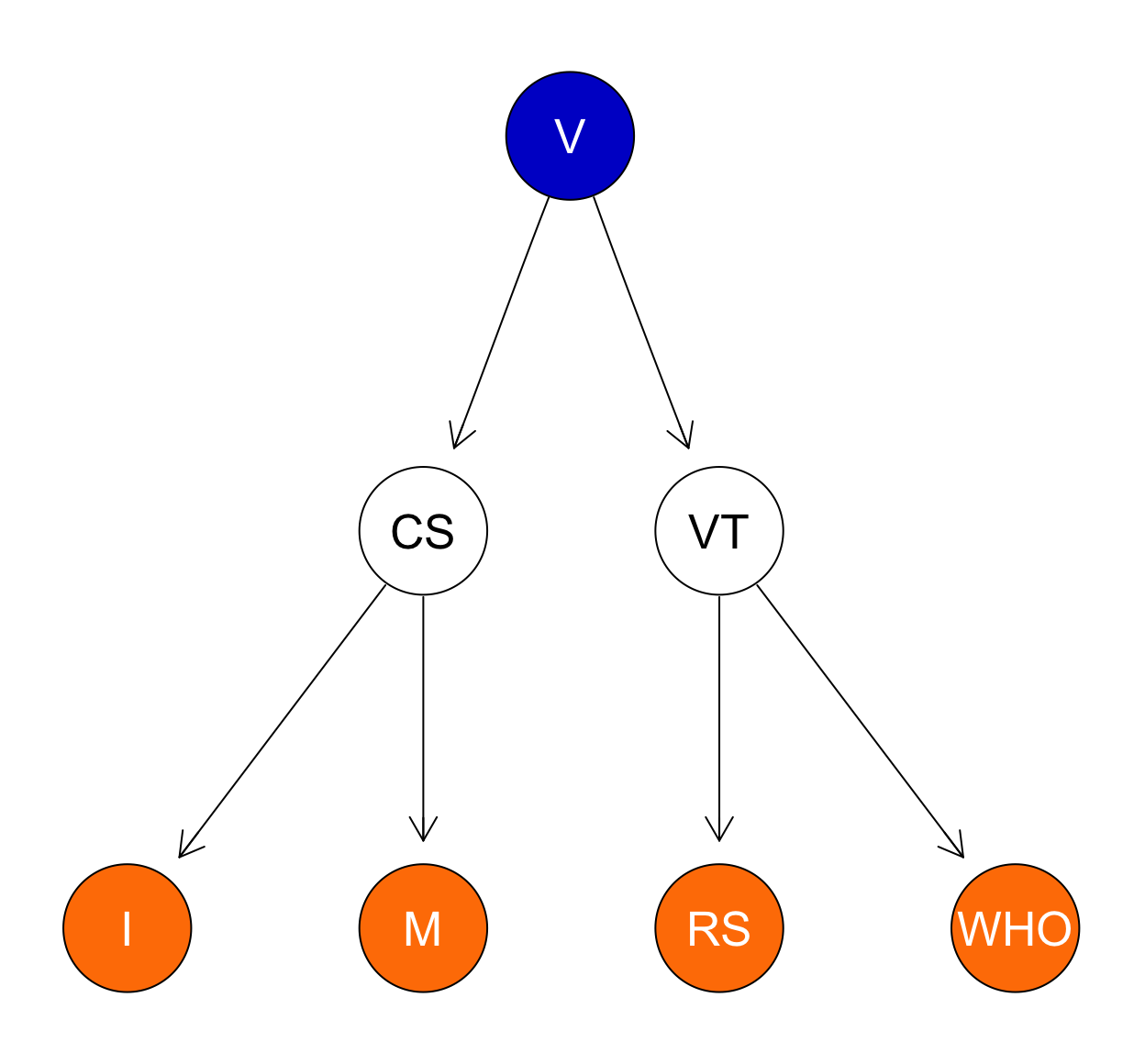}
  \caption{`Small Net' Bayesian network.}
  \Description{`Small Net' Bayesian network.}
  \label{fig:smallnet}
\end{figure}

\newpage
\section{Simulation Details}

The following contains additional information for the case studies.

\subsection{Case Study 1}
\label{section:case1}
\subsubsection{Setting Out the Research Question}

For our basic simulation, we used the 'Big Net'  network from the Norman version 1.0 model presets (see \ref{fig:bignet}). To set up the simulation we initialised 800 model runs, up to a maximum time step of 25 (0-25). The mean belief of the agent population was recorded at each time step, 0 reflecting the pre-deliberation average degree of belief within the group (given an initial evidence draw of 1 and a prior of 0.5). There were 8 conditions, 4 (number of agents: 10,50,100 and 500) X 2 (network type: small world and complete), with 100 runs in each condition. After the simulation was run, groups were classified as `for' and `against' based on the initial direction of belief at time step 0, with `for' being >0.5 and `against' being <0.5. Total group (N=800) means and standard error range were calculated for the `for' and `against' groups at time steps 0 and 25 to produce a bar chart, shown in Figure \ref{fig:shift}. Means and standard error range were calculated for each run, for each of the 8 conditions, at all time steps (number of time steps =26), shown in Figure \ref{fig:shift_runs}.

This simulation was also repeated with the Asia/Lung Cancer BN network, shown in Figure \ref{fig:asianet}. Again, after the simulation was run, groups were classified as `for' and `against' based on initial direction of belief at time step 0, with `for' being >0.5 and `against' being <0.5. Means and standard error range were calculated for each run, for each of the 8 conditions, at all time steps (number of time steps = 26), shown in Figure \ref{fig:shift_runs_asia}.

\subsubsection{Simulation Parameters}

The following list specifies the actual parameters used in the simulations summarised in Figures \ref{fig:shift} and \ref{fig:shift_runs}: 

Model variables:["number-of-agents" 10 50 100 500];["network" "complete" "small-world"];["causal-structure" "big net"];["initial-draws" 1];["maximum-draws" 1];["share" "random"];["Seed" 2];["chattiness" 0.5];["conviction-threshold" 0.5];["curiosity" 0];
["k" 2];["rewiring-probability" 0.2];["hypothesis-probability" 0.5];["show-me" false];["stop-at-full-information" false]; and,["approximation" "seed"]. Measures/Reporters:mean [agent-belief] of turtles; optimal-posterior; and, evidence-list. Repetitions: 100. Time limit: 25.

The actual data produced by these simulation runs can be found on OSF: \newline
\url{https://osf.io/6atr2/?view_only=dadfa771b5de4c86b3408c4bbba40e67}. 

\subsubsection{Supplementary Figures}

Figure \ref{fig:shift_runs_valence} is a further illustration of aspects of the shift to extremity \ref{section:case1}. Fig. \ref{fig:shift_runs_valence} takes the same data as Fig.  \ref{fig:shift_runs} but rather than the splitting runs by whether the group initially leaned `for' or 'against' the target hypothesis, individual runs are split by whether or not the optimal posterior of a given run is `for' or `against' the target hypothesis (termed `valence'). Cases where the optimal prior was neutral (exactly equal 0.5) were excluded. Then, as before, means and standard error range were calculated for each run, for each of the 8 conditions, at all time steps (number of time steps =26). This illustrates the connection between the `sample' of arguments present in a deliberating group, and the underlying structure of the world, which will determine what evidence could be available in principle. 

\begin{center}
  \centering
  \includegraphics[keepaspectratio=true,scale=0.16]{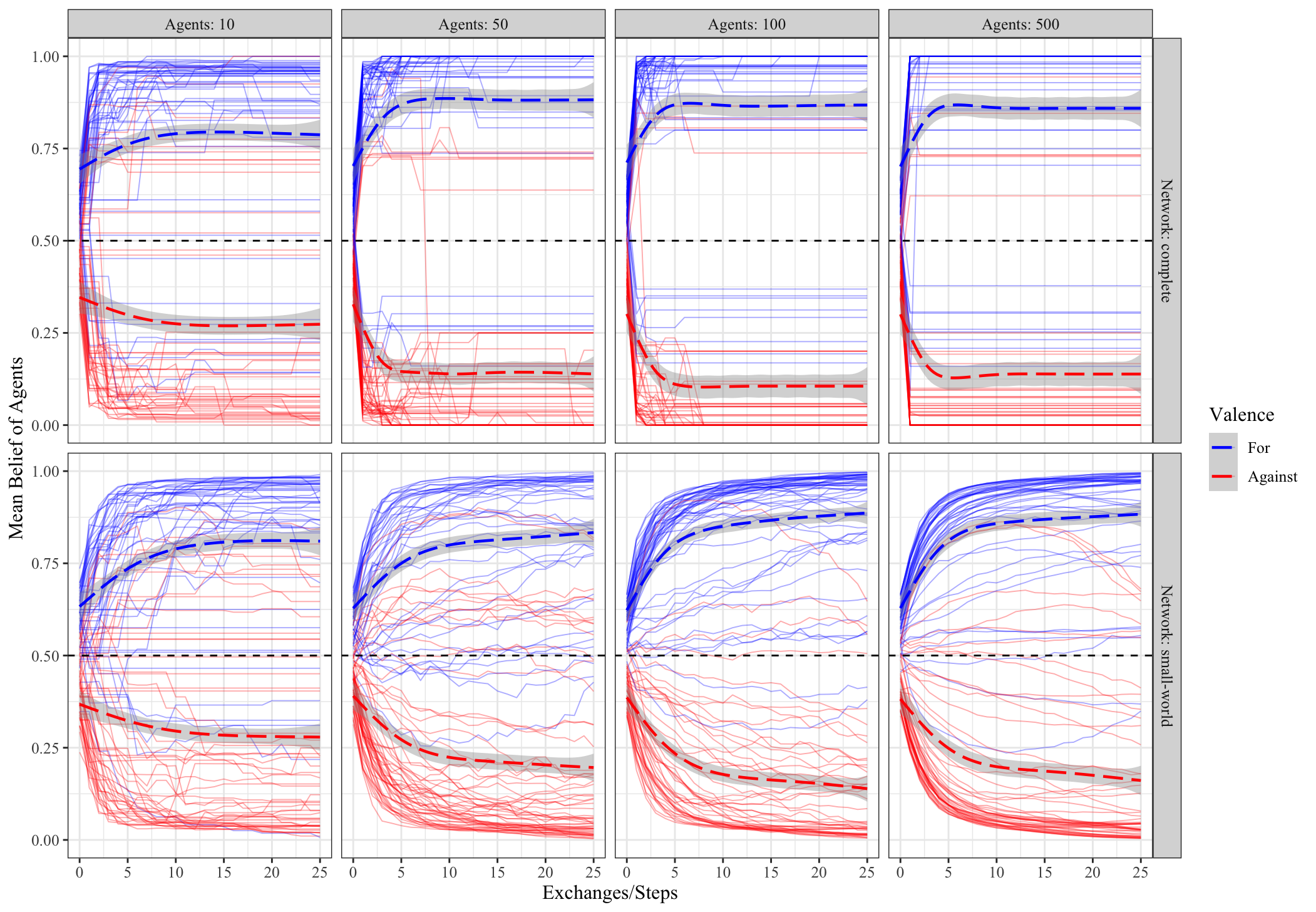}
  \captionof{figure}{As in Fig. \ref{fig:shift_runs} main text, the Figure plots the trajectories of mean group beliefs across time (thin lines) and the average of those means (dashed lines) with standard error (grey shaded area). Again, the top row shows the simulation results for complete networks of various sizes (10,50,100, and 500 agents) and the bottom row shows the same dynamics in small world networks with corresponding numbers of agents. Note that the `pre-evidence' neutral point in these simulations is .5 (base rate of hypothesis), as indicated by the black dotted line. Runs are split by whether or not the optimal posterior of a given run is `for' (blue) or `against' (red) the target hypothesis (termed `valence').}
  \Description{This is the Shift to Extremity}
  \label{fig:shift_runs_valence}
\end{center}

Finally, Fig. \ref{fig:shift_runs_asia} examines the  shift to extremity seen in  Figure \ref{fig:shift_runs} above using the Asia/Lung Cancer Network of Fig. \ref{fig:asianet} in the main text. For this supplementary experiment the following model parameters were used: ["number-of-agents" 10 50 100 500]; ["network" "complete" "small-world"]; ["causal-structure" "asia"]; ["initial-draws" 1]; ["maximum-draws" 1]; ["share" "random"]; ["Seed" 2]; ["chattiness" 0.5]; ["conviction-threshold" 0.5]; ["curiosity" 0]; ["k" 2]; ["rewiring-probability" 0.2]; ["hypothesis-probability" 0.5]; ["show-me" false]; ["stop-at-full-information" false]; and,["approximation" "seed"]. Measures or reporters were:mean [agent-belief] of turtles; optimal-posterior; and, evidence-list. This simulation, again, used 100 repetitions and a time limit of 25. The actual data produced by these simulation runs can be found on OSF:
\newline{\url{https://osf.io/6atr2/?view_only=dadfa771b5de4c86b3408c4bbba40e67}.}

\begin{center}
  \centering
  \includegraphics[keepaspectratio=true,scale=0.16] {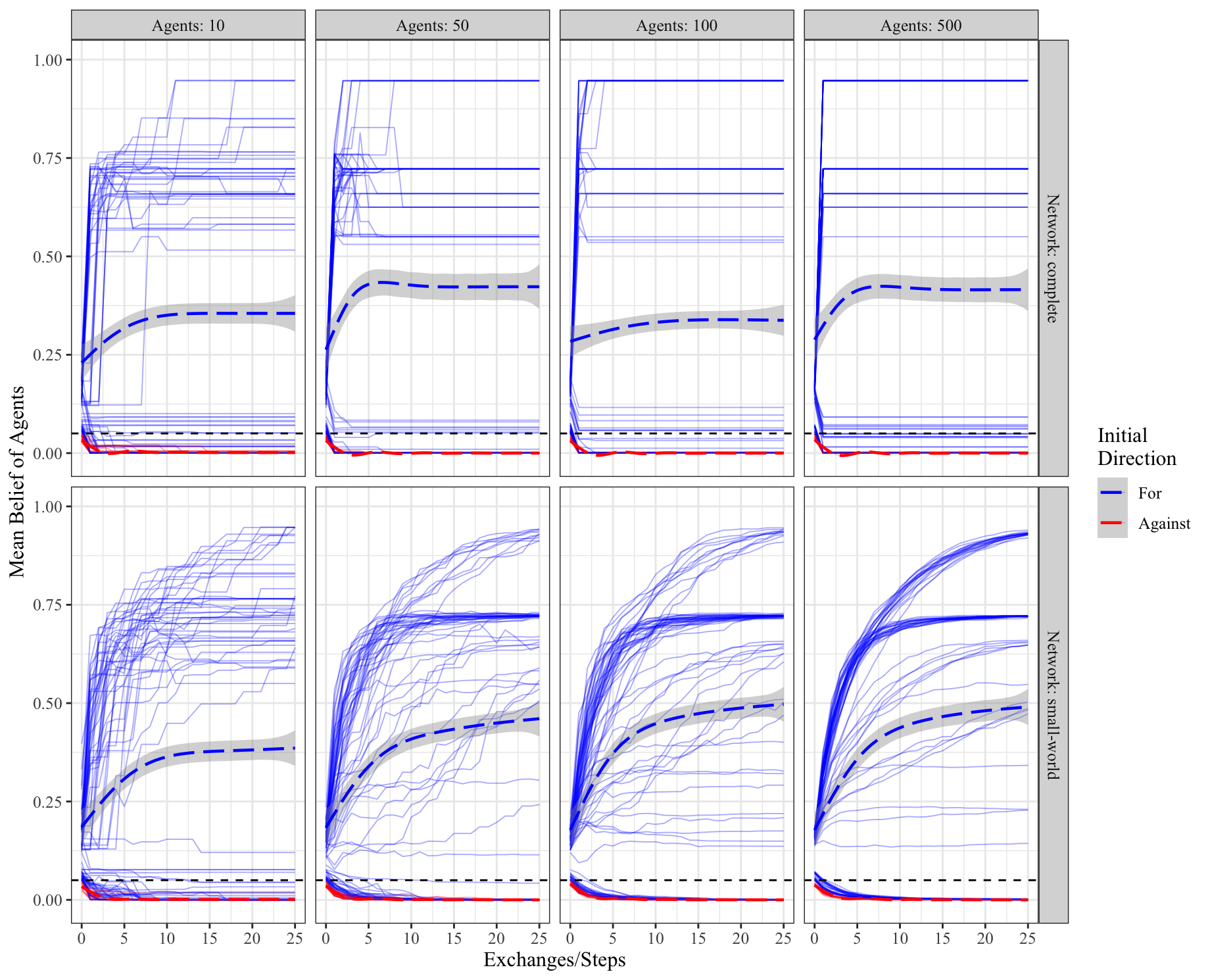}
  \captionof{figure}{Simulation using the Asia/Lung Cancer Network \ref{fig:asianet}. Shown are the trajectories of mean group beliefs across time (thin lines) and the average of those means (dashed lines) with standard error (grey shaded area), split by whether the group initially leaned `for' (blue) or `against' (red). The top row shows the simulation results for complete networks of various sizes (10,50,100, and 500 agents). The bottom row shows the same dynamics in small world networks with corresponding numbers of agents. Note that the `pre-evidence' neutral point in these simulations is .05 (base rate of hypothesis, in this instance the base rate of lung cancer), as indicated by the black dotted line.  Note that the `pre-evidence' neutral point in these simulations is .055 (base rate of lung cancer), not .5. Results could be transformed to a .5 initial mid-point without loss.}
  \Description{Shift to Extremity, Asia Network}
  \label{fig:shift_runs_asia}
\end{center}

Fig. \ref{fig:shift_runs_asia} shows the same `shift to extremity' characteristics described in \ref{section:case1}. Note that in advance of receiving any evidence about the world, agents have a prior of .055, so leaning `for' or `against' is determined relative to that. This is a function of the fact that for NormAN 1.0, agents' subjective model of the world is the same as the objective BN. The values shown could be re-scaled to show what would obtain if agents' pre-evidence prior was .5, without affecting anything about the pattern seen.   

\subsection{Case Study 2}
\label{section:case2}

\subsubsection{Research Question and Simulation Parameters}
 For the second case study, we simply showcased single, representative model runs using three different sharing rules (\texttt{RANDOM-SHARE}, \texttt{RECENT-SHARE}, \texttt{IMPACT-SHARE}), as visualized in Figures \ref{fig:arguments}, \ref{fig:beliefscomplete}, \ref{fig:smallworlddynamics}. The purpose was to highlight how each of these sharing rules may lead to convergence or to scattering of the agents' beliefs (\texttt{agent-belief}), as well as to illustrate the sharing frequency of each piece of evidence.
 
 The simulations' initial conditions were the same for each run, except for the social network and the sharing rule. Each agent is initialized as knowing one piece of evidence, and throughout the simulation, they only learn through communication. The exact parameter values are: \texttt{causal-structure}= Vole (see Appendix, Section \ref{section:BNworlds} for a description), \texttt{chattiness}=0.5, \texttt{conviction-threshold}=0, \texttt{curiosity}=0, \texttt{initial-draws}=1, \texttt{max-draws}=1, \texttt{social-network}=complete/small-world, \texttt\texttt{number-of-agents}=50. The prior \texttt{initial-belief}$\approx 0.3$.
 
\subsubsection{Supplementary Figures}

\begin{figure}[H]

         \centering
         \includegraphics[width=0.48\textwidth]{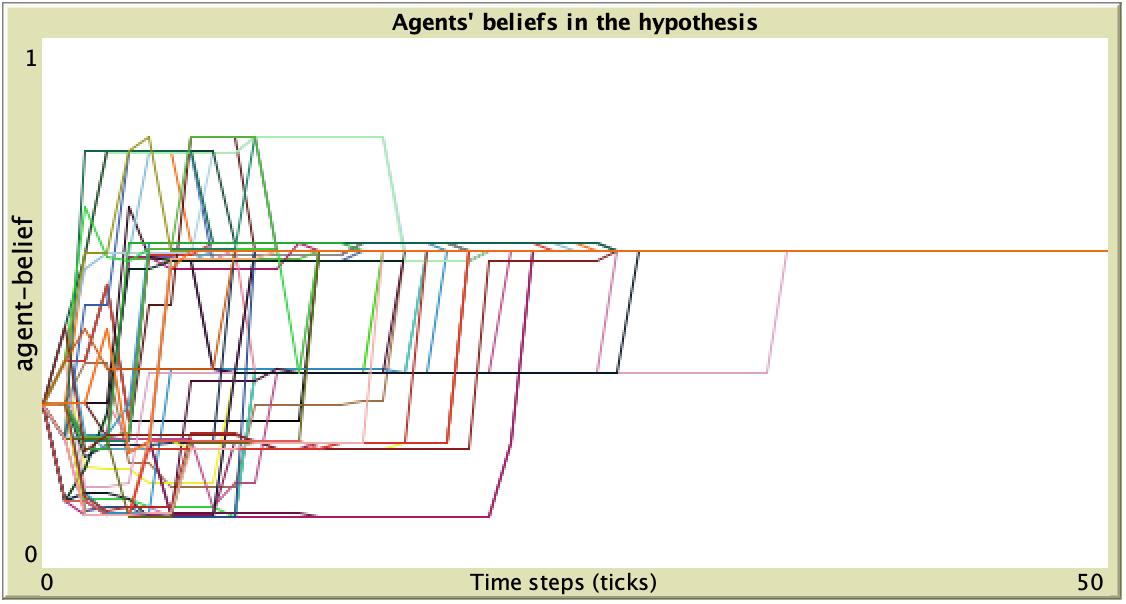}
         \includegraphics[width=0.48\textwidth]{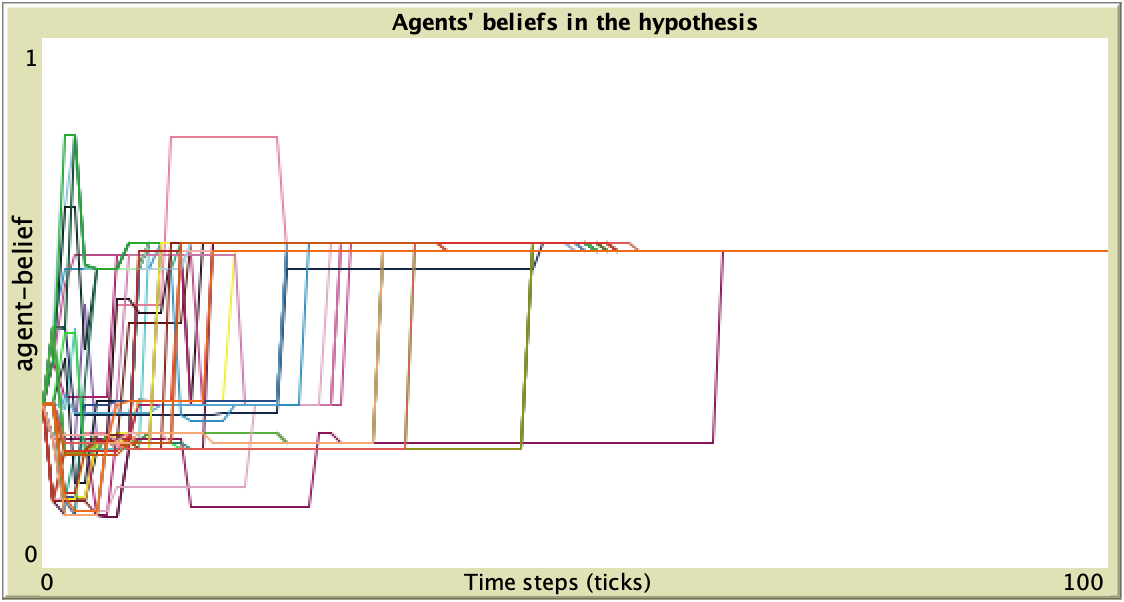}

     
         \centering
         \includegraphics[width=0.48\textwidth]{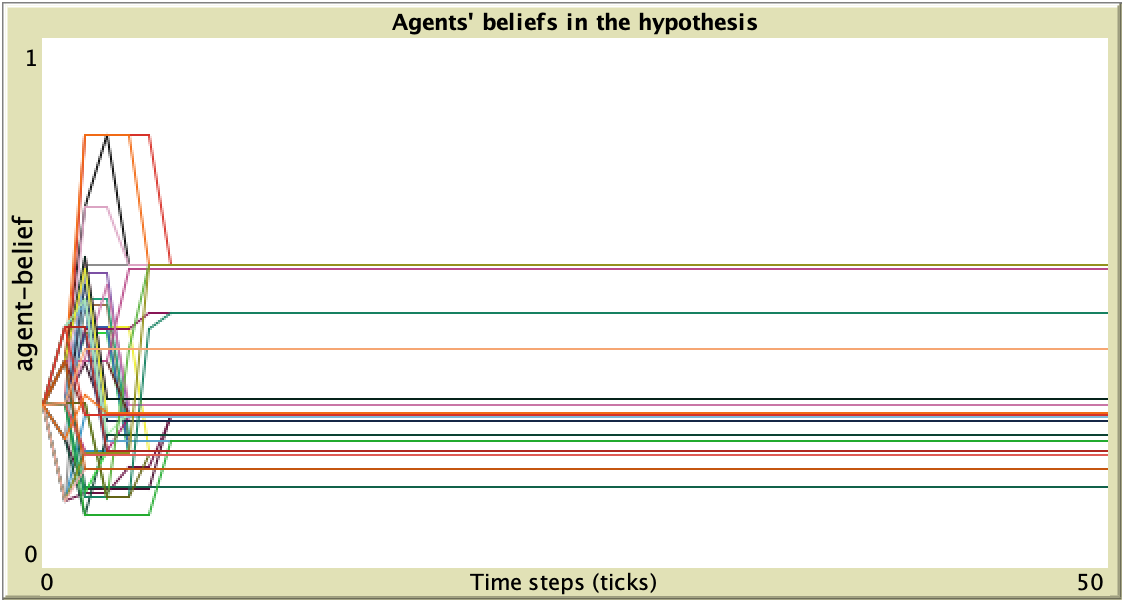}
         \caption{The beliefs of a population over time, for three sharing rules: random (top left), recency (top right) and impact (below). This time, the agents are connected via a small-world network. Parameter values: \texttt{causal-structure}= Vole (see Appendix, Section \ref{section:BNworlds} for a description), \texttt{chattiness}=0.5, \texttt{conviction-threshold}=0, \texttt{curiosity}=0, \texttt{initial-draws}=1, \texttt{max-draws}=1, \texttt{social-network}=small-world,\texttt{rewiring-probability}=0.2, \texttt{k}=2,  \texttt\texttt{number-of-agents}=50. The prior \texttt{initial-belief}$\approx 0.3$.}
         \label{fig:smallworlddynamics}
\end{figure}

Supplementary Fig. (\ref{fig:smallworlddynamics}) showcases a model run of each sharing rule, carried out over a small-world network. The parameters are in the caption. 

\section{Getting Started}
\begin{landscape}
\begin{figure}
    \centering
    \includegraphics[width=20cm]{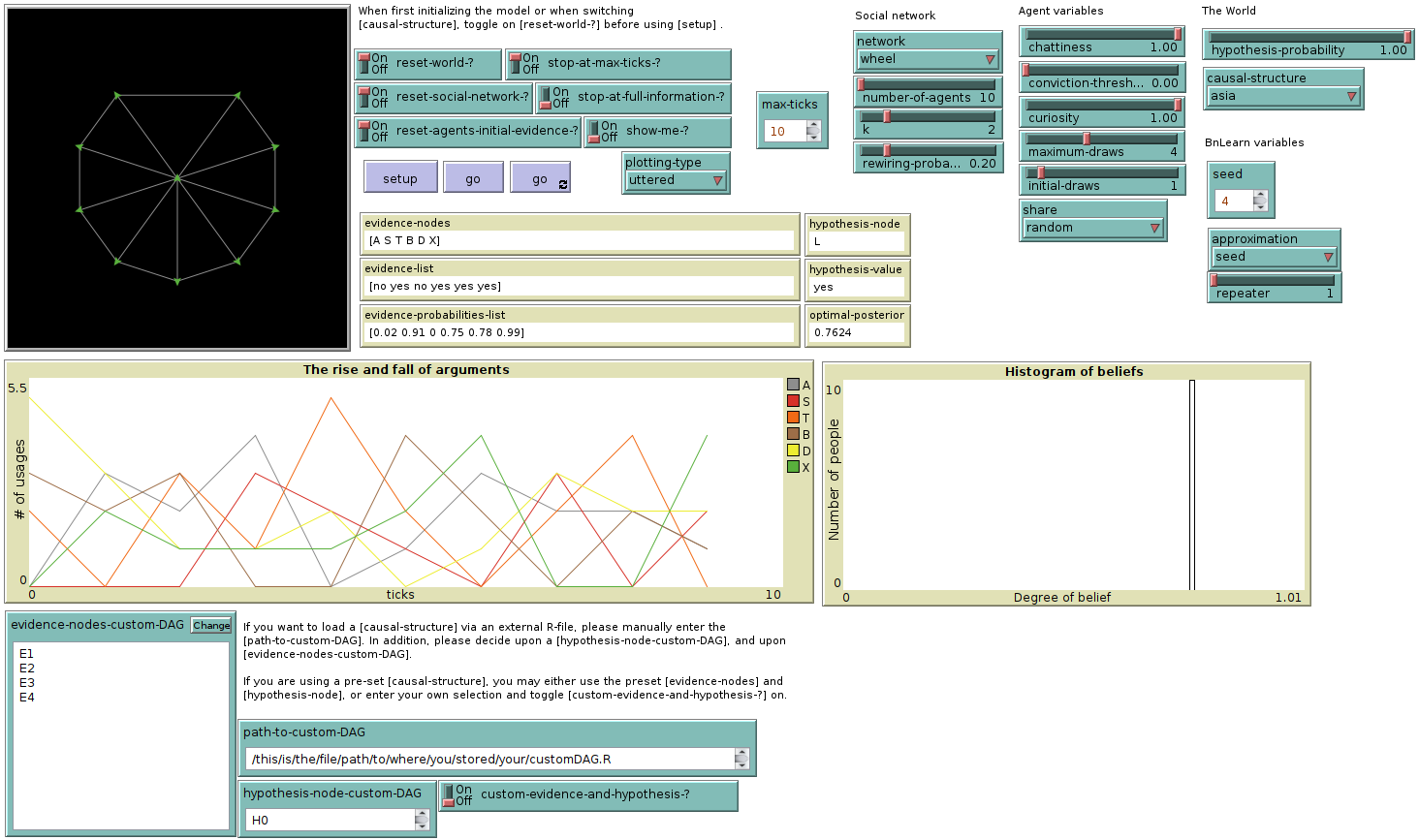}
    \caption{User interface of NormAN 1.0.}
    \label{fig:interface} 
\end{figure}
\end{landscape}

For users who want to quickly start the NetLogo model, here are the relevant steps (all mentioned variables are in the user interface, cf. Fig. \ref{fig:interface}): 

\begin{enumerate}
\item[(1)] When first opening the model, make sure \texttt{reset-world-?} and \texttt{reset-social-network-?} are \textit{on}.
\item[(2)] The World: choose a \texttt{causal-structure} (chooser).
\item[(3)] The social network: choose a \texttt{social-network} (chooser) and a \texttt{number-of-agents } (slider).
 
\item[(4)] Click \texttt{setup}: the social network will appear in the interface, the right bottom monitor will show a histogram of agent-beliefs and the middle output will show which pieces of evidence are true
\item[(5)] Press \texttt{go} to start the simulation 
\item[(6)] If you would like to monitor what each agent does each round (in the command center), toggle \texttt{show-me-?} \textit{on}.
\end{enumerate}

\label{section: Getting Started}

\section{Online Resources}

\label{section:online}

The code for NormAN version 1.0 can be found here:
\url{https://github.com/NormAN-framework/base-model}.

\end{document}